\documentclass[epj]{svjour}

\usepackage{graphics}

\pdfoutput=1  

\usepackage[T1]{fontenc}  
\usepackage{amsmath}
\usepackage{amssymb}
\usepackage{graphicx}
\usepackage{epstopdf}
\usepackage{bm}
\usepackage{epsfig}
\usepackage{graphics}
\usepackage{xspace}
\usepackage{hyperref}
\usepackage{slashed}
\usepackage{xcolor}
\usepackage{longtable}
\usepackage[small]{subfigure}
\usepackage[normalem]{ulem}
\usepackage{multirow}

\newcommand{\nl}{n_{\ell}}

\newcommand{\mbar}{\overline{m}}
\newcommand{\MSR}{\mathrm{MSR}}

\newcommand{\MSb}{\overline{\mathrm{MS}}}
\newcommand{\df}{{\rm d}}

\newcommand{\LQCD}{\Lambda_\mathrm{QCD}}

\begin{document}
\title{Calibrating the Na\"\i ve Cornell Model with NRQCD}

\author{Vicent~Mateu\inst{1,2}\thanks{vmateu@usal.es} \and Pablo~G.~Ortega\inst{1}\thanks{pgortega@usal.es} 
\and David~R.~Entem\inst{1}\thanks{entem@usal.es} \and Francisco~Fern\'andez\inst{1}\thanks{fdz@usal.es}
}                      
\institute{Departamento de F\'isica Fundamental and IUFFyM, Universidad de Salamanca.\\
Plaza de la Merced S/N, E-37008 Salamanca, Spain.\vspace*{0.15cm} \and
Instituto de F\'isica Te\'orica UAM-CSIC. C/ Nicol\'as Cabrera 13-15,\\
Campus de Cantoblanco, E-28049 Madrid, Spain.}
\date{Received: date / Revised version: date}
\abstract{
Along the years, the Cornell Model has been extraordinarily successful in describing hadronic 
phenomenology, in particular in physical situations for which an effective theory of
the strong interactions such as NRQCD cannot be applied. As a consequence of its achievements,
a relevant question is whether its model parameters can somehow be related to fundamental
constants of QCD. We shall give a first answer in this article by comparing the predictions of
both approaches.
Building on results from a previous study on heavy meson spectroscopy, we
calibrate the Cornell model employing NRQCD predictions for the lowest-lying bottomonium
states up to N$^3$LO, in which the bottom mass is varied within a wide range. We find that the
Cornell model mass parameter can be identified, within perturbative uncertainties, with the
MSR mass at the scale $R = 1\,$GeV. This identification holds for any value of $\alpha_s$ or the
bottom mass, and for all perturbative orders investigated. Furthermore, we show that\,: a)~the
``string tension'' parameter is independent of the bottom mass, and b)~the Coulomb
strength $\kappa$ of the Cornell model can be related to the QCD strong coupling constant
$\alpha_s$ at a characteristic non-relativistic scale. We also show how to remove the $u=1/2$
renormalon of the static QCD potential and sum-up large logs related to the renormalon subtraction
by switching to the low-scale, short-distance MSR mass, and using R-evolution. Our R-improved
expression for the static potential remains independent of the heavy quark mass value and
agrees with lattice QCD results for values of the radius as large as $0.8\,$fm, and with the Cornell
model potential at long distances.
Finally we show that for moderate values of $r$, the R-improved NRQCD and Cornell static potentials are in 
head-on agreement.
}  
\maketitle
\section{Introduction}\label{intro}
The discovery of the $J/\psi$ in 1974~\cite{Augustin:1974xw,Aubert:1974js} caused a revolution in hadron spectroscopy,
because the large mass of the $c$ quark made a non-relativistic description feasible.
However, the limited development of Quantum Chromodynamics (QCD) for heavy quarkonium systems at that time did not
provide analytical expressions for the binding forces among quarks, in particular for the confinement.
This is the reason why people were forced to resort to models that, retaining as many QCD characteristics as possible,
allowed to perform calculations susceptible to be compared with experimental results.

One of the most popular (and simple) model was the Cornell potential~\cite{Eichten:1978tg,Godfrey:1985xj,Stanley:1980zm}.
Within this model, quarks are assumed to be bounded due to flavor-independent gluonic degrees of freedom. Perturbative 
dynamics dominate at short distances, while at long distances non-perturbative effects become manifest. The short-distance 
interaction is assumed to be dominated by a single t-channel gluon exchange,  
what leads to a Coulomb interaction proportional to the strong coupling constant $\alpha_s$, in which the electric 
change is replaced by the first $SU(N_c)$ Casimir $C_F= (N_c^2-1)/(2N_c) =4/3$ for $N_C=3$. The long range linear confining 
interaction has been confirmed by lattice QCD calculations~\cite{Bali:1994de,Bali:2000vr}, and can be understood 
under the flux-tube picture where the confinement arises from the chromoelectric potential. Recent investigations
also show that it can be understood purely in terms of perturbation theory~\cite{Sumino:2003yp,Sumino:2004ht}. 
Then, the Cornell potential (sometimes dubbed funnel potential) has the expression
\begin{align}\label{eq:Cornellpot}
    V_{\rm Cornell}(r)=\sigma\,r-\frac{\kappa}{r}\,,
\end{align}
where $\kappa$ and $\sigma$ are purely phenomenological constants of the model.  
In any case, to ease a comparison with the $\mathcal{O}(\alpha_s)$ expression
for the QCD static potential we define $\kappa \equiv C_F\alpha_s^{\rm Cornell}$.

This model has been successful in describing a huge amount of experimental data including masses, widths,
radiative and strong transitions, etc. Recently~\cite{Eichten:2005ga}, due to its flexibility to describe 
coupled-channels problems, the model has been extended above the charm threshold where a great amount of new
experimental information exists. Quark models become specially relevant to describe states that do not admit a treatment in
terms of effective theories of QCD (e.g.\ molecular, hybrid or highly excited states), simply because there is no hierarchy of
scales to be exploited. They are also crucial to guide experimental searches of new bound states, which shade light in the
way quarks couple into colorless bound states. Finally, they provide useful hints to build novel approaches within QCD to tackle
the treatment of these hard-to-describe states. Given the success of such a simple model, a pressing question is whether its
parameters can be related to fundamental QCD constants. Indeed, to our knowledge, no work has ever studied the
connection between quark model potentials and the non-relativistic limit of QCD. For the many reasons presented in this
paragraph, establishing a connection between them and QCD appears certainly warranted.

The small velocity of the charm and bottom quarks in $Q\overline{Q}$ bound states enables the use of non-relativistic
effective theories within QCD to study heavy quarkonia. The use of non-relativistic approaches implies that heavy
quarkonia bound states can be organized in terms of the non-relativistic quantum numbers $(j,\ell,s)$ and the radial
excitation number $n$, while the hyperfine splittings are power corrections, starting at order ${\cal O}(m_Q^{-2})$. For
heavy quarkonia we can distinguish three well defined scales\,: the heavy quark mass $m_Q$ acting as the \emph{hard
scale}, the \emph{soft scale} determined by the relative momentum of the $Q\overline{Q}$ system ($p\sim m_Q v$ with
$v\ll c$) and the \emph{ultrasoft scale} marked by the average kinetic energy of the heavy quarkonia ($E\sim m_Qv^2$).
These scales have a well-defined hierarchy ($m_Q\gg p\gg E$), which allows for significant simplifications.

NRQCD~\cite{Lepage:1987gg} is obtained from QCD by integrating out the heavy quark mass $m_Q$, and one can exploit the
fact that $m_Q\gg\Lambda_{\rm QCD}$ to perform a perturbative matching. This implies that NRQCD inherits all the light 
degrees of freedom from QCD. The NRQCD Lagrangian is thus expanded as a series in $1/m_Q$ powers, factorizing the terms
which contribute at the hard scale as Wilson coefficients. Even though the hard scale lays in the perturbative domain,
the soft and ultrasoft scales are probed e.g.\ when building up $Q\overline{Q}$ systems, which in some cases could make
computations in terms of partonic degrees of freedom unreliable. Additionally, there is still a fundamental problem
in NRQCD\,: it does not distinguish soft and ultrasoft scales, which complicates the power-counting.

A solution to this problem arrived with the construction of EFTs such as velocity NRQCD (vNRQCD)~\cite{Luke:1999kz} and
potential NRQCD (pNRQCD)~\cite{Pineda:1997bj,Brambilla:1999xf}, which describe the interactions of a non-relativistic
system with ultrasoft gluons, organizing the perturbative expansions in $\alpha_s$ and the velocity of heavy quarks
systematically. Such EFTs only include degrees of freedom relevant for $Q\overline Q$ systems near threshold, while the
rest of degrees of freedom are integrated out. Indeed, pNRQCD is obtained from NRQCD by integrating out soft and potential gluons,
and soft quarks, under the assumption $p\gg \Lambda_{\rm QCD}$.

The specific treatment of the remaining degrees of freedom will depend on the relation between $E$ and the 
scale $\Lambda_{\rm QCD}$. At long distances QCD becomes strongly coupled and hadronic degrees of freedom emerge.
The $Q\overline Q$ system is assumed to be dominated by perturbative physics and non-perturbative corrections are taken
into account by local or non-local condensates. Under these assumptions, the static potential of QCD for color-singlet states
can be computed, which is by itself an interesting field of research. On the one hand, it is fundamental to study the energy
spectrum and, on the other, it is a good tool to explore weak and strong coupling regimes and analyze phenomena such as
confinement. In fact, the static potential can be directly compared with lattice QCD simulations~\cite{Pineda:2003jv}.

In this article we attempt to calibrate the simplest realization of the Cornell model against NRQCD. To that end
we compare observables that can be reliably predicted both in the theory and the model, namely the mass of the lowest-lying
$Q\overline{Q}$ bound states, varying the quark mass and the strong coupling constant. This exercise is inspired by a
similar analysis carried out in Ref.~\cite{Butenschoen:2016lpz} for the top quark parameter embedded in the parton-shower
Monte Carlo \textsc{Pythia}~\cite{Sjostrand:2006za,Sjostrand:2014zea}. As a byproduct, we show that the Cornell
potential agrees for large values of $r$ with the QCD static potential once the latter is expressed in terms of the MSR
mass and improved with all-order resummation of large renormalon-related logs via R-evolution. Our R-improved static
potential also compares nicely with lattice QCD simulations from Refs.~\cite{Bazavov:2014pvz,Bazavov:2017dsy}.

This article is organized as follows\,: In Sec.~\ref{sec:Numerov} we discuss the solution of the two-body problem for the Cornell potential. A fit for the parameters of the Cornell model to experimental
bottomonium and charmonium data is carried out in Sec.~\ref{sec:fit}. Section~\ref{sec:MSR} introduces the concepts of the
MSR mass and R-evolution. A discussion of the QCD static potential, and how its leading renormalon is canceled is
presented in Sec.~\ref{sec:StaticPot}. Sec.~\ref{sec:lattice} contains a comparison to lattice results. NRQCD
analytic results for the mass of $Q\overline{Q}$ bound states are compiled in Sec.~\ref{sec:massless}. The calibration
procedure is explained in Sec.~\ref{sec:strategy}, while the results are presented in Sec.~\ref{sec:results}.
Sec.~\ref{sec:conclusions} contains our conclusions. In Appendix~\ref{sec:NumerovAppendix} we discuss the 
numerical solution of the Schr\"odinger equation for the Cornell potential using the Numerov method.

\section[The Cornell Potential for the $Q\overline Q$ system]
{The Cornell Potential for the $\mathbf{Q\overline Q}$ system}
\label{sec:Numerov}
\begin{figure*}[tbh!]
	\center
	\subfigure[]
	{\label{fig:Pot0}\includegraphics[width=0.375\textwidth]{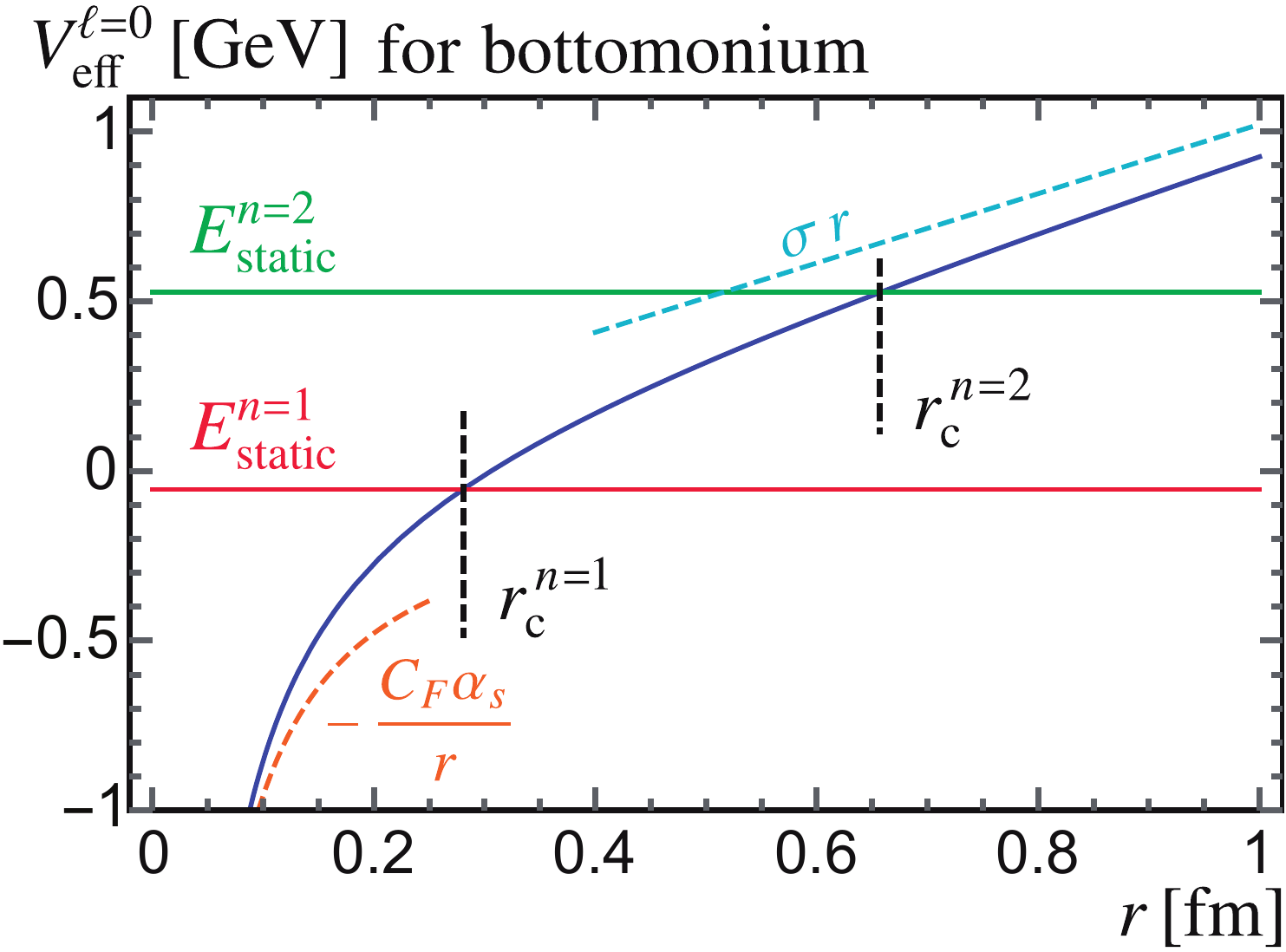}~~~~}
	\subfigure[]
	{\label{fig:Pot1}\includegraphics[width=0.373\textwidth]{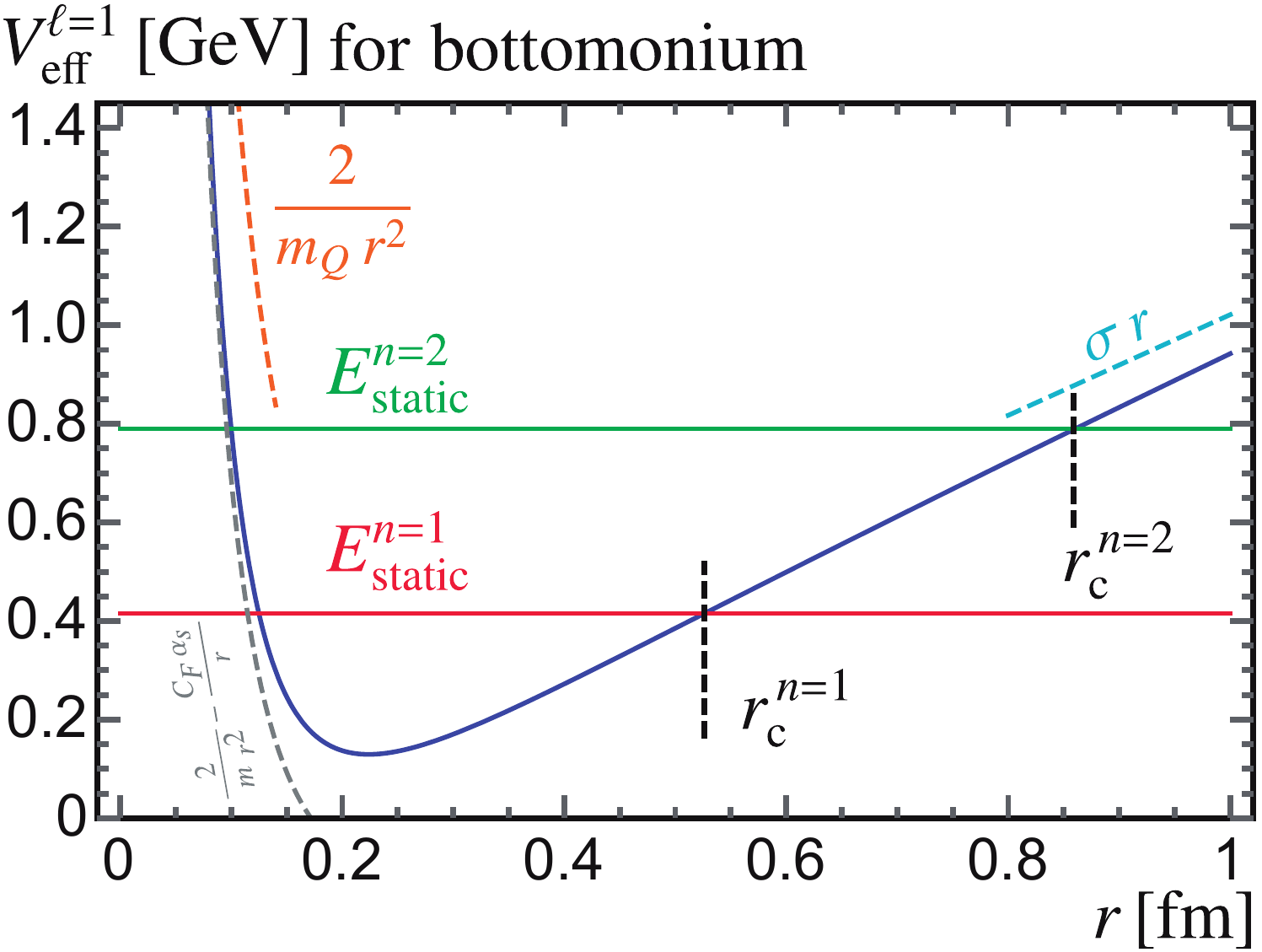} }
	\subfigure[]
	{\label{fig:Pot0-charm}\includegraphics[width=0.378\textwidth]{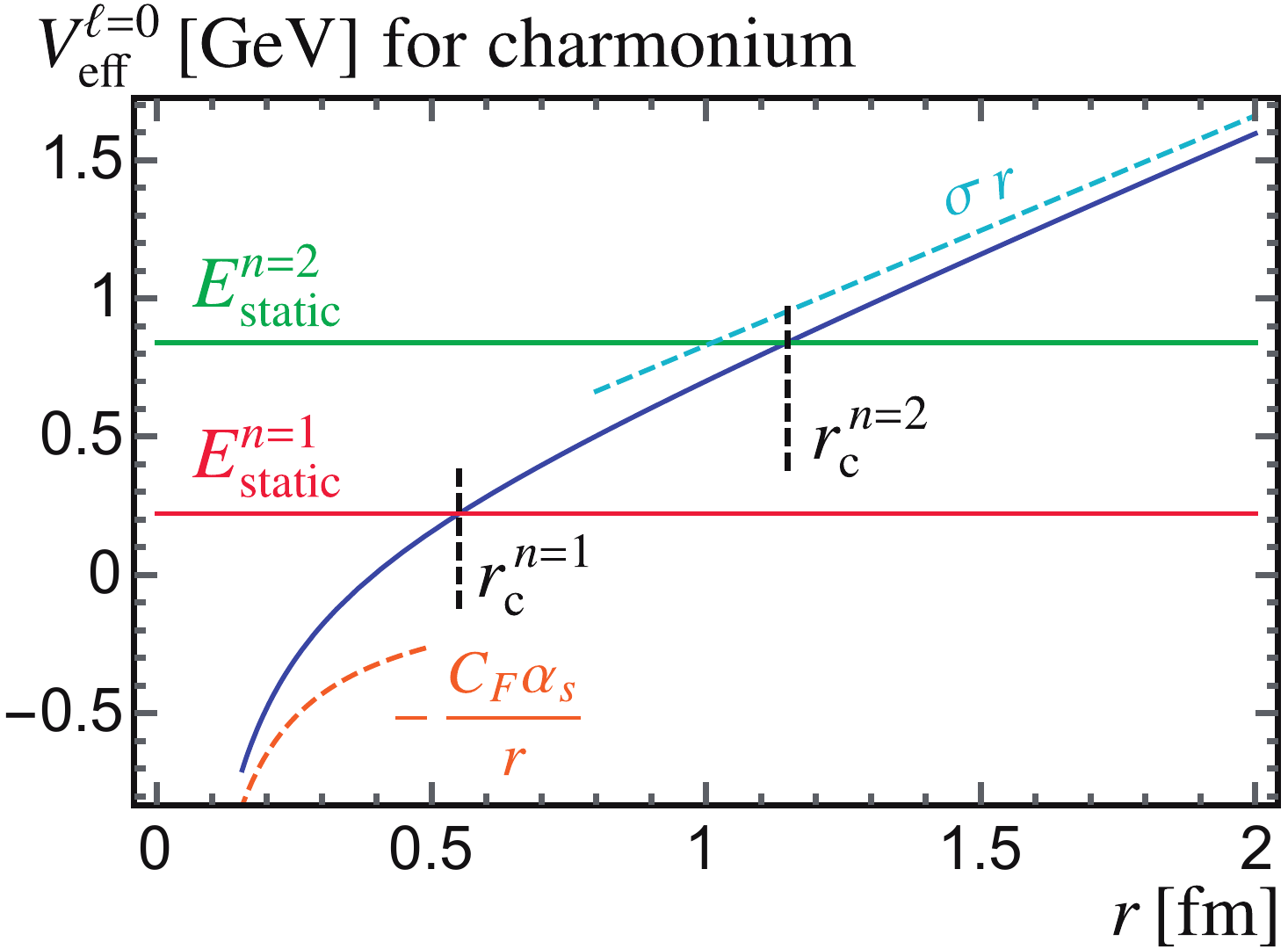}~~~~}
	\subfigure[]
	{\label{fig:Pot1-charm}\includegraphics[width=0.37\textwidth]{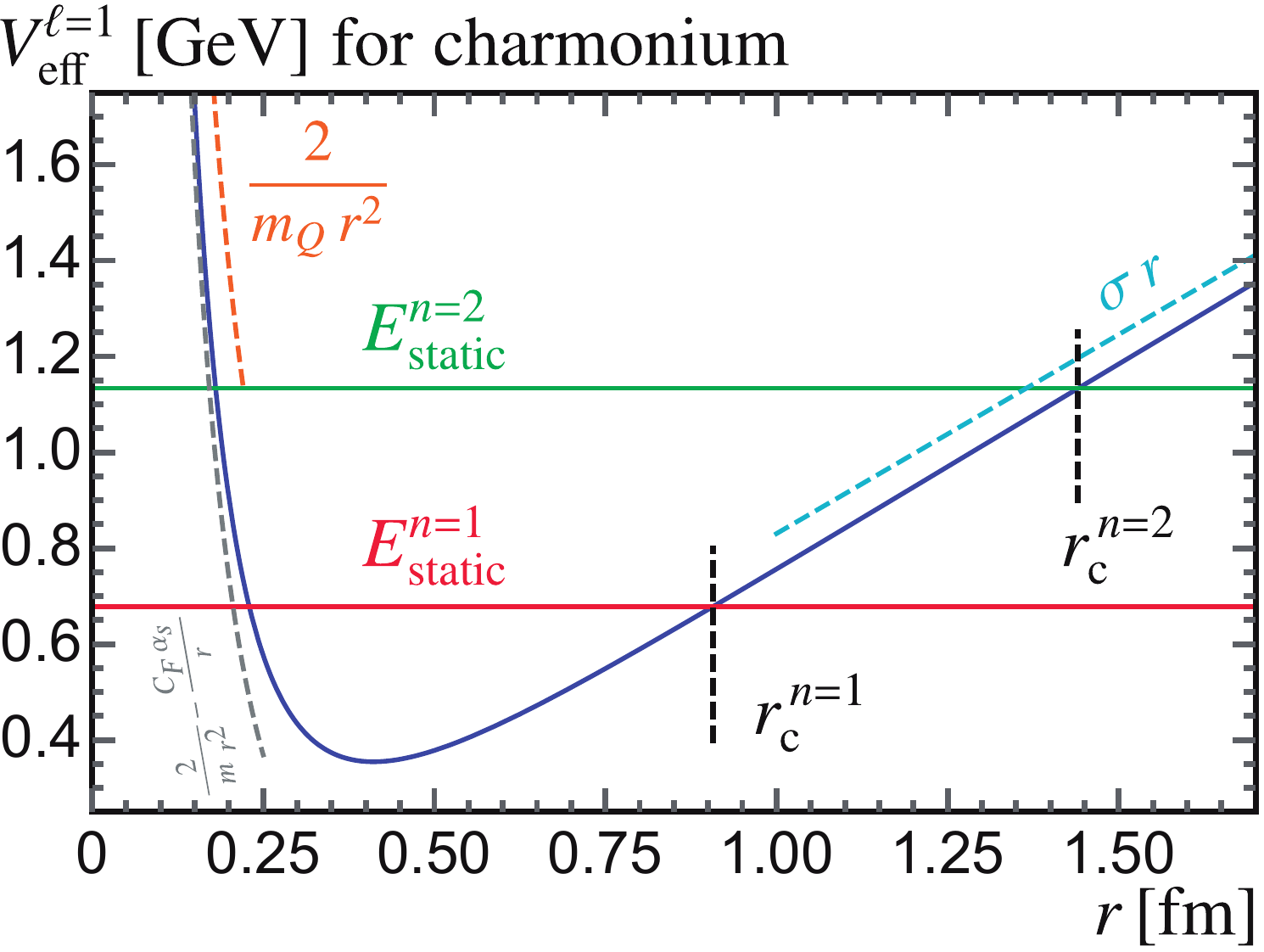} }
	\caption{\label{fig:Veff} Effective static Cornell potential for $\ell = 0$ (left panels) and $\ell = 1$
		(right panels), for bottom (upper panels) and charm (lower panels), as defined in Eq.~\eqref{eq:Veff}.
		The first two energy eigenvalues are shown as horizontal
		lines in red (ground state) and green (first excited state). The corresponding classical radius for these
		energies are signaled with vertical, black, dashed lines. We also show the asymptotic behavior for large
		$r$ in dashed cyan, and the small $r$ dominant terms in dashed orange (Coulombic for left plot and centrifuge for
		right plot), and for the right plots the Coulombic plus centrifuge terms added together, in dashed gray.}
\end{figure*}
As already mentioned, our aim is to obtain the spectrum of the Cornell potential
and compare it with NRQCD predictions to extract relations between the Cornell model parameters and
QCD fundamental constants. Therefore, we need to solve a two-body problem with a central potential.

The time-independent Schr\"odinger equation for a two-body problem reads
\begin{equation}\label{eq:schrodinger}
\bigg[\!-\frac{\nabla^2}{m_Q^{\rm Cornell}}+V_{\rm Cornell}(r)\bigg]\psi(\vec r)=E\,\psi(\vec r)\,,
\end{equation}
where $\vec r$ is the relative spatial coordinate between the two quarks, $m_Q^{\rm Cornell}$
(twice the reduced mass of the system) is the Cornell mass parameter and $V_{\rm Cornell}$ is defined in Eq.~\eqref{eq:Cornellpot}.
Here $E$ is the binding energy, and the mass of the quarkonium bound
state is given by \mbox{$M = 2\,m_Q^{\rm Cornell} + E$}.\footnote{For conciseness we
will drop the mass superscript ``Cornell'' in the remainder of this section and in Appendix \ref{sec:NumerovAppendix}.}
 Since we are dealing with a central potential,
the relative wave function of the quark-antiquark pair $\psi$ can be factorized in an angular part, expressed in
terms of spherical harmonics, and the radial wave function $R_{n\ell}(r)$\,:
\begin{align}\label{eq:angular}
\psi(r,\theta,\phi)=R_{n\ell}(r)\,Y_{\ell m}(\theta,\phi)\,.
\end{align}
Here $\ell$ is an non-negative integer number which represents the orbital angular momentum, and $n$ is a natural
number accounting for the radial excitation. The latter simply accounts for the infinitely many (but denumerable) bound
states that can be found for a given value of $\ell$. It should not be confused with the principal quantum number
$n_p = n + \ell>0$, which bounds the possible values of the orbital angular momentum $0\leq\ell\leq n_p-1$.
Using the factorization shown in~\eqref{eq:angular}, Eq.~\eqref{eq:schrodinger} can be simplified if written in terms
of the reduced wave function $u_{n\ell} = r\, R_{n\ell}(r)$, yielding an ordinary differential equation for $u_{n\ell}(r)$,
\begin{align}\label{eq:ODE}
u_{nl}^{\prime\prime}(r)+k(r)\, u_{n\ell}(r)=0\,,
\end{align}
with\,\footnote{In the subsequent sections of this article the eigenvalues of the Cornell potential will be refereed to as
static energies, denoted as $E_{\rm static}^{n,\ell}$.}
\begin{align}\label{eq:Veff}
k(r)&=m_Q\,[\,E_{n\ell}-V_{\rm eff}^\ell(r)\,]\,,\\
V_{\rm eff}^\ell(r)& = V_{\rm Cornell}(r)+\frac{\ell\,(\ell+1)}{m_Q\, r^2}\,.\nonumber
\end{align}
Note that we are using $c=\hbar=1$ units. In Fig.~\ref{fig:Veff} we show the effective Cornell potential
for $n=1,2$ and $\ell = 0,\,1$, for charmonium and bottomonium, using for the parameters the values shown in
Eqs.~\eqref{eq:Charmonium-Fit} and \eqref{eq:Cornell-Fit}, respectively.  
The figure also shows the approximations at large and small $r$,
as well as the binding energies and classical radii.

The wave function has to be normalized to one, implying that the reduced radial wave function should be
square integrable on the positive real axis\,:
\begin{align}\label{eq:Norm}
\int_0^\infty\!\df r\, |R_{n\ell}(r)|^2\, r^2 =\! \int_0^\infty\!\df r\, |u_{n\ell}(r)|^2  = 1\,,
\end{align}
where we have used
$\int\df\Omega\, Y_{\ell m}(\Omega)Y_{\ell^\prime m^\prime}(\Omega) = \delta_{m,m^\prime}\,\delta_{\ell,\ell^\prime}$,
the orthogonality property of the spherical harmonics. This condition is satisfied if at large distances $u(r)$ falls
off sufficiently rapidly.

\begin{figure}[tbh!]\centering
\includegraphics[width=0.95\linewidth]{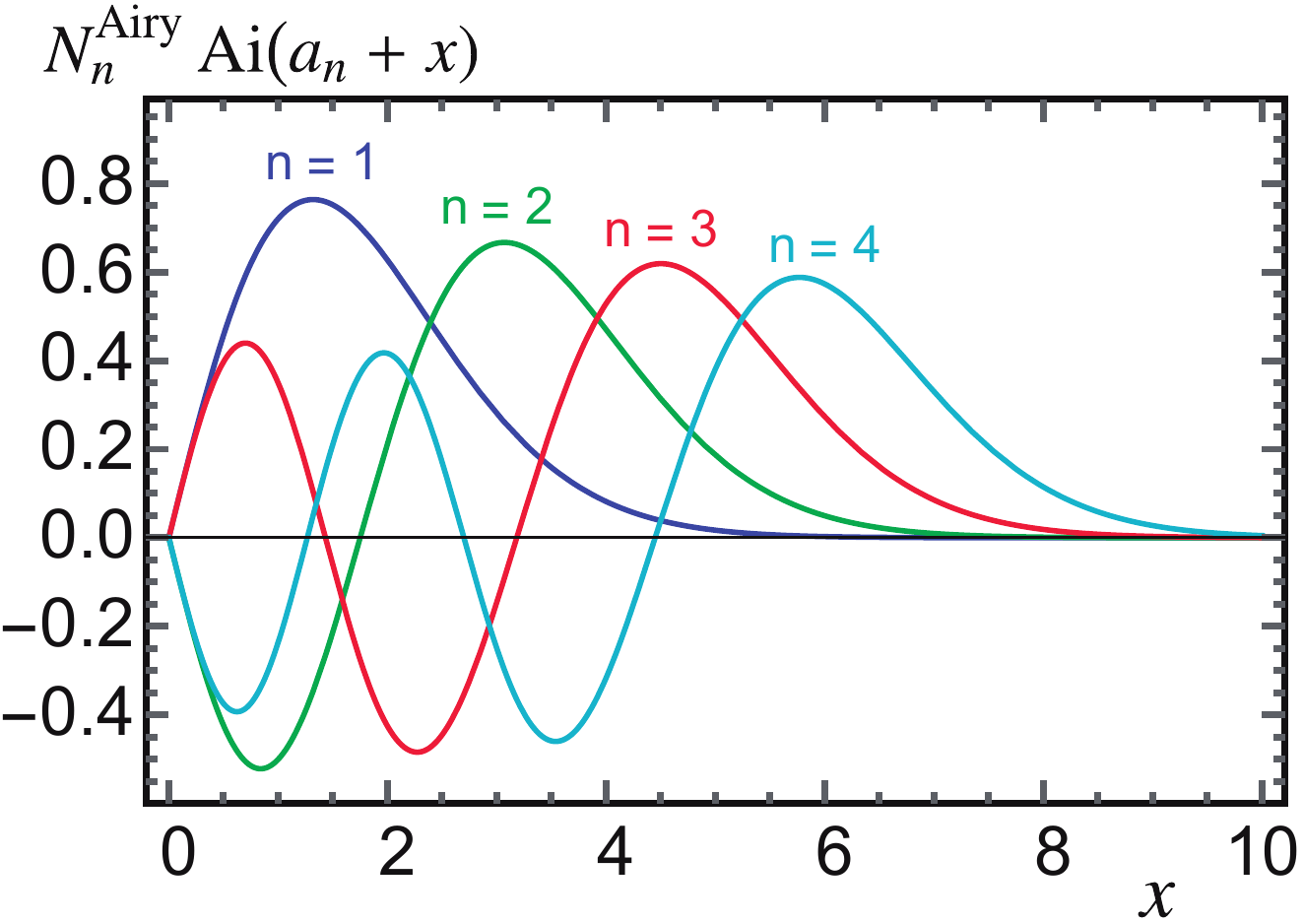}
\caption{Normalized solutions to the Airy Schr\"odinger equation in Eq.~\eqref{eq:AiryExact} in arbitrary
units of length.}
\label{fig:Airy}
\end{figure}

The solution of Eq.~\eqref{eq:ODE} requires the proper determination of two boundary conditions. On the one hand, for
small $r$ Eq.~\eqref{eq:ODE} becomes $u_{nl}^{\prime\prime}(r)+\ell(\ell + 1) u_{n\ell}(r)/r^2=0$, which
implies a power-like behavior of the reduced wave function with exponent $\ell + 1$,
\begin{align}\label{eq:boundary1}
u_{n\ell}(r\to0)\to r^{\ell+1}\,.
\end{align}
For large $r$, the Coulomb and centrifugal terms are largely suppressed, and we get the following asymptotic  
equation for the radial wave function\,:
\begin{align}\label{eq:airySE}
u_{n\ell}^{\prime\prime}(r) + m_Q (E_{n\ell} -\sigma\, r) u_{n\ell}(r)=0\,,
\end{align}
which is independent of $\ell$ and can be analytically solved in terms of Airy function
\begin{align}\label{eq:Airy}
u(r\to\infty) \propto {\rm Ai}
\biggl(\frac{m_Q(\sigma\, r - |E_{n\ell}|)}{(m_Q\sigma)^{2/3}}\biggr)\,.
\end{align}
The energy quantization of Eq.~\eqref{eq:airySE} is achieved simply imposing the boundary condition
at small $r$ for $\ell = 0$, that is, that the reduced wave function vanishes at $r=0$. The Airy function
has an infinite denumerable number of zeros which happen to be all negative. These can be computed
numerically and we denote them by $a_n$ such that ${\rm Ai}(a_n)=0$ and $|a_n| > |a_{n^\prime}|$ if
$n > n^\prime$. With this at hand we find for the Airy potential self-energies\,:
\begin{align}\label{eq:Airy-Energy}
E^{\rm Ai}_n = -\,\biggl(\frac{\sigma^2}{m_Q}\biggr)^{\frac{1}{3}}\,a_n\,.
\end{align}
Therefore the exact solution of the Airy Schr\"odinger equation takes a very simple form\,:
\begin{align}\label{eq:AiryExact}
u_{n}^{\rm Airy}(r) = N^{\rm Airy}_n\, {\rm Ai}\bigl(a_n + r\,(m_Q\sigma)^{1/3}\bigr)\,.
\end{align}
Indeed Eq.~\eqref{eq:AiryExact} behaves linearly near $r = 0$, and the $n$-th state crosses the horizontal axis exactly
$n\,-\,1$ times, as expected. This behavior can be seen in Fig.~\ref{fig:Airy} for the first four states.
The wave functions for different excited states are related, up to normalization,
by shifts in their argument. The normalization factor $N^{\rm Airy}_n$ is independent of $m_Q$ and $\sigma$,
and can be easily computed numerically. It decreases as $n$ increases, and we find that for $n <6$ it takes
the following values\,: $\{ 1.426, 1.245, 1.156, 1.098, 1.056\}$. The exact solutions for the wave function
and self-energies of Eq.~\eqref{eq:airySE}, in Eqs.~\eqref{eq:Airy-Energy} and \eqref{eq:AiryExact}, respectively,
will serve as a sanity check on our numerical program. Before we move on, let us insist that Eqs.~\eqref{eq:Airy}
and \eqref{eq:Airy-Energy} correspond to the exact solution of the Cornell potential if both $\kappa$ and
$\ell$ are zero. If we have $\kappa = 0$ but $\ell>0$ there is no exact solution to the Schr\"odinger equation.

From the asymptotic behavior of the Airy functions we can obtain a simpler boundary condition at
large $r$\,: 
\begin{equation}\label{eq:boundary2}
u_{n\ell}(r\to\infty)\propto \frac{\exp^{-\frac{2}{3}x^\frac{3}{2}}}{x^\frac{1}{4}}\,,
\qquad x = \frac{\sigma\,r-E_{n\ell}}{\sigma^\frac{2}{3}}\,.
\end{equation}
For the Cornell potential in Eq.~\eqref{eq:Cornellpot} it is not possible to find analytic expressions for
the energy eigenvalues and eigenfunctions, so a numerical approach must be employed.
There are many numerical methods to solve the Cornell potential in the literature. For its simplicity
and robustness, we will use the Numerov algorithm~\cite{Numerov1,Numerov2}  (also called Cowell's method).
Further details can be found in Appendix~\ref{sec:NumerovAppendix}. 
We have implemented the Numerov algorithm and the matrix element computations in a Fortran 2008 code
\cite{gfortran}, which contains both public and in-house routines, and is used to predict the mass 
of the bound states as well as to perform fits to data. 

The potential in Eq.~\eqref{eq:Cornellpot} by itself is nonetheless unable to describe the $\Upsilon(1S)-\eta_b(1S)$
mass splitting, as it does not depend on the spin of the quarkonium state. Furthermore, it provides degenerate
masses for the $\chi_{bJ}$ multiplet (with $J=\{0,1,2\}$). Hence, in order to describe with higher accuracy
the low-lying bottomonium spectrum, one has to add $1/m_Q^2$ terms to the Cornell static potential that take
into account the spin-spin, spin-orbit and tensor interactions, breaking the present degeneracy~\cite{DeRujula:1975qlm}.
Such contributions arise from  
the leading relativistic corrections of the t-channel gluon
and confinement interactions, giving the following terms~\cite{Eichten:1979pu,Godfrey:1985xj}\,:\,\footnote{Here the LS
term from confinement is obtained by the exchange of a scalar particle.}\,\footnote{We denote with the superscript ``OGE'' the terms arising from gluon exchanges and with ``CON'' those coming from the confinement interaction.}
\begin{align}
V_{\rm SS}^{\rm OGE}(r)&=\frac{8\alpha_s^{\rm Cornell}}{9m_Q^2\,r^2}
(\vec S_1\cdot \vec S_2)\,\delta(r)\,,\label{eq:subleading}\\
V_{\rm LS}^{\rm OGE}(r)&=\frac{2\alpha_s^{\rm Cornell}}{m_Q^2\,r^3}\,(\vec L\cdot\vec S)\,,\nonumber\\
V_{\rm LS}^{\rm CON}(r)&=-\frac{\sigma}{2m_Q^2\,r}\,(\vec L\cdot \vec S)\,,\nonumber\\
V_{\rm T}^{\rm OGE}(r)&=\frac{\alpha_s^{\rm Cornell}}{3m_Q^2\,r^3}\,S_{12}\,,\nonumber
\end{align}
where $\vec S=\vec S_1+\vec S_2$ is the total spin, $\vec L$ the relative orbital momentum and $S_{12}$ the
tensor operator of the $Q\overline Q$ bound state, defined as
\begin{equation}
S_{12}=2\,(\vec S_1\cdot\hat r)(\vec S_2\cdot\hat r)-(\vec S_1\cdot\vec S_2)\,,
\end{equation}
with $\hat r= {\vec r}/r$. Analytical expressions can be found for the spin-spin, spin-orbital and tensor operators,
\begin{subequations}\label{eq:SU2}
\begin{align}
\langle\vec S_1\cdot \vec S_2\rangle &= \frac{1}{4}\,[\,2\, s\,(s+1)-3\,]\,\label{eq:spinspin},\\
\langle\vec L\cdot\vec S\rangle&=\frac{1}{2}\,[\,j\,(j+1)-\ell\,(\ell+1)-s\,(s+1)\,]\,\label{eq:spinorbit},\\
\langle S_{12}\rangle &= \frac{2\, [\,2\,\ell\,(\ell+1)\,s\,(s+1)-
3\,\langle\vec L\cdot\vec S\rangle-6\,\langle\vec L\cdot\vec S\rangle^2\,]}
{(2\,\ell-1)\, (2\,\ell+3)}\label{eq:tensor}\,,
\end{align}
\end{subequations}
with $(j, \ell,s)$ the quantum numbers of the $Q\overline Q$ state. The specific values for the
quantum numbers considered in this article are given in Tab.~\ref{tab:SpinCoefficients}.

Given the assumed large mass of the heavy quark, the above terms are expected to be small. Therefore, their
contribution to the bottomonium mass will be incorporated to the model using first-order perturbation theory.
For consistency, if perturbation theory is to be used to second order, one needs to include the $1/m_Q^4$
potential. In practice one needs to take matrix elements of the operators in Eq.~\eqref{eq:subleading}. To that
end we use the angular decomposition in Eq.~\eqref{eq:angular} and write the three-dimensional integration in
spherical coordinates $\df^3\,{\vec r}=r^2\,\df r\,\df \Omega$ such that the angular integrations are carried out with
ordinary angular momentum algebra [\,see Eq.~\eqref{eq:SU2}\,] and the radial integral becomes a one-dimensional
matrix element of the reduced wave function $u_{n\ell}(r)$, given that the Jacobian factor $r^2$ cancels when writing
$|R_{n\ell}(r)|^2=|u_{n\ell}(r)|^2/r^2$ as in Eq.~\eqref{eq:Norm}. 
\begin{table}
\caption{Spin-spin, spin-orbital and tensor coefficients for the states considered in our study. Analytical
expressions can be found in Eq.~\eqref{eq:spinspin} for the spin-spin, Eq.~\eqref{eq:spinorbit} for the
spin-orbital and Eq.~\eqref{eq:tensor} for the tensor operators.}
\label{tab:SpinCoefficients}
\centering
\begin{tabular}{|l|rrrrrrr|}
\hline 
& $^1S_0$ & $^3S_1$ & $^1P_1$ & $^3P_0$ & $^3P_1$ & $^3P_2$ & $^3D_2$  \\
\hline
$\langle\vec S_1\cdot \vec S_2\rangle$ & $-\frac{3}{4}$ & $\frac{1}{4}$ & $-\frac{3}{4}$ & $\frac{1}{4}$ & $\frac{1}{4}$ & $\frac{1}{4}$ & $\frac{1}{4}$  \\
$\langle\vec L\cdot\vec S\rangle$ &  $0$ & $0$ & $0$ & $-2$ & $-1$ & $1$ & $-1$  \\
$\langle S_{12}\rangle$ &  $0$ & $0$ & $0$ & $-4$ & $2$ & $-\frac{2}{5}$ & $2$ \\
\hline
\end{tabular}
\end{table}
\section{Fitting the Cornell Model to Experimental data}\label{sec:fit}
In this section we present the results from $\chi^2$ fits of the Cornell model parameters to charmonium and bottomonium
experimental data. We  restrict ourselves to \mbox{$n_p \le 2$}, since for higher values of the principal quantum
number one needs to include string-breaking effects, as can be seen in Fig.~\ref{fig:Cornell-Fit}. The case of
bottomonium will
serve as a proof of concept for our calibration, that is, it will show that the three parameters of the model
can be determined from fits to the $8$ lowest-lying bottomonium bound states. This is crucial since in QCD we
can only reliably predict these within perturbation theory, as argued in Ref.~\cite{Mateu:2017hlz}. In many phenomenological
applications of the quark model approach, in which more sophisticated Hamiltonians are used, the parameters of the potential
are not obtained by a full fledged $\chi^2$ minimization,
but simply adjusted by a rough comparison to data plus physically motivated priors. This procedure is in many cases justified,
since there is a large amount of data that the model aims to describe, but its accuracy might vary widely for different observables.
Given that it is very hard to add a theoretical covariance matrix to the $\chi^2$ (the model does not provide a method to quantify
its ``modeling'' error), such a fit could lead to biased model parameter values. For the simpler situation of the na\"ive Cornell
model, which seeks to describe only the low-lying quarkonium spectrum, we show that the fit is indeed possible, and we shall
see that no bias is observed.

The minimization of the $\chi^2$ is carried out using the Fortran~77 package MINUIT~\cite{James:310399}. We have
checked that the algorithm is very effective in finding the minimum, but due to the strong correlation between the various
fit parameters [\,see Eq.~\eqref{eq:correlation}\,] it does not estimate the covariance matrix correctly. To solve this
problem we compute a three-dimensional grid of $5^3$ elements centered around the minimum found by MINUIT,
which is then adjusted to a 3D quadratic polynomial by a linear regression. We find that the $\chi^2$ function clearly
behaves in Gaussian way near the minimum, which permits an estimate of the covariance matrix in the Hessian
approximation, that is, from the quadratic terms of our regression.

\subsection{Bottomonium Fits}\label{sec:bottomonium-fits}
We take the experimental values from the PDG~\cite{Tanabashi:2018oca}, which are collected in Table~3 of
Ref.~\cite{Mateu:2017hlz}. Since experimental data is extremely precise, we find $\chi^2_{\rm min}/{\rm d.o.f.} = 258$,
much larger than unity. As a penalty for this deficiency we rescale the uncertainties on the parameters that come out
from the fit by the square root of the reduced $\chi^2$ at its minimum. With this procedure we get\,:
\begin{align}\label{eq:Cornell-Fit}
m_b^{\rm Cornell} & = 4.733 \pm 0.018\,{\rm GeV}\,,\nonumber\\
\sigma & = 0.207 \pm 0.011\,{\rm GeV}^2\,,\\
\alpha_s^{\rm Cornell} & = 0.356 \pm 0.015\,.\nonumber
\end{align}
The quoted uncertainties correspond to $1$ standard deviation for each parameter (i.e.\ it corresponds to a $68\,\%$
confidence level in each of them), as obtained from the intersection of the $\chi^2$ function with the horizontal hyperplane
$\chi_{\rm min}^2+1$. The $68\,\%$ confidence level in the space spanned by these three parameters is obtained
by combining the correlation matrix in Eq.~\eqref{eq:correlation} with the uncertainties given in Eq.~\eqref{eq:Cornell-Fit}
rescaled by the factor $1.878$. It corresponds to the intersection of the $\chi^2$ function with the hyperplane
$\chi_{\rm min}^2+3.53$. However, the uncertainties as given in Eq.~\eqref{eq:Cornell-Fit} together with the
correlation matrix are used to compute the incertitude of any function of these parameters (e.g.\ the masses of the
bound states).

Eq.~\eqref{eq:correlation} shows that indeed there is a very strong
correlation between the three parameters, very close to $100\,\%$ (anti-)correlation. We find a very strong negative
correlation between $m_Q^{\rm Cornell}$ and $\sigma$ as well as between $\sigma$ and $\alpha_s^{\rm Cornell}$.
However $m_b^{\rm Cornell}$ and $\alpha_s^{\rm Cornell}$ are strongly positively correlated\,:
\begin{align}\label{eq:correlation}
R = \left( {\begin{array}{ccc}
1 & -0.977199 & 0.970264 \\
-0.977199 & 1 & -0.906111 \\
0.970264 & -0.906111 & 1 \\
\end{array} } \right),
\end{align}
where the order of columns and rows is the same as in Eq.~\eqref{eq:Cornell-Fit}. These results can be easily
interpreted\,: an increase of either $m_Q^{\rm Cornell}$ or $\sigma$ make the mass of all bound states larger, while
one has to decrease $\alpha_s^{\rm Cornell}$ produce the same effect.

\begin{figure*}[tbh!]
\center
\subfigure[]
{\label{fig:bottom-wave}\includegraphics[width=0.4\linewidth]{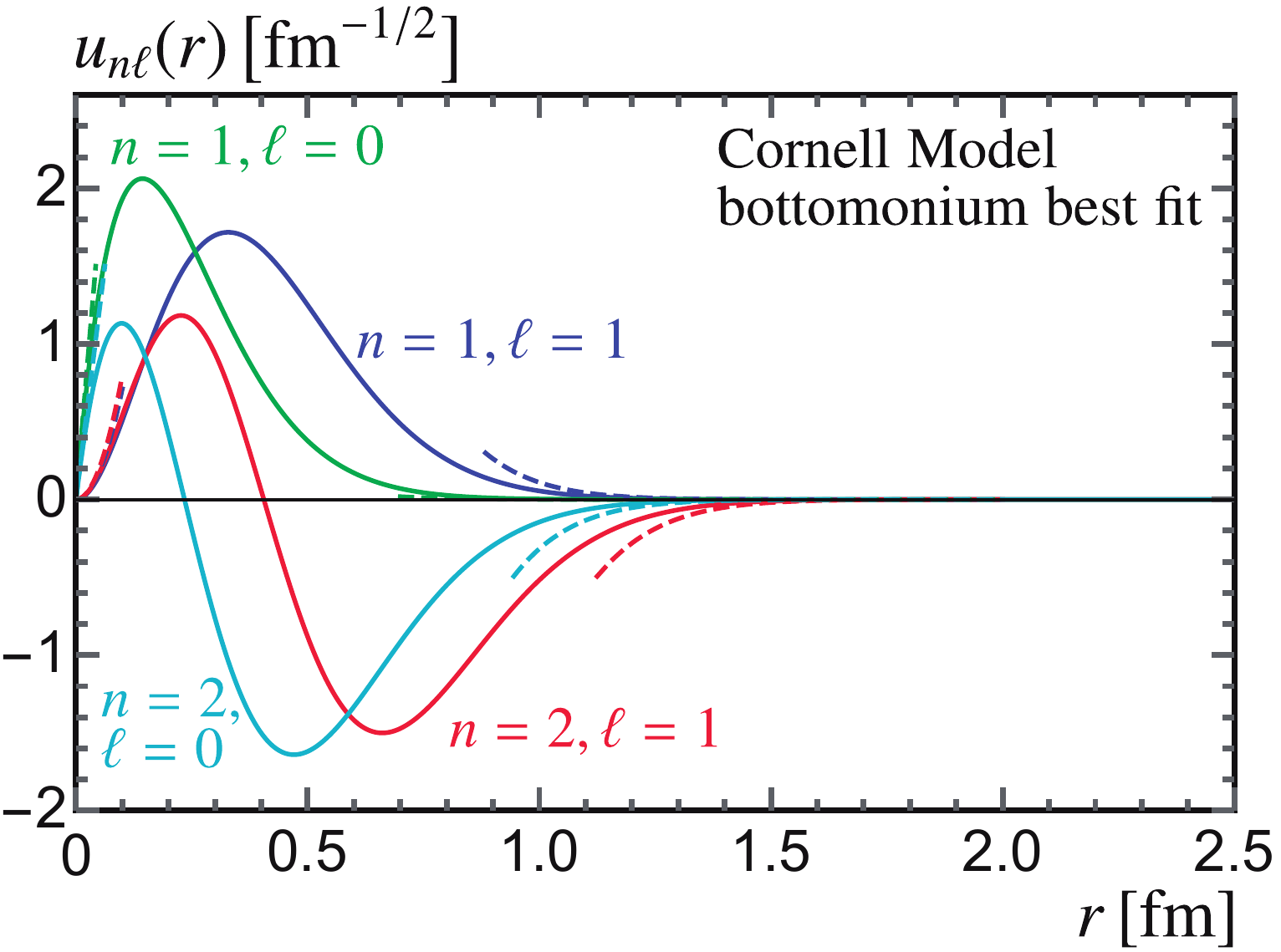}~~}
\subfigure[]
{\label{fig:charm-wave}\includegraphics[width=0.4\linewidth]{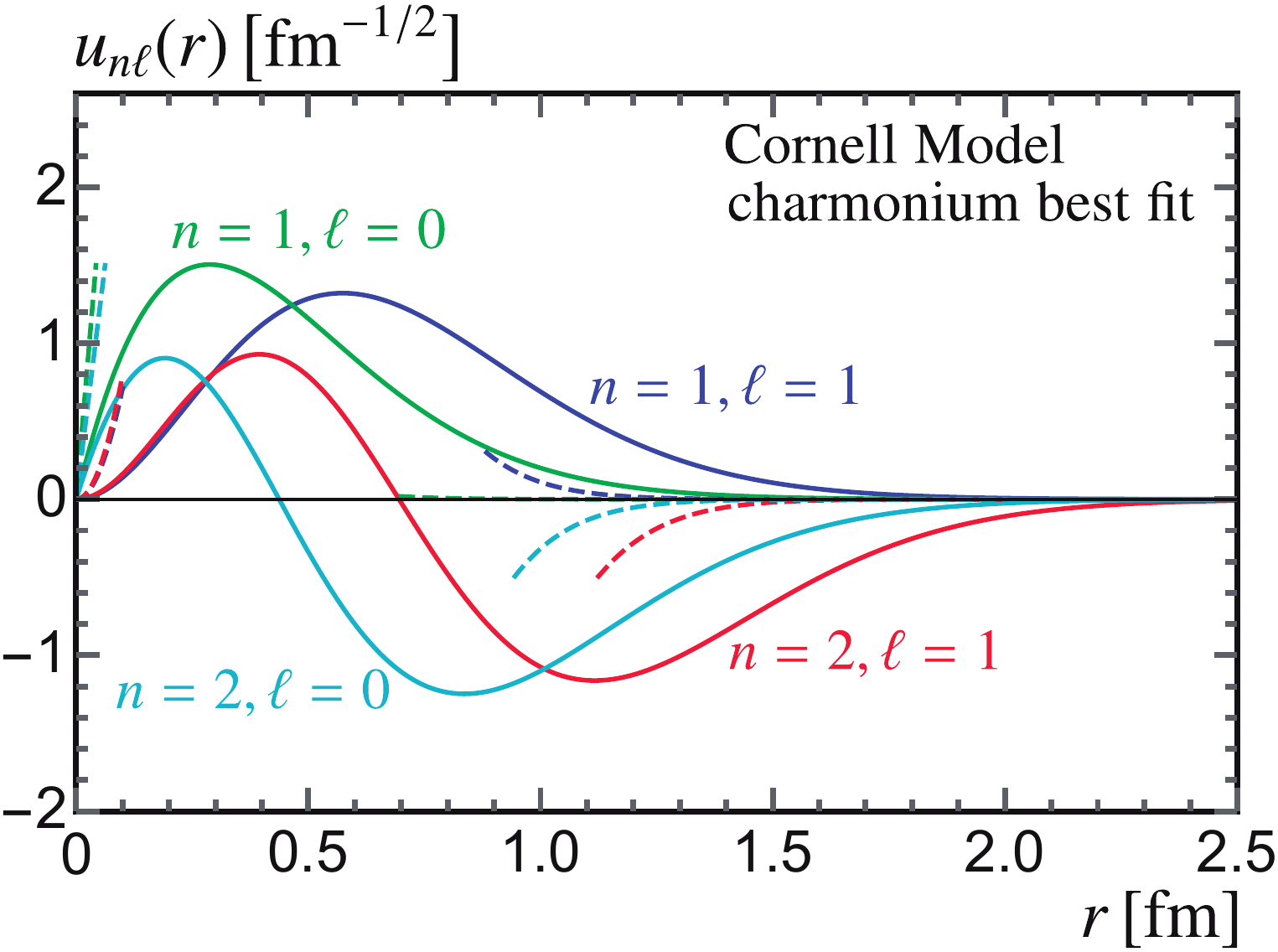} }
\caption{Eigenfunctions of the static Cornell potential, shown as the reduced wave functions $u_{n\ell}(r)$, obtained using
the central values of our fit to bottomonium [\,panel~(a)\,] and charmonium [\,panel~(b)\,] experimental data. Green and
blue correspond to $n = 1$, while cyan and red have $n = 2$. Green and cyan have $\ell = 0$ while blue and red correspond
to $\ell=1$. The dashed lines show the asymptotic behavior both at large and small $r$.}
\label{fig:Wave-functions}
\end{figure*}

In Fig.~\ref{fig:bottom-wave} we show the reduced wave functions, eigenfunctions of the static Cornell potential,
for $n = 1,2$ and $\ell=0,1$, values that correspond to the states included in the
fit. For a given $n$ value, the peak is shifted to the right for larger values of $\ell$. The exponential
falloff starts at larger $r$ values for higher values of $n$, and is roughly $\ell$ independent. As expected,
for $n=2$ the wave function crosses the $x$ axis once. In all cases the wave function is strongly suppressed
already at $r = 1.5\,$fm, therefore justifying our choice \mbox{$r_{\rm max} = 4\,$fm}. Taking $N = 5000$
corresponds to a numerical uncertainty in the solution of the Schr\"odinger equation of $0.08\,\%$ and $0.0004\,\%$
for $n = 0,~\ell = 0,1$ states, respectively, and $0.034\,\%$ and $0.0002\,\%$ for $n=1,~\ell = 0,1$, 
being $N$ the number of steps in the Numerov method (see Appendix~\ref{sec:NumerovAppendix}).
This choice of $N$ inflicts an uncertainty on the extraction $m_b^{\rm Cornell}$, $\sigma$, and
$\alpha_s^{\rm Cornell}$ from a fit to data of $ 0.013\,\%$, $0.22\,\%$ and $0.017\,\%$, respectively.\footnote{To
figure out this precision we have compared to a numerical computation with \mbox{$N=200\,000$}, which is taken as
exact.} These are a factor of $28$, $24$ and $237$ smaller than the fit uncertainties, and therefore negligible.
The numerical uncertainty associated to $r_{\rm max} = 4\,$fm is even smaller. As an additional
check on the accuracy of our numerical method, we compute the $\ell=0$ states with
$\alpha_s^{\rm Cornell}=0$, for which we know the exact solution, shown in Eq.~\eqref{eq:Airy-Energy}.
The numeric solutions reproduce the exact result with an accuracy of $2\times 10^{-6}\,\%$. At this point
it is also worth mentioning that neglecting the Coulomb-like interaction (that is, setting
$\alpha_s^{\rm Cornell} = 0$) overshots the exact result of the static energies by $5.5\,\%$ and $3\,\%$ for
$\ell = 0$ and $n=1,2$, respectively, and $0.6\,\%$ and $0.4\,\%$ for $\ell = 1$ and $n=1,2$. Likewise, setting
$\sigma = 0$ one can use the known solution for the hydrogen atom bound states, which can be obtained from
Eq.~\eqref{eq:EXpole} truncated at $\mathcal{O}(\alpha_s^2)$, which undershoots the exact
results by $2.4\,\%$ and $6.1\,\%$, for $\ell = 0$ and $n=1,2$, respectively, and $5\,\%$ and $8.2\,\%$ for $\ell = 1$
and $n=1,2$. It might seem that either of these approximations is good, but it only appears so because most of the
contribution comes from the quark mass. If we compare binding energies then the approximations become much
worse, rising up to $600\,\%$ for the Coulomb limit and $1400\,\%$ for the Airy limit. Therefore none of the two terms
in the Cornell potential can be treated as a perturbation with respect to the other for bottomonium.

We have checked the robustness of our fits by removing one, two, or three points from our dataset. It tuns out that
if one keeps the two states with $n_p=1$ and the highest mass state $\Upsilon(3^3S_1)$, the largest variation from
the default is always smaller than $10\,$MeV for the Cornell mass parameter, and below $2\,\%$ for either $\sigma$
or $\alpha_s^{\rm Cornell}$.

\begin{figure}[tbh!]\centering
\includegraphics[width=0.95\linewidth]{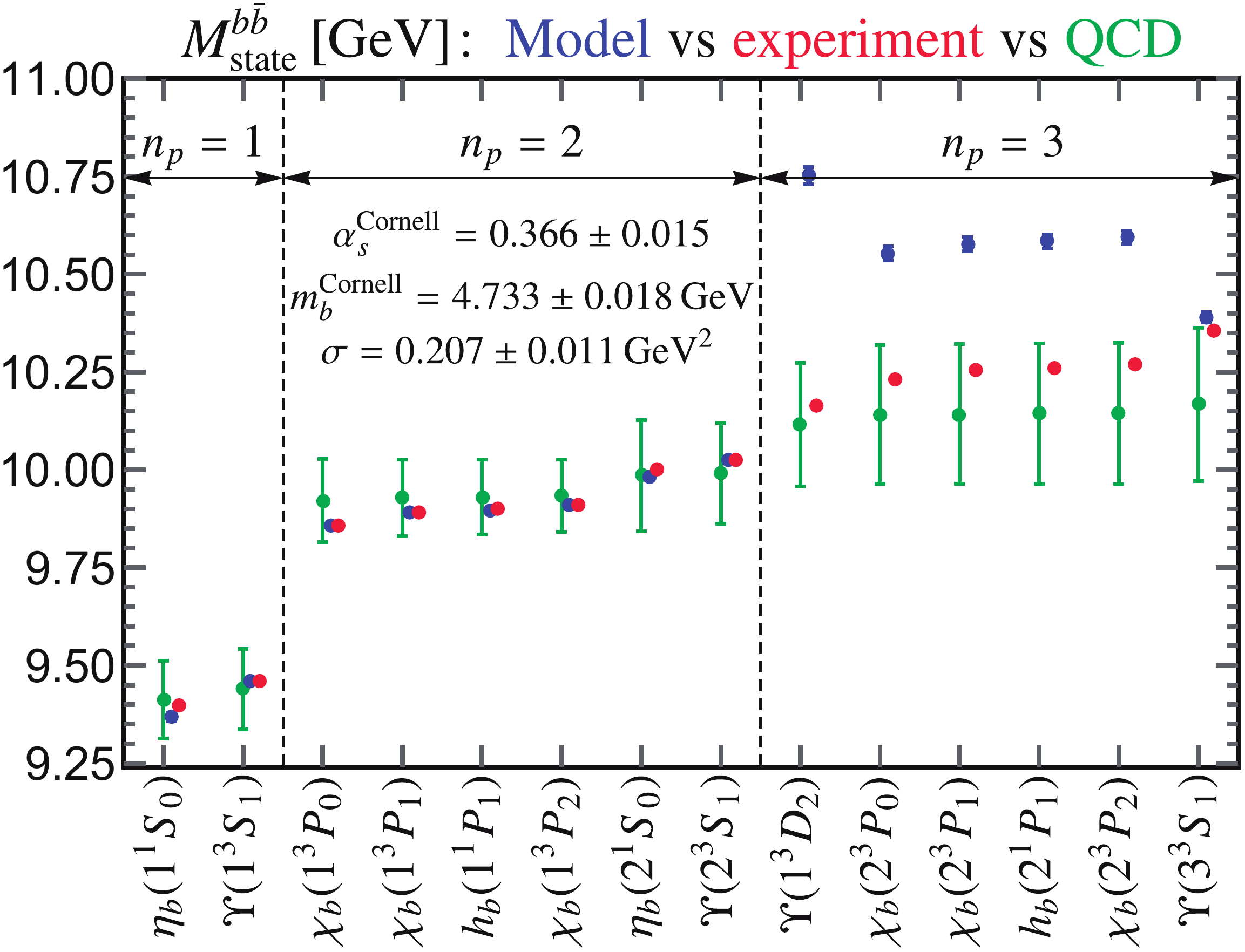}
\caption{Comparison of the Cornell Model (blue) and QCD (green) predictions to experimental bottomonium data (red).
Both QCD and the Cornell model have been fitted to the same experimental data with $n_p\le2$. QCD error bars
correspond to perturbative uncertainties, while the Cornell-Model uncertainties come only from the fit.}
\label{fig:Cornell-Fit}
\end{figure}

In Fig.~\ref{fig:Cornell-Fit} we compare the predictions of the Cornell model for the $8$ states that have been
used in the fit, employing the best-fit values and propagating the fit uncertainties into the masses through the
covariance matrix. We include experimental data with $n_p = 3$, which have not been included in the fit, but serve as
an illustration for the limitations of this simple model. States with $n_p > 2$ can only bee described if
some sort of ``string breaking'' is implemented~\cite{Born:1989iv,Bali:2005fu}. The N$^3$LO QCD predictions of
Ref.~\cite{Mateu:2017hlz} are also shown for illustration.
\subsection{Charmonium Fits}\label{sec:charmonium-fits}
\begin{table}[t!]
\begin{center}
\caption{\label{tab:exp-charm} Experimental masses of the charmonium states, up to $n=2$. Data from 
Ref.~\cite{Tanabashi:2018oca}}
\begin{tabular}{|l|ccl|}
\hline  
State & $n$ & $^{2s+1}\ell_j$ & $M_{\rm exp}$ [GeV] \\
\hline
$\eta_c(1S)$    & $1$ & $^1S_0$ & $2.9839(5)$   \\
$J/\psi$        & $1$ & $^3S_1$ & $3.096900(6)$ \\
$\chi_{c0}(1P)$ & $2$ & $^3P_0$ & $3.41471(30)$ \\
$\chi_{c1}(1P)$ & $2$ & $^3P_1$ & $3.51067(5)$  \\
$h_c(1P)$       & $2$ & $^1P_1$ & $3.52538(11)$ \\
$\chi_{c2}(1P)$ & $2$ & $^3P_2$ & $3.55617(7)$  \\
$\eta_c(2S)$    & $2$ & $^1S_0$ & $3.6376(12)$  \\
$\psi(2S)$  & $2$ & $^3S_1$ & $3.686097(25)$    \\
\hline  
\end{tabular}
\end{center}
\end{table}
Experimental data for charmonium has been taken from the PDG~\cite{Tanabashi:2018oca}, and for the reader's
convenience is collected in Tab.~\ref{tab:exp-charm}. Given the astonishing precision of some charmonium mass
measurements, the reduced $\chi^2$ at the minimum is in this case even larger than for bottomonium. We find
\mbox{$\chi^2_{\rm min}/{\rm d.o.f.} = 1.3\times 10^5$} and, consequently, apply the same penalty as was done in
Sec.~\ref{sec:bottomonium-fits}. We obtain\,:
\begin{align}\label{eq:Charmonium-Fit}
m_c^{\rm Cornell} & = 1.406 \pm 0.043\,{\rm GeV}\,,\nonumber\\
\sigma & = 0.164 \pm 0.011\,{\rm GeV}^2\,,\\
\alpha_s^{\rm Cornell} & = 0.491 \pm 0.080\,.\nonumber
\end{align}
Two comments are in order\,: a)~the value of $\alpha_s^{\rm Cornell}$ is, as expected, larger than for bottomonium. To
the extent that this model parameter can be related to the QCD strong coupling at some characteristic energy, one expects
the typical scale for charmonium smaller than for bottomonium, which translates into a larger $\alpha_s$ value for the
former; b)~the uncertainties for the quark mass and $\alpha_s$ are larger than for bottomonium, while remain the same
for $\sigma$. The larger errors can be explained by the huge penalty factor applied here, namely $363$, which is $23$
times larger than for bottomonium. Long-distance interactions matter more in charmonium, since it is a more extended system
in which softer gluons are more often exchanged. Accordingly, the confining parameter parameter is fixed less ambiguously
than the rest.

The correlation among the three parameters is even stronger than for bottomonium, and follows the exact same pattern\,:
\begin{align}\label{eq:charm-correlation}
R = \left( {\begin{array}{ccc}
1 & -0.959388 & 0.991643  \\
-0.959388 & 1 & -0.916766 \\
0.991643 & -0.916766 & 1  \\
\end{array} } \right).
\end{align}
Similar to what we found in bottomonium, the static binding energies are not well reproduced if any of the two terms
in the potential is set to zero, but neglecting the Coulomb term is a much better approximation. Setting $\sigma = 0$
produces binding energies which are incorrect by up to $2300\,\%$, but setting $\alpha_s = 0$ overshoots the exact
answer by $60\,\%$ at worst. This confirms the expectation that charmonium bound states are more afflicted by
long-distance effects, parametrized by $\sigma$ in the Cornell model, and also explains why this parameter is
determined more accurately in Eq.~\eqref{eq:Charmonium-Fit}.

In Fig.~\ref{fig:charm-wave} we show the charmonium reduced wave functions. They have identical shapes to those
of bottomonium, but as anticipated from dimensional arguments, the wave functions extend to larger distances as
compared to bottomonium states, and have accordingly lower (shallower) peaks (valleys). However, it is still justified to
use $r_{\rm max} = 4\,$fm as at $2\,$fm the wave function is already negligibly small. We have checked that using
$r_{\rm max} = 6\,$fm produces a shift in the $4$-th or $5$-th decimal place in any of the fit parameters shown in
Eq.~\eqref{eq:Charmonium-Fit}. Since we will not calibrate the Cornell model for charmonium, we do not study the
optimal choice of $N$, the number of steps in the Numerov method. We simply take $200\,000$ for which the associated
numerical errors are several orders of magnitude smaller than the fit uncertainties.

We have performed a similar robustness check in our charmonium fits, that is, selectively remove one, two, or three points from our dataset to analyze the consequent variations in the Cornell parameters, and we conclude that the stability is not as good as for
bottomonium. Removing a single point from our dataset, and always keeping the two lowest-mass states
and the highest-mass state, we find that the impact on the Cornell mass parameter can be as large as $70\,$MeV,
while in the other two parameters can reach deviations of $6\,\%$ and $17\,\%$ for $\sigma$ and $\alpha_s^{\rm Cornell}$,
respectively. Since we are not going to calibrate the charm mass, these findings are only a minor concern, which again seem
to signal that the non-relativistic approximation is worse for charmonium.

\begin{figure}[t!]\centering
\includegraphics[width=0.95\linewidth]{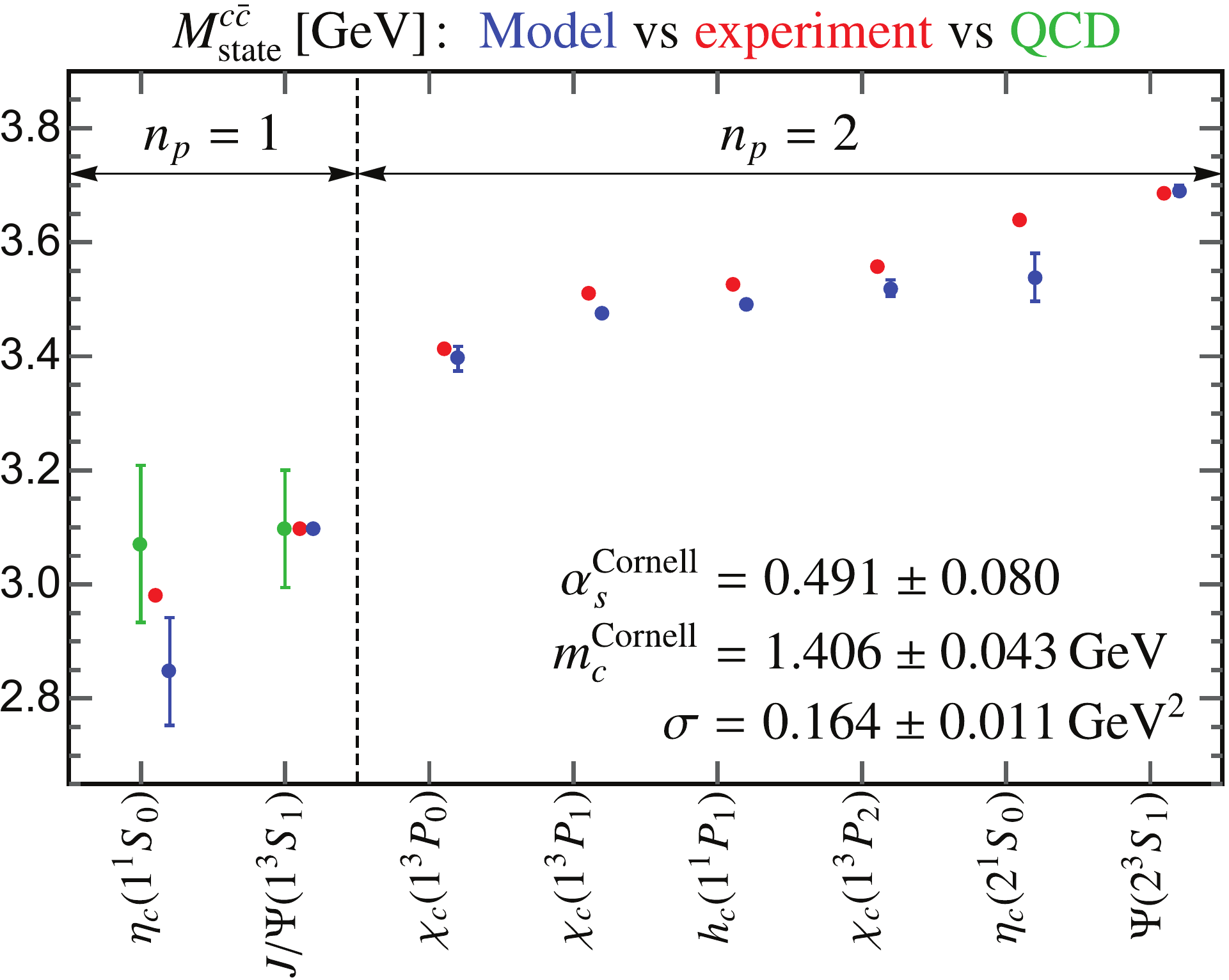}
\caption{Same as Fig.~\ref{fig:Cornell-Fit} for charmonium bound states. Here QCD has been fitted to $n_p=1$ experimental
data, for which non-perturbative effects are quite small, while the Cornell model has been fitted to states with
$n_p\le 2$.} \label{fig:Cornell-Fit-charm}
\end{figure}

In Fig.~\ref{fig:Cornell-Fit-charm} we again confront the predictions of the Cornell model after the fit with the $8$
masses that enter our $\chi^2$ function. In this case we refrain from showing results for $n_p = 3$, since the
$D\bar D$ threshold is quite low in charmonium. Similarly, we only show QCD predictions (taken from
Ref.~\cite{Mateu:2017hlz}) for $n_p = 1$, since higher $n_p$ states cannot be reliably predicted in perturbation theory.
It is worth noting that the Cornell model prediction for the $J/\Psi$ is extremely precise, with an error bar which seems
unnaturally small given the uncertainties quoted in Eq.~\eqref{eq:Charmonium-Fit}. The explanation for this
apparent contradiction is the large cancellations that happen between the different error sources due to the very
strong correlations among the three parameters. The $J/\Psi$, having an experimental uncertainty $40$ times smaller
than the next most precise measurement, overly drives the fit and renders the correlation such that this particular
mass is predicted with an uncertainty roughly $40$ times smaller than for the rest of states. Such a situation does
not happen for bottomonium, therefore all states are predicted with similar incertitudes.

\section{The MSR mass and R-evolution}\label{sec:MSR}
The $\MSR$ scheme (see Refs.~\cite{Hoang:2008yj,Hoang:2009yr,Hoang:2017suc} for a review), can be seen as the
natural extension of the $\MSb$ mass for renormalization scales below the heavy quark mass. It relies on the fact
that the renormalon ambiguity in the $\MSb$-pole relation is independent of the value of the mass. The $\MSR$ mass is
then directly defined from this relation, and depends on an infrared scale $R$. The MSR scheme absorbs into
the mass definition quark self-energy fluctuations above the scale $R$.\,\footnote{In contrast, the pole and $\MSb$ masses absorb
all fluctuations above $0$ and $\mbar(\mbar)$, respectively.} The optimal choice of $R$ depends on the observable,
and can be chosen such that large logs do not appear in perturbative series. In addition, the subtraction
series relating the MSR and pole masses can be expressed in terms of $\alpha_s(\mu)$, with $\mu$ the $\MSb$
renormalization scale. This is essential to cancel the renormalon in other series when the pole mass expressed in terms
of the MSR mass\,:\,\footnote{In this article we use only the natural version of the MSR mass, in which heavy quark
virtual effects are integrated out.}
\begin{align}\label{eq:MSRdef}
\!\!\delta m_Q^\MSR &\equiv
 m_Q^{\rm pole}\!-m_Q^\MSR(R) = R\!\sum_{n=1}^\infty\! d_{n,0}(\nl)\!\biggl(\frac{\alpha^{(\nl)}_s(R)}{4\pi}\biggr)^{\!\!n}\nonumber\\
& =R\sum_{n=1}^\infty \sum_{k=0}^n d_{n,k}(\nl)\biggl(\frac{\alpha^{(\nl)}_s(\mu)}{4\pi}\biggr)^{\!\!n}
\ln^k\Bigl(\frac{\mu}{R}\Bigr)\,.
\end{align}
Exploiting the $\mu$ independence of the $\MSR$ mass, the $d_{n,k}$ coefficients can be expressed as a function of $d_{k,0}$
through the following recursion relation\,:
\begin{equation}\label{eq:recursion}
d_{n,k}(\nl) = \frac{2}{k}\,\sum_{i=k}^{n-1}\,i\,d_{i,k-1}(\nl)\,\beta_{n-1-i}(\nl)\,,
\end{equation}
with $\beta_i$ the coefficients of the QCD beta function, defined as
\begin{align}\label{eq:alphaRGE}
\frac{{\df} \alpha_s^{(n_f)}}{\df \ln \mu}=  
-\,2\,\alpha_s^{(n_f)}(\mu)\sum_{n=0}^\infty
\beta_n(n_f)\biggl(\frac{\alpha_s^{(n_f)}}{4\pi}\biggr)^{\!\!n+1}.
\end{align}
The renormalization group equation of the $\MSR$ mass is given by\,:
\begin{equation}\label{eq:Revol}
-\frac{\df}{\df R}m_Q^{\MSR}(R)=  
\sum_{n=0}^\infty \gamma_n^R(\nl) \biggl(\frac{\alpha_s^{(\nl)}(R)}{4\pi}\biggr)^{\!\!n+1},
\end{equation}
with $\gamma_n^R$ the R-anomalous dimension coefficients~\cite{Hoang:2017suc}, which are renormalon free. The
$\gamma_n^R$ can be easily calculated from the first line of Eq.~\eqref{eq:MSRdef} using that the pole mass is
$R$ independent
\begin{align}
\!\!\!\gamma_n^R(\nl) = d_{n+1,0}(\nl)-2\!\sum_{j=0}^{n-1} (n-j)\beta_j(\nl) d_{n-j,0}(\nl),
\end{align}
and the renormalon cancels between the first term and the sum. The solution of the RGE in Eq.~\eqref{eq:Revol} sums up
powers of $\log(R_1/R_2)$ to all orders in perturbation theory\,:
\begin{align}\label{eq:MSRsum}
\Delta^{\rm MSR}(\nl,R_1,R_2)&\equiv m_Q^{\rm MSR}(R_2) - m_Q^{\rm MSR}(R_1) \\
&=\!\int_{R_1}^{R_2}\!\df R\,\gamma_n^R[\nl,\alpha_s^{(\nl)}(R)]\,.\nonumber
\end{align}
The MSR mass can be easily matched to the $\MSb$ mass at the scale \mbox{$R = \mbar_Q(\mbar_Q)$}, and then
run down from there to any value of $R<\mbar_Q$.\\[0.05cm]

The $\MSb$ and MSR bottom masses (as well as the QCD static potential and the bottomonium masses) receive corrections
from the finite mass of the charm quark. Here and in what follows we assume the charm quark is massless, as its mass
plays no role in the calibration of the Cornell model but might add complications when scanning over the bottom
quark mass over a large range. We close this section noting that if we compute the MSR mass from the $\MSb$ employing
the world average values $\mbar_b = 4.18\pm0.03\,$GeV and $\alpha_s^{(n_f=5)}(m_Z) = 0.1181$ we obtain
$m_b^{\rm MSR}(R=1\,{\rm GeV}) = 4.679\,$GeV, which is only $1.54$ standard deviations away from $m_b^{\rm Cornell}$
as found in Eq.~\eqref{eq:Cornell-Fit}.

\section{The QCD Static Potential}\label{sec:StaticPot}
As in the Cornell model, to compute the energy levels of $Q\overline{Q}$ states in NRQCD at
leading order in $1/m_Q$ one needs the static QCD potential. It is defined as the potential between two static
quarks, that is, the color-neutral interaction between two infinitely heavy color-triplet states. It is well
defined up to $\mathcal{O}(\alpha_s^4)$, where the static approximation breaks down and the potential becomes
time-dependent. This feature manifests itself in dimensional regularization as an $1/\epsilon$ pole, that once regulated
leaves behind a dependence on the ``ultrasoft'' scale $\mu_{\rm us}$. The pole and $\mu_{\rm us}$ logarithms
cancel in observables such as quarkonium masses when adding ultrasoft effects from wave-function renormalization.
Nevertheless, to define the static potential at four loops we simply subtract the $1/\epsilon$ pole.

In position space the perturbative contribution to the static QCD potential can be written in a compact form as
follows\,:\,\footnote{As already mentioned, we omit the corrections coming form the finiteness of the charm
quark, which start at $\mathcal{O}(\alpha_s^2)$.}
\begin{align}\label{eq:static}
V^{(\nl)}_{\rm QCD}(r,\mu) =& - C_F\,\frac{\alpha^{(\nl)}_s(\mu)}{r}\sum_{i=0}\sum_{j=0}^{i}
\biggl(\frac{\alpha^{(\nl)}_s(\mu)}{4\pi}\biggr)^{\!\!i}\\
&\times a_{i,j}(\nl)\,\log^{\,j}(r\,\mu\, e^{\gamma_E})+
V_{\rm QCD}^{\rm us}(\nl,r)\,.\nonumber
\end{align}
The coefficients $a_{i,0}$ are known to four loops~\cite{Fischler:1977yf,Billoire:1979ih,Schroder:1998vy,Pineda:1997hz,Brambilla:1999qa,Kniehl:2002br,Penin:2002zv,Smirnov:2008pn,Smirnov:2009fh,Anzai:2009tm}
and $a_{i,j>0}$ can be derived from the former requiring that $V_{\rm QCD}$ does not depend on $\mu$,
with a recursion relation identical to Eq.~\eqref{eq:recursion}\,:
\begin{equation}
a_{i,j}(\nl) = \frac{2}{j}\sum_{k=j}^{i}\,k\,a_{k-1,j-1}(\nl)\,\beta_{i-k}(\nl)\,.
\end{equation}
For charmonium (bottomonium) one has $n_\ell = 3(4)$ and the coefficients $a_{i,0}$ take the following numerical values\,:
\begin{align}
a_{0,0} & = 1\,, & a_{1,0}& =7(5.88889)\,,\\
a_{2,0} & = 535.277(439.548)\,, &a_{3,0}& = 30374.4 (22666.1)\,.\nonumber
\end{align}
The $V_{\rm QCD}^{\rm us}(r)$ term in Eq.~\eqref{eq:static} depends on the ultrasoft factorization scale $\mu_{\rm us}$
and takes the following form\,:
\begin{align}\label{eq:usPot}
V_{\rm QCD}^{\rm us}(\nl,r)=-\frac{9\,C_F}{4\pi}\,\alpha_s^{(\nl)}(\mu)^4\,\frac{1}{r}\log\,(\mu_{\rm us}\,r)\,.
\end{align}
Following e.g.\ Ref.~\cite{Tormo:2013tha}, for the ultrasoft factorization scale we take the expression
\begin{align}
\mu_{\rm us}=\frac{N_c}{2}\,\mu\,\alpha^{(\nl)}_s(\mu)\,,
\end{align}
that takes into account the power counting of pNRQCD~\cite{Pineda:1997bj,Brambilla:1999xf}. In this article we
do not perform any resummation of large ultrasoft logarithms besides those that can be absorbed in the running
of $\alpha_s$. This ultrasoft resummation has been performed at N$^3$LL in Ref.~\cite{Brambilla:2009bi}.

In Refs.~\cite{Sumino:2003yp,Sumino:2004ht} it is shown from renormalon dominance arguments and in the framework of
the operator product expansion of pNRQCD, that perturbation theory alone should be capable of describing both the
Coulomb and linear behavior of the static potential, and that nonperturbative corrections start at
$\mathcal{O}(\LQCD^3r^2)$. In the following we shall confirm that claim numerically using the MSR scheme.

\begin{figure}[tbh!]\centering
\includegraphics[width=0.8\linewidth]{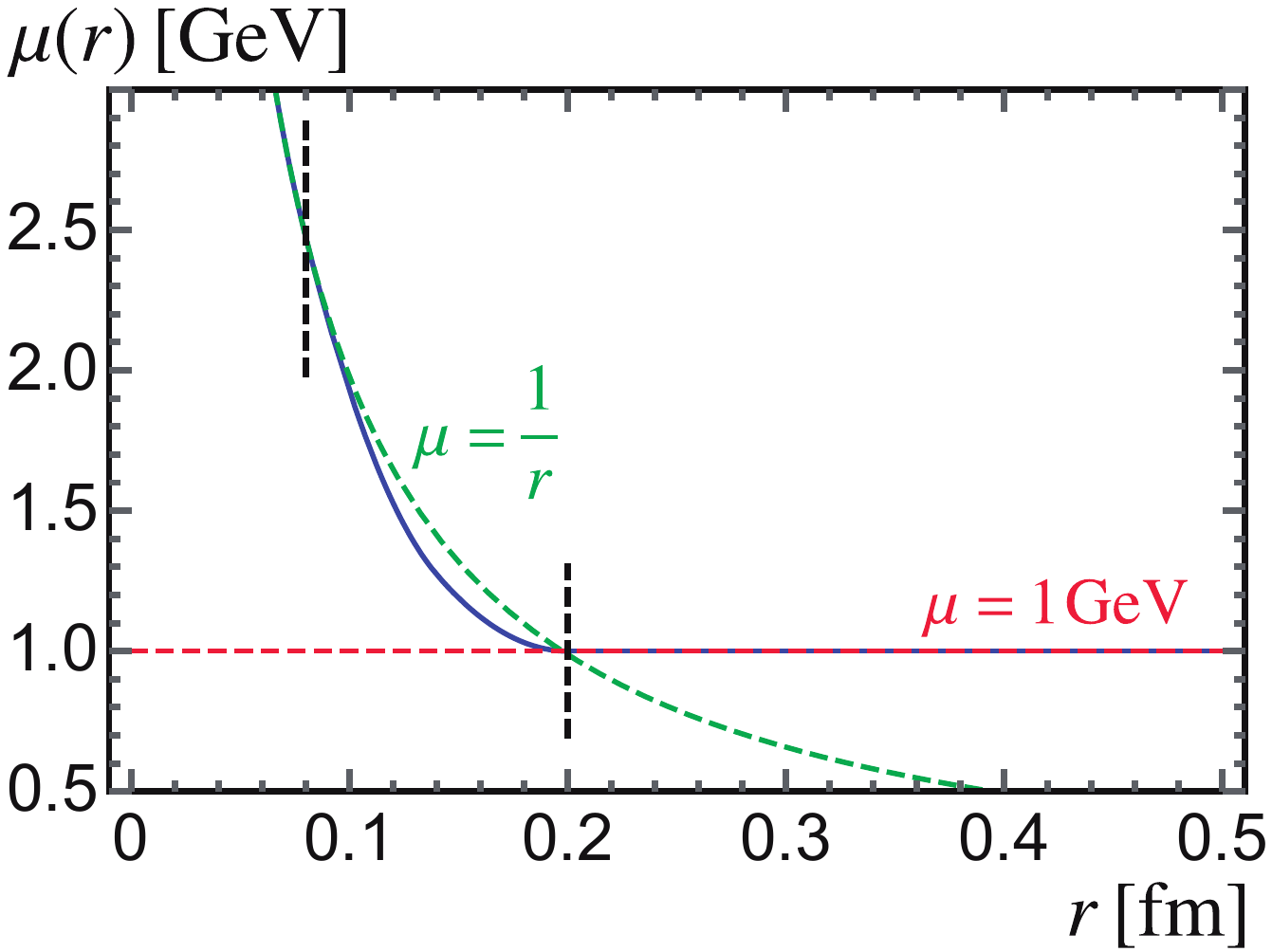}
\caption{Dependence of the static potential renormalization scale in Eq.~\eqref{eq:static} with $r$. In solid
blue the profile function of Eq.~\eqref{eq:profile} is shown, while dashed green and dashed red show the behavior
for small ($1/r$) and large (constant) $r$, respectively. The positions in which the piece-wise function changes its
functional form are signaled by vertical dashed black lines.}\label{fig:Profile}
\end{figure}

\begin{figure*}[tbh!]
\center
\subfigure[]
{\label{fig:polebottom}\includegraphics[width=0.39\textwidth]{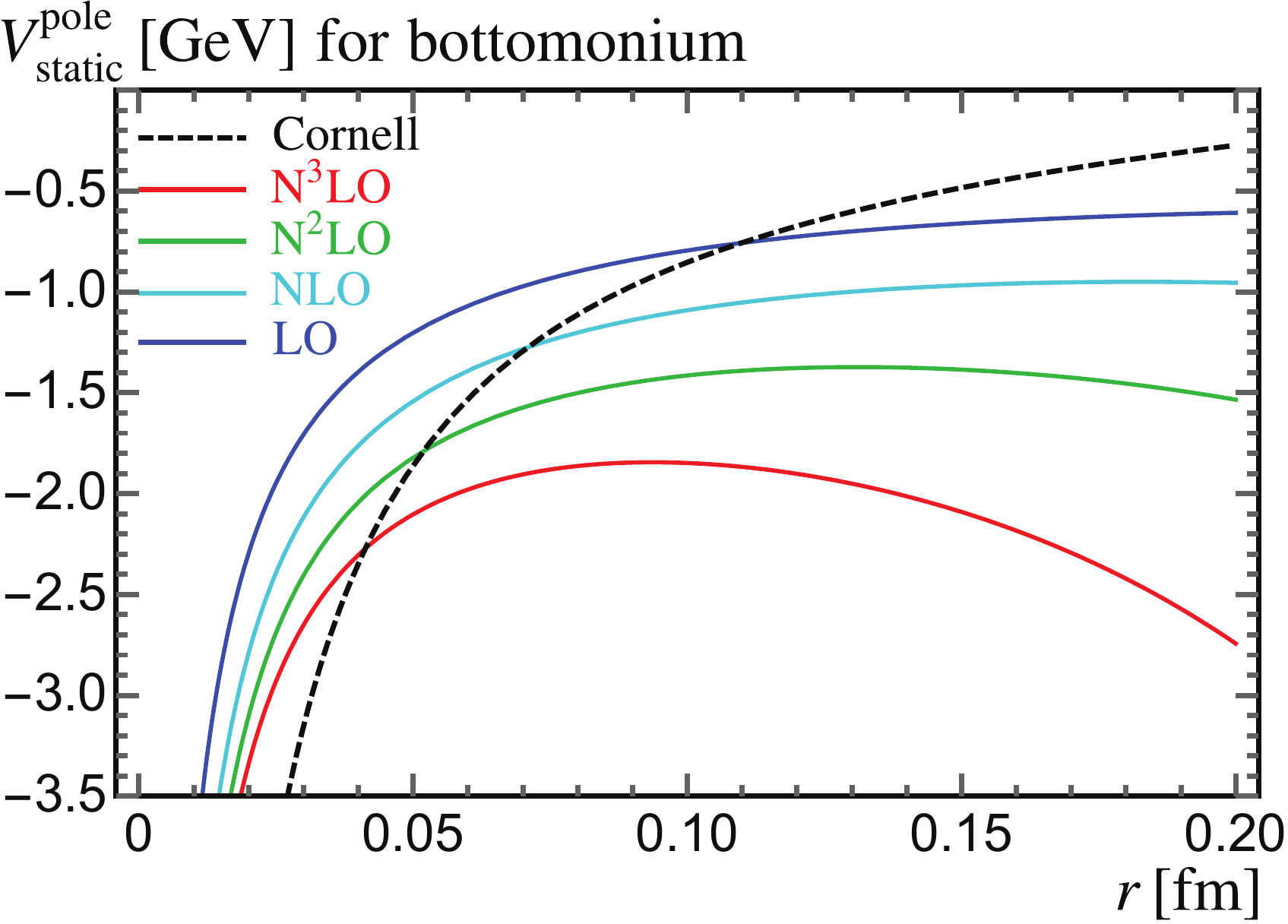}~~~~}
\subfigure[]
{\label{fig:MSRnbottom}\includegraphics[width=0.37\textwidth]{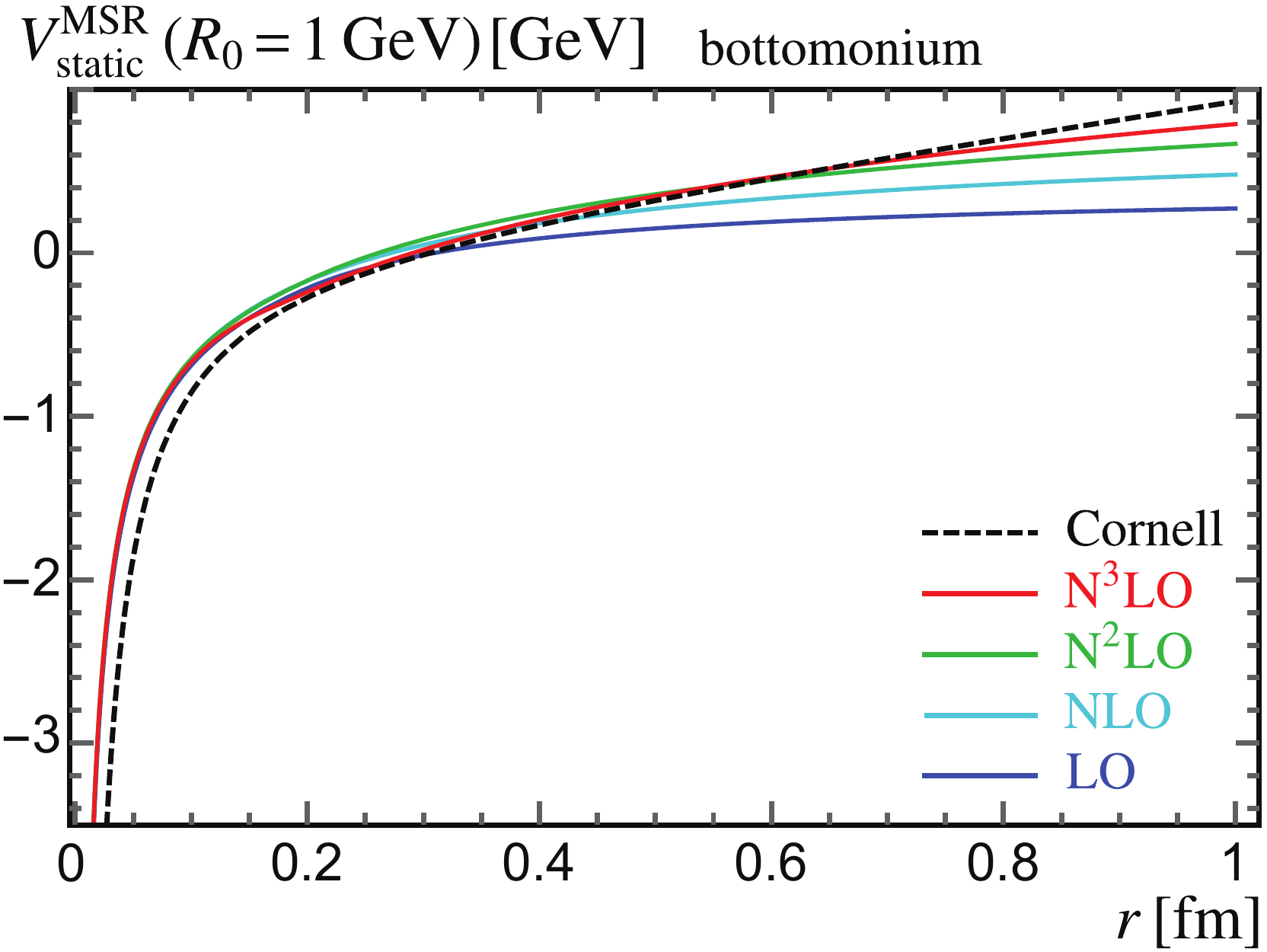} }
\subfigure[]
{\label{fig:polecharm}\includegraphics[width=0.39\textwidth]{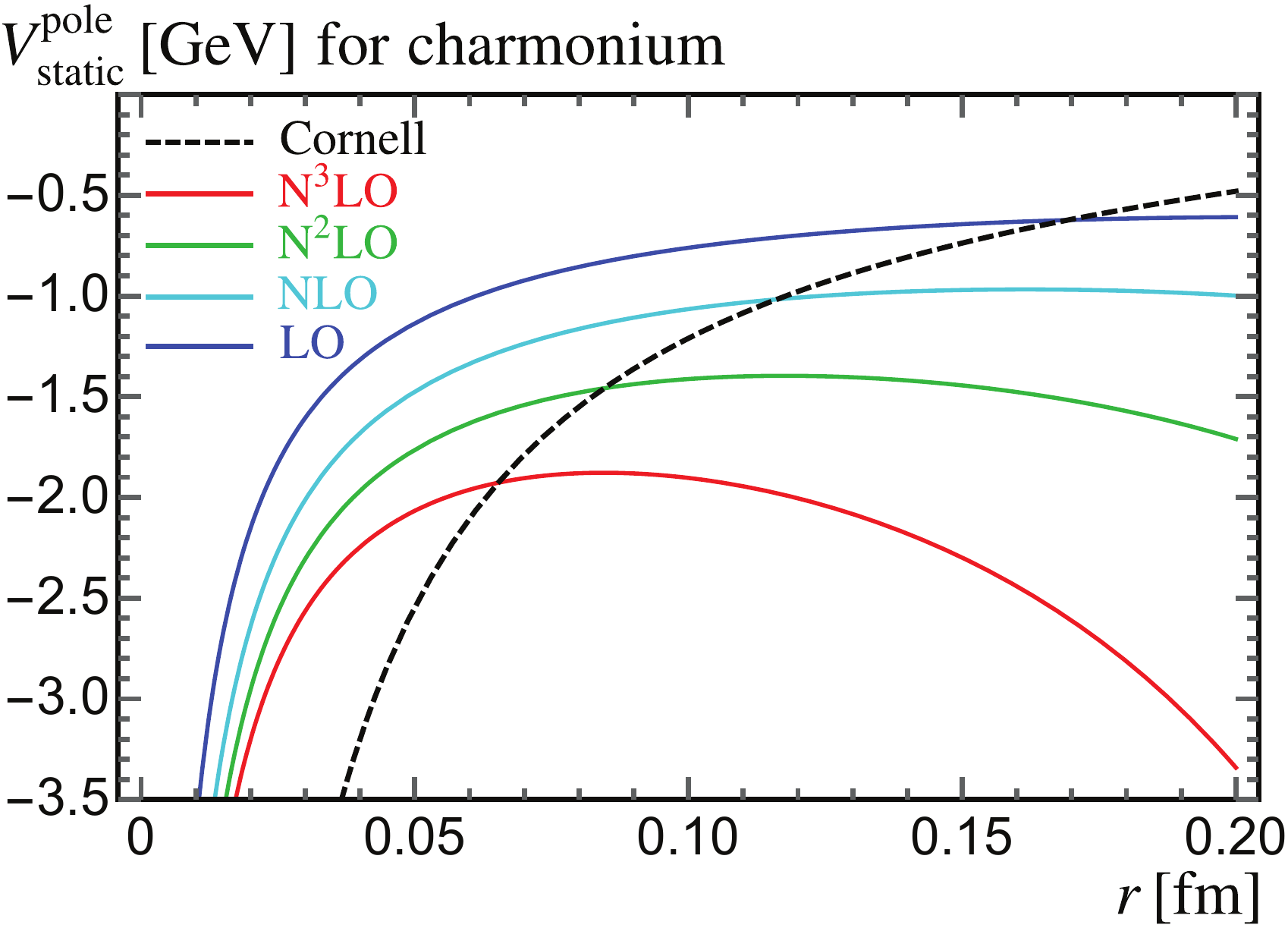}~~~~}
\subfigure[]
{\label{fig:MSRncharm}\includegraphics[width=0.37\textwidth]{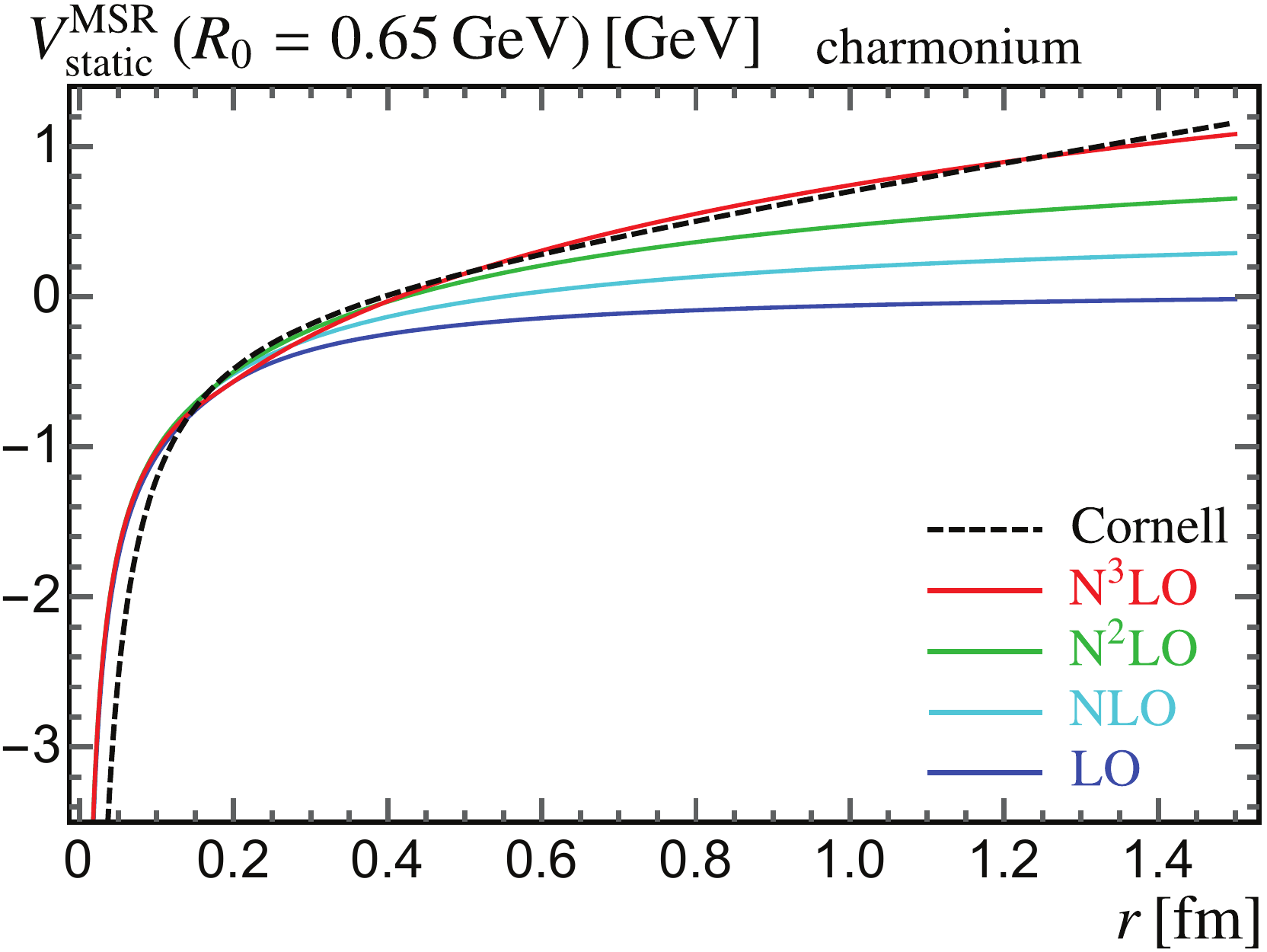} }
\caption{\label{fig:Static} Comparison of the Cornell potential (dashed black line) with the static QCD potential at
N$^3$LO (red), N$^2$LO (green), NLO (cyan) and LO (blue). Upper (lower) panels correspond to bottomonium (charmonium).
The left panels show the static QCD potential in the pole scheme, and the right panels use the MSR scheme with the
reference scales $R_0=1\,$GeV and $0.65\,$GeV for bottomonium and charmonium, respectively. The Cornell potential uses
the parameters of Eqs.~\eqref{eq:Cornell-Fit} and \eqref{eq:Charmonium-Fit}, while the Static QCD potential uses the
world average value $\alpha_s(m_Z) = 0.1181$. The left plot uses the canonical scale $\mu = 1/r$, but the right plot uses
our profile function shown in Fig.~\ref{fig:Profile}.}
\end{figure*}

The static QCD potential suffers from a factorially divergent growth at large orders, also known as a $u=1/2$ renormalon,
which translates into an $\mathcal{O}(\Lambda_{\rm QCD})$ ambiguity. This ambiguity happens to be $r$-independent,
and its nature depends only on the coefficients of the QCD beta function. In Figs.~\ref{fig:polebottom} and
\ref{fig:polecharm} this bad perturbative behavior manifests itself as a (roughly constant) vertical
shift of the potential in the region between $0.05\,$fm and $0.2\,$fm every time a new order is
included.\,\footnote{In this plot we choose a canonical scale for the renormalization scale $\mu=1/r$
because $\mu$ never becomes smaller than $1\,$GeV if $r\leq 0.2\,$fm, see Fig.~\ref{fig:Profile}.}
Furthermore, none of the orders makes the QCD static potential close to the Cornell model. It is well
understood~\cite{Pineda:1998id,Hoang:1998nz,Beneke:1998rk} that the ambiguity exactly matches that of
the pole mass except for a factor of $-2$, such that the static energy
$E_{\rm stat}(r,\mu) = 2\,m_Q^{\rm pole} + V_{\rm QCD}(r,\mu)$ is renormalon free. Since the pole mass ambiguity
is independent of the quark mass itself, the cancellation happens irrespectively of the specific numeric
value for the mass. Therefore one only needs to re-write the pole mass in terms of a short-distance scheme
to make the cancellation manifest. For the cancellation to take place one also needs to express the
perturbative series $\delta m_Q^{\rm SD}$ that relates the pole mass with a short-distance mass $m_Q^{\rm SD}$
in terms of $\alpha_s(\mu)$, as done in the second line of Eq.~\eqref{eq:MSRdef}.
There are  powers of $\log(\mu/R)$ in $\delta m_Q^{\rm SD}$ that
may become large if $\mu$ and $R$ are very different. Since one has to choose $\mu$ such that
$\log(r\,\mu\, e^{\gamma_E})\sim\mathcal{O}(1)$, that is $\mu$  should depend on $r$, renormalization schemes
with a fixed value of $R$ such as the $\MSb$, are disfavored. Following Ref.~\cite{Mateu:2017hlz} we use
the MSR mass and choose $\mu = R$ to simultaneously minimize logs in the potential and in $\delta m_Q^{\rm MSR}$.
Since the canonical choice $\mu=1/r$ quickly dives into non-perturbative values, we freeze it to $1\,$GeV once it
reaches this value. To avoid a kink in the potential we smoothly convert the $1/r$ behavior into a constant
employing a transition function between $r=0.08\,$fm and $0.2\,$fm. This function is composed of two quadratics
smoothly connected with each other and with $1/r$ and $1\,$GeV at the junction points, as has been employed
for instance in Ref.~\cite{Hoang:2014wka} in the context of event shapes. A graphical representation of this
piecewise form, which will be referred to as  ``profile function'', can be seen in Fig.~\ref{fig:Profile}, and has the
following analytical form
\begin{align}\label{eq:profile}
\!\!\!\mu(r)=\left\{\begin{array}{ll}
\frac{1}{r}                   &~~~~~~~~~\;\; r\le r_0\\[0.15cm]
f_{t1}(r)=c_1+c_2\,r+c_3\,r^2 &~~~ r_0\;<r<r_m\\[0.15cm]
f_{t2}(r)=c_4+c_5\,r+c_6\,r^2 &~~~ r_m<r<r_1\\[0.15cm]
\mu_c                         &~~~~~~~~~~\, r\ge r_1
\end{array}\right.,
\end{align}
where we use $r_m=(r_0+r_1)/2$ and $\mu_c$ is the constant renormalization scale at long distances, 
taken as $1\,$GeV.
The six coefficients $c_i$ in $f_{t1}(r)$ and $f_{t2}(r)$ are obtained by imposing continuity
of the functions and their derivatives at the transition points $r_0$, $r_m$ and $r_1$.
Consequently, we obtain
\begin{align}
c_1 &= \frac{4\,\mu_c\,r_0^3+ (4\,r_1^2-3\,r_0^2- 5\,r_0\,r_1)}{ 2\,r_0(r_0-r_1)^2}\,,\\
c_2 &= \frac{(6\,r_0^2-r_0\,r_1-r_1^2)-4\,\mu_c\,r_0^3}{r_0^2(r_0-r_1)^2}\,,\nonumber\\
c_3 &= \frac{4\,\mu_c\,r_0^2+(3\,r_1-7\,r_0)}{2\,r_0^2(r_0-r_1)^2}\,,\nonumber\\
c_4 &= \mu_c\,-\frac{4\,\mu_c\,r_0^2-\,(5\,r_0-r_1)}{2\,r_0^2(r_0-r_1)^2}\,r_1^2\,,\nonumber\\
c_5 &= \frac{4\,\mu_c\,r_0^2\,r_1-(5\,r_0-r_1)}{r_0^2(r_0-r_1)^2}\,,\nonumber\\
c_6 &= -\frac{4\,\mu_c\,r_0^2-(5\,r_0-r_1)}{2\,r_0^2 (r_0-r_1)^2}\,.\nonumber
\end{align}
For our specific choice $r_0=0.08$\,fm and $r_1=0.2\,$fm, the coefficients $c_{i}$ take the following numerical
values\,:
\begin{align}
c_{1} &= 30.8935\,{\rm fm}^{-1}\,, &c_{2} &= -\,303.588\,{\rm fm}^{-2}\,,\nonumber\\
c_{3} &= 920.865\,{\rm fm}^{-3}\,, &c_{4} &= 20.3164\,{\rm fm}^{-1}\,,\\
c_{5} &= -152.487\,{\rm fm}^{-2}\,,&c_{6} &= 381.218\,{\rm fm}^{-3}\,.\nonumber
\end{align}
Using the above values in the $\mu(r)$ profile [\,Eq.~\eqref{eq:profile}\,], the ultrasoft logarithm in
Eq.~\eqref{eq:usPot} remains between $-2$ and $-0.4$ for radii above $0.0005$\,fm making sure that
$V_{\rm QCD}^{\rm us}$ does not become unnaturally large. A similar scale setting is implemented in
Refs.~\cite{Pineda:2013lta,Peset:2018ria,Peset:2018jkf} in the  renormalon-subtracted scheme, but with a somewhat
more abrupt transition between the canonical and frozen regimes.

Our next goal is to define a short-distance potential which is renormalon free and independent of the quark mass.
This is a sensible criterion, since the static potential is defined in the limit of infinitely heavy quarks. It will
however depend on the scale $R_0$ at which the renormalon is subtracted. The concept of R-evolution comes in
very handy to accomplish our task. Some basic algebra brings the static energy into a convenient form\,:
\begin{align}\label{eq:MSRstaticE}
E_{\rm stat}(r,\mu) \,&\,=\, 2\,m_Q^{\rm pole} + V_{\rm QCD}(r,\mu) \\
\,&\,=\, 2\,m_Q^{\rm MSR}(R_0) + 2\,\delta m_Q^{\rm MSR}(R_0,\mu)
+ V_{\rm QCD}(r,\mu)\nonumber\\
\,&\,\equiv\,2\,m_Q^{\rm MSR}(R_0) + V^{\rm MSR}_{\rm QCD}(r,\mu,R_0)\,.\nonumber
\end{align}
Which allows us to define a renormalon-free potential at a given scale $R_0$. To avoid the occurrence large logs of $R_0/\mu$
in $\delta m_Q^{\rm MSR}(R_0,\mu)$ we express it in terms of $\delta m_Q^{\rm MSR}(R,\mu)$ and use R-evolution to
sum up large logs\,:
\begin{align}
\delta m_Q^{\rm MSR}(R_0) &= \delta m_Q^{\rm MSR}(R_0) + \delta m_Q^{\rm MSR}(R) - \delta m_Q^{\rm MSR}(R)\nonumber \\
& = \delta m_Q^{\rm MSR}(R) + m_Q^{\rm MSR}(R) - m_Q^{\rm MSR}(R_0) \nonumber\\
& = \delta m_Q^{\rm MSR}(R) + \Delta^{\rm MSR}(R,R_0)\,.
\end{align}
The term $\Delta^{\rm MSR}$, defined in Eq.~\eqref{eq:MSRsum}, is mass independent and sums up large logs of
$R/R_0$ to all orders. Therefore we can write the following R-improved expression for the MSR-scheme Static QCD
Potential\,:
\begin{align}\label{eq:MSRStat}
V^{\rm MSR}_{\rm QCD}(r,\mu,R_0) = &\,V_{\rm QCD}(r,\mu) + 2\, \delta^{\rm MSR}(R,\mu) \\
&+ 2\, \Delta^{\rm MSR}(R,R_0)\,. \nonumber
\end{align}
The R-improved static potential is similar to the Renormalon Subtracted scheme used in Ref.~\cite{Brambilla:2009bi},
based on \cite{Pineda:2001zq}, and to the analysis of Ref.~\cite{Sumino:2003yp} based on renormalon dominance.
Different choices of $R_0$ simply shift the potential vertically, as can be inferred from Eq.~\eqref{eq:MSRsum}.
Therefore $R_0$ can be used to parameterize the arbitrary origin of the potential energy. Given that the
Lattice QCD static potential is arbitrary up to a constant (in the Cornell potential this arbitrariness is absorbed into
the Cornell mass definition), we can use $R_0$ as a fit parameter when comparing it with our R-improved potential.
For the numerics shown in this article we use the canonical choice $R=\mu(r)$ that sets to zero all
logs related to the renormalon subtraction.

The result in Eq.~\eqref{eq:MSRStat} is shown for bottomonium in Fig.~\ref{fig:MSRnbottom} setting \mbox{$R_0=1\,$GeV},
and for charmonium in Fig. \ref{fig:MSRncharm} setting \mbox{$R_0=0.65\,$GeV}. We take $R_0$
similar to the value at which the renormalization scale freezes. This choice for the reference scale also makes
the bottom MSR mass $m_b^{\rm MSR}(R_0)$ very similar to the Cornell model parameter $m_Q^{\rm Cornell}$
in Eq.~\eqref{eq:Cornell-Fit}. Other values should simply shift the potential vertically, but for this specific choice
we observe that the static MSR potential converges very nicely towards the Cornell model for moderate values of $r$.
When more perturbative orders are added, the agreement becomes better over larger distances. On the other hand,
for high values of $r$, since the renormalization scale is frozen, $\log(r\mu)$ becomes large, which makes
perturbation theory unreliable. At small distances all orders agree very well because $\alpha_s$ becomes very small,
but disagree with the Cornell model. So we can conclude that the Cornell model and QCD agree for moderate values of
$r$, but disagree in the ultraviolet, as the model does not incorporate logarithmic modifications due to the running
of $\alpha_s$. For bottomonium the two potentials start disagreeing at a distance of approximately $r_0\sim0.2\,$fm,
which in natural units corresponds to a scale of roughly $1\,$GeV, in head on agreement with our choice of $R_0$.
A legitimate question is then if this difference in the UV can be absorbed in the definition of the
quark mass. We shall answer this question in Sec.~\ref{sec:results} of this article.

The ultrasoft potential in Eq.~\eqref{eq:usPot} is only a small contribution of the total MSR \mbox{R-improved} potential. At
N$^3$LO we find its weight is only $0.6\,\%$ at short distances, quickly becoming below the per-mil for
$r\gtrsim1\,$fm. To fully asses the impact of this term one would need to do a thorough study of perturbative
uncertainties through scale variation, which is beyond the scope of this article.
\section{Comparison to Lattice QCD}\label{sec:lattice}
In this section we perform a comparison to lattice QCD results. We start by comparing
the Cornell model parameters as obtained in fits to data to specific lattice studies that determine them. Finally
we compare our R-improved QCD static potential to lattice simulations.

Our result for $\sigma$ in Eq.~\eqref{eq:Cornell-Fit} is in very nice agreement with two lattice determinations\,:
Ref.~\cite{Koma:2006fw} uses Wilson loops and quotes $\sigma = 0.2098 \pm 0.0009\,$GeV$^2$, while
Ref.~\cite{Kawanai:2013aca} uses a relativistic heavy quark action and quotes $0.206 \pm 0.010\,$GeV$^2$. The comparison
to the result in Eq.~\eqref{eq:Charmonium-Fit} is worse, and this fact will be discussed in the next paragraph. When it comes to the
$\alpha_s^{\rm Cornell}$ parameter our bottomonium (charmonium) result is roughly a factor of $2$ larger than what
is found in these lattice analyses. Given that the static potential at short distances can be described in perturbation
theory, we know that loop corrections modify the short-distance $1/r$ behavior. Therefore fitting a
Coulomb-like function to lattice output might be meaningless, and the discrepancy should be of little concern.

Lattice simulations for the static potential use three dynamical quarks, and therefore their results should be
interpreted as the charmonium potential. Hence, it is confusing that we find better agreement for the linear
confining parameter in the fits to the bottomonium spectrum. However, as pointed out in Ref.~\cite{Ayala:2014yxa},
the charm quark is effectively decoupled for the low-lying bottomonium states. Therefore the same static potential
should be used for charmonium and bottomonium, as long as some (small) charm mass corrections are included in the
latter. Therefore one could think that also the same Cornell potential should be used for charmonium and bottomonium
systems, and given that the non-relativistic approximation is more accurate for the latter, the right comparison
might be between lattice and Eq.~\eqref{eq:Cornell-Fit}. Let us emphasize again that our Cornell model parameters are
not obtained from a direct comparison to the static QCD potential, but from fits to experimental data on quarkonium
bound states. At the sight of Figs.~\ref{fig:MSRnbottom} and \ref{fig:MSRncharm} it is clear that a smaller value of
$\alpha_s^{\rm Cornell}$ would improve the agreement with $V_{\rm static}^{\rm MSR}$.

\begin{figure}[t!]\centering
\includegraphics[width=0.9\linewidth]{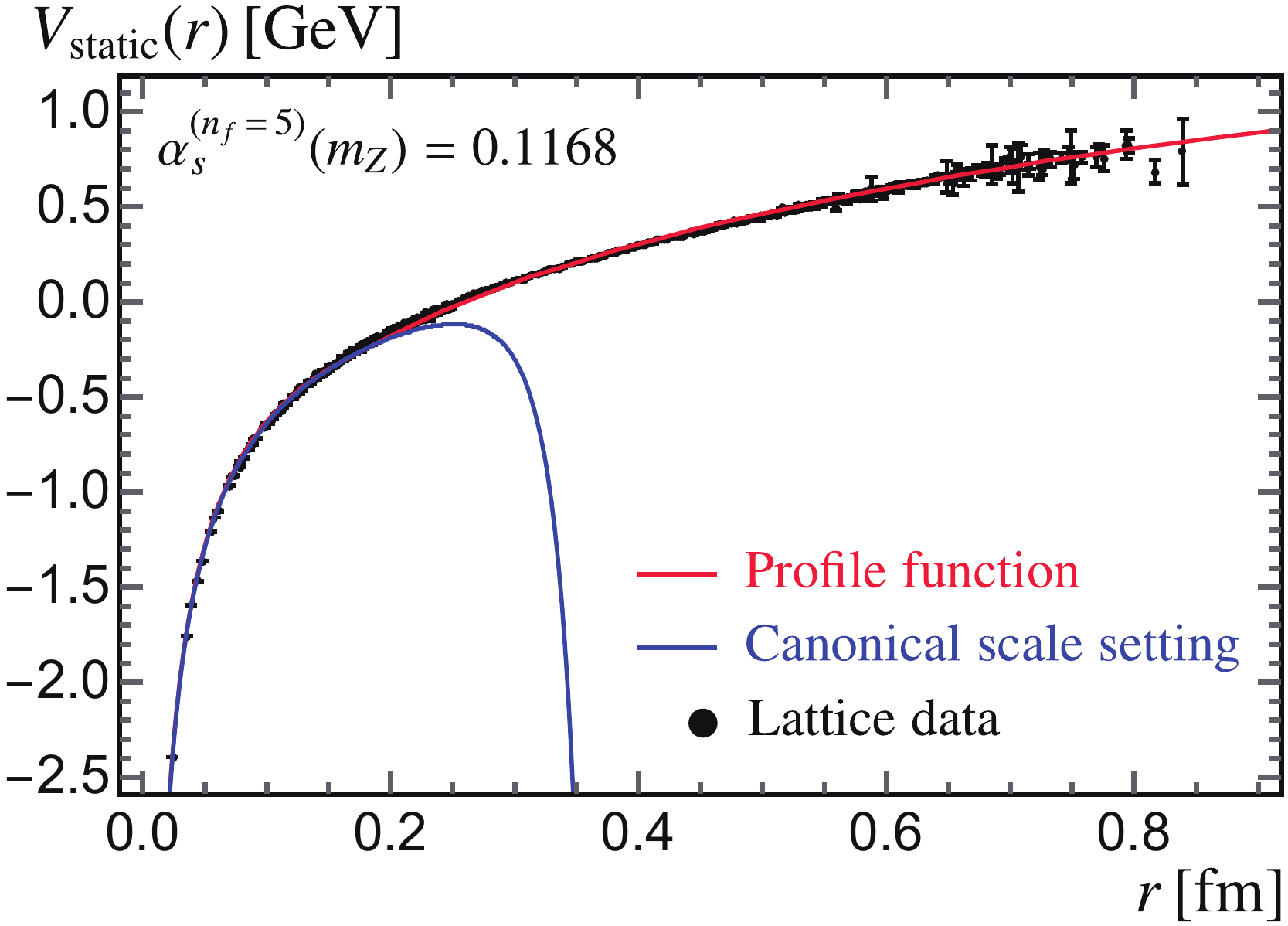}
\caption{Comparison of the R-improved perturbative static potential with lattice QCD results for $n_f = 3$
dynamical flavors. We use $\mu=R$, with $\mu$ set to the profile shown in Fig.~\ref{fig:Profile}. We determine
$\alpha_s$ and $R_0$ fitting our pNRQCD theoretical expression to the lattice
simulations.}\label{fig:Lattice}
\end{figure}

We turn now our attention to lattice computations of the static potential. On the lattice, the static approximation also
breaks down, and this manifests as a linearly divergent term. This divergence is removed by additive renormalization, as
studied in Ref.~\cite{Bazavov:2016uvm}. We will use the results of Ref~\cite{Bazavov:2014pvz} for a lattice
spacing $a=0.04\,$fm. These cover the range $0.039\,{\rm fm} \leq r \leq 0.84\,$fm, and have an average relative precision
of $2.6\,\%$. This precision seems to be roughly proportional to $1/r^2$, but shows a very pronounced peak around
\mbox{$r = 0.25\,$fm}. We complement this dataset with results from \cite{Bazavov:2017dsy} with lattice spacing
$a = 0.025\,$fm, which cover values of the radius as small as $0.024\,$fm. Since uncertainties in \cite{Bazavov:2017dsy}
are larger for higher values of $r$, we only consider data with $r\le0.25\,$fm. Moreover, in the range
\mbox{$0.024\,{\rm fm}\leq r \leq0.25\,$fm} \cite{Bazavov:2017dsy} has more density of points than \cite{Bazavov:2014pvz}.
Our complete dataset is plotted in Fig.~\ref{fig:Lattice} as black dots with error bars. The HotQCD collaboration has
results for larger lattice spacing, with predictions for the potential covering values for the radius as large as $1.59\,$fm.
However these do not extend as much into the small-$r$ region where perturbation theory dominates, and therefore we do not
take them into account in this simple comparison.

Since the potential is arbitrary up to an additive constant (that is, a vertical offset), which in its R-improved version
can be parametrized by the subtraction scale $R_0$ [\,see Eq.~\eqref{eq:MSRStat}\,],  and otherwise dependent only on
$\alpha_s$, we perform a two-parameter fit of our \mbox{R-improved} static QCD potential with the scale setting
shown in Eq.~\eqref{eq:profile} to the lattice data. Since a detailed analysis is beyond the scope of this article, we do
not include resummation of large ultrasoft logarithms and do not attempt to estimate perturbative uncertainties.
Furthermore, we do not study the dataset selection dependence and simply include all lattice points in
our $\chi^2$ function. Finally, since the correlation matrix is currently unknown, we assume individual lattice measurements
are statistically independent. Within this approximation we find\,:
\begin{align}\label{eq:alphaSfit}
\alpha_s^{(n_f = 5)}(m_Z) = 0.1168\,,\qquad R_0 = 1.024\,{\rm GeV}.
\end{align}
The result for $R_0$ is in remarkable agreement with the value used to compare with the Cornell model potential. We use the
five-loop QCD beta function to perform the $\alpha_s$ running. We cross the charm and bottom quark thresholds using the
four-loop matching relation. The matching from $n_f=3$ to $n_f=4$ is performed at the scale $\mbar_c = 1.3\,$GeV, and the
matching from $n_f=4$ to $n_f=5$ at the scale \mbox{$\mbar_b = 4.2\,$GeV}.
We refrain from showing fit uncertainties as they do not reflect the actual accuracy one can reach by this procedure.
The resulting QCD potential which uses the best-fit value for $\alpha_s$ and the vertical offset is shown as a red
line in Fig.~\ref{fig:Lattice}. This plot also shows as a blue line the R-improved static QCD potential with the same
values for the parameters, but using a fully canonical profile $\mu = 1/r$. Whereas our profile function seems to
capture some of the infrared physics, completely agreeing with lattice QCD results up to distances of approximately
$1\,$fm, the canonical profile behaves in an unphysical manner for $r\gtrsim 0.2\,$fm. We believe our findings can
have an impact on the precise determination of the strong coupling constant from fits to lattice simulations on the QCD
static potential by enlarging the range of data than can be utilized in constructing the $\chi^2$. Such determinations
have been carried out in Refs.~\cite{Bazavov:2012ka,Bazavov:2014soa}. To the best of our knowledge, the maximum
value of $r$ considered in those fits is $r_{\rm max} \simeq 0.23\,$fm.

\section{Analytic NRQCD Formula for Bottomonium with Massless Light and Charm Quarks}\label{sec:massless}
The complete bottomonium spectrum up to $n_p = 2$ for arbitrary \mbox{$\mbar_b\equiv\mbar_b(\mbar_b)$}
will be constructed using Non-Relativistic Quantum Chromodynamics (NRQCD) up to N$^3$LO~\cite{Kiyo:2014uca},
which will be briefly explained below (further details can be found e.g.\ in Ref.~\cite{Mateu:2017hlz}).

In the pole mass scheme~\cite{Penin:2002zv,Beneke:2005hg,Kiyo:2014uca}, the energy of a non-relativistic $Q\overline{Q}$
bound state, characterized by the quantum numbers $(n_p,j,\ell,s)$ and with $\nl$ massless active flavors, can be written
as\,:\,\footnote{Again we neglect the corrections coming from a finite charm quark mass, which start at
$\mathcal{O}(\varepsilon^2)$. We also work in the $n_\ell$ scheme, see Ref.~\cite{Mateu:2017hlz} for more details.}
\begin{align}\label{eq:EXpole}
 M^{(\nl)}_{n_p,j,\ell,s}(\mu) \,=\, &2\,m_Q^{\rm pole}\!
\Bigg[1-\frac{C_F^2\,\alpha^{(\nl)}_s(\mu)^2}{8n_p^2}\\
&\times\sum_{i=0}^{\infty}\bigg(\frac{\alpha^{(\nl)}_s(\mu)}{4\pi}\bigg)^{\!\!i}\,
\varepsilon^{i+1}P_i(L_{\nl})\Bigg],\nonumber
\end{align}
where
\begin{align}\label{eq:NRlog}
 L_{\nl}&=\log\!\bigg(\frac{n_p\mu}{C_F\alpha_s^{(\nl)}(\mu)m_Q^{\rm pole}}\bigg)+ H_{n_p+\ell}\,,\\
P_i(L) &= \sum_{j=0}^i\,c_{i,j}\,L^j\,,\nonumber
\end{align}
with $H_{n}$ the $n$-th harmonic number. In Eq.~\eqref{eq:EXpole} $\varepsilon$ acts as a bookkeeping
parameter that properly implements the so called $\Upsilon$-expansion~\cite{Hoang:1998ng,Hoang:1998hm}.
The $c_{i,0}$ coefficients have been computed up to $i=3$~\cite{Brambilla:2001qk,Kiyo:2013aea},
while the $c_{i,j>0}$ coefficients can be directly obtained from the latter $c_{i,0}$ imposing $\mu$
independence of the quarkonia mass. We denote the sum in Eq.~\eqref{eq:NRlog} truncated to
$\mathcal{O}(\varepsilon^{n+1})$ as the N$^n$LO result. This formula does not include the resummation of
large ultrasoft logarithms. Recently this summation has been carried to N$^3$LL precision for P-wave states
in Ref.~\cite{Peset:2018jkf}.

The $u=1/2$ renormalon in the QCD static potential discussed in Sec.~\ref{sec:StaticPot} is inherited by the
perturbative expansion of the quarkonia mass in Eq.~\eqref{eq:EXpole}. Exactly as it happened for the potential,
such renormalon is canceled by expressing the pole mass in terms of a short-distance mass. In the past the
$\MSb$ mass has been employed, although the natural scenarios for this scheme are processes where the involved energy
scale is much larger than the heavy quark mass. Even though the QCD static potential starts off at $\mathcal{O}(\alpha_s)$,
the first correction to the quarkonia masses is $\mathcal{O}(\alpha_s^2)$.\,\footnote{This extra factor of $\alpha_s$ comes
from solving the Schr\"odinger equation, and it has been argued it should be counted as $\mathcal{O}(1)$.} When switching
the pole mass to a short-distance scheme in Eq.~\eqref{eq:EXpole} the first correction from this conversion will be
$\mathcal{O}(\alpha_s)$, which appears to follow a different pattern. It is at this point when the use of the $\varepsilon$
parameter becomes relevant. Since the renormalon cancellation happens already in the static potential, it is crucial to treat
the short-distance mass subtractions in Eq.~\eqref{eq:EXpole} exactly as in the static energy, and therefore terms
proportional to $\alpha_s^n$ in the subtraction are regarded as $\mathcal{O}(\varepsilon^n)$ in the
$\Upsilon$-expansion counting.

Heavy quarkonium mases probe energy scales below $m_Q$, and therefore the relativistic logs showing up in the
$\MSb$-pole relation series $\delta^{\rm SD}$, namely $\log(\mu/\mbar_Q)$, may become large if $\mu$
is chosen to minimize the non-relativistic logarithm that appears in Eq.~\eqref{eq:NRlog}. Therefore it would be
better to simultaneously minimize logarithms appearing in $\delta m_Q^{\rm SD}$ and in $L_{\nl}$, whose argument is
the ratio of a non-relativistic scale and $\mu$. This is in full
analogy with the static potential analysis, where $r$ is being replaced by the Bohr radius in bound states masses.
Therefore, a low-scale short-distance mass is advisable. Following the results of Sec.~\ref{sec:StaticPot} and the
analysis in Ref.~\cite{Mateu:2017hlz} we will employ the MSR mass~\cite{Hoang:2017suc} for our analysis.
Therefore we express $m_Q^{\rm pole}$ in Eqs.~\eqref{eq:EXpole} and \eqref{eq:NRlog} in terms of the
MSR mass as shown in the second line of Eq.~\eqref{eq:MSRdef}, and re-expand the result in powers of $\alpha_s(\mu)$
respecting the $\Upsilon$ counting scheme. In this way the quarkonia masses depend on two renormalization scales,
namely $\mu$ and $R$.

\section{Bottomonium masses with a floating bottom mass}\label{sec:strategy}
The ultimate aim of this article is to calibrate the Cornell model constants (especially the Cornell mass) in terms
of the QCD fundamental parameters $\alpha_s$ and $m_b$. To achieve this goal one has to scan over these two parameters.
Specifically, we need QCD predictions for the eight lowest-lying bottomonium resonances in a reasonable range of values for
the bottom mass and the strong coupling constant. For the former we will consider the recent determination of
Ref.~\cite{Mateu:2017hlz}, $\mbar_b= 4.2\,$GeV, and vary the $\MSb$ mass between $4$ and $8\,$GeV in steps of $500\,$MeV.
Since the value of $\alpha_s$ that enters the perturbative expansion in Eq.~\eqref{eq:EXpole} has $n_\ell=4$ active
flavors, if computed from the reference value $\alpha_s^{(n_\ell=5)}(m_Z)$, it retains a (small) residual dependence on $m_b$
from threshold contributions. To make sure no bottom mass effects come from these matching corrections, we keep fixed the
value of $\alpha_s^{(n_\ell=4)}(1.3\,{\rm GeV})$. We consider the current world average for $\alpha_s$, which translates
into $\alpha_s^{(n_\ell=4)}(1.3\,{\rm GeV}) = 0.38331$, plus an evenly spaced grid between $0.34$ and $0.41$ in steps
of $0.01$, which corresponds to varying $\alpha_s^{(n_\ell=5)}(m_Z)$ between $0.114$ and $0.12$ for $\mbar_b = 4.2\,$GeV.

\begin{figure}[t!]\centering
\includegraphics[width=0.85\linewidth]{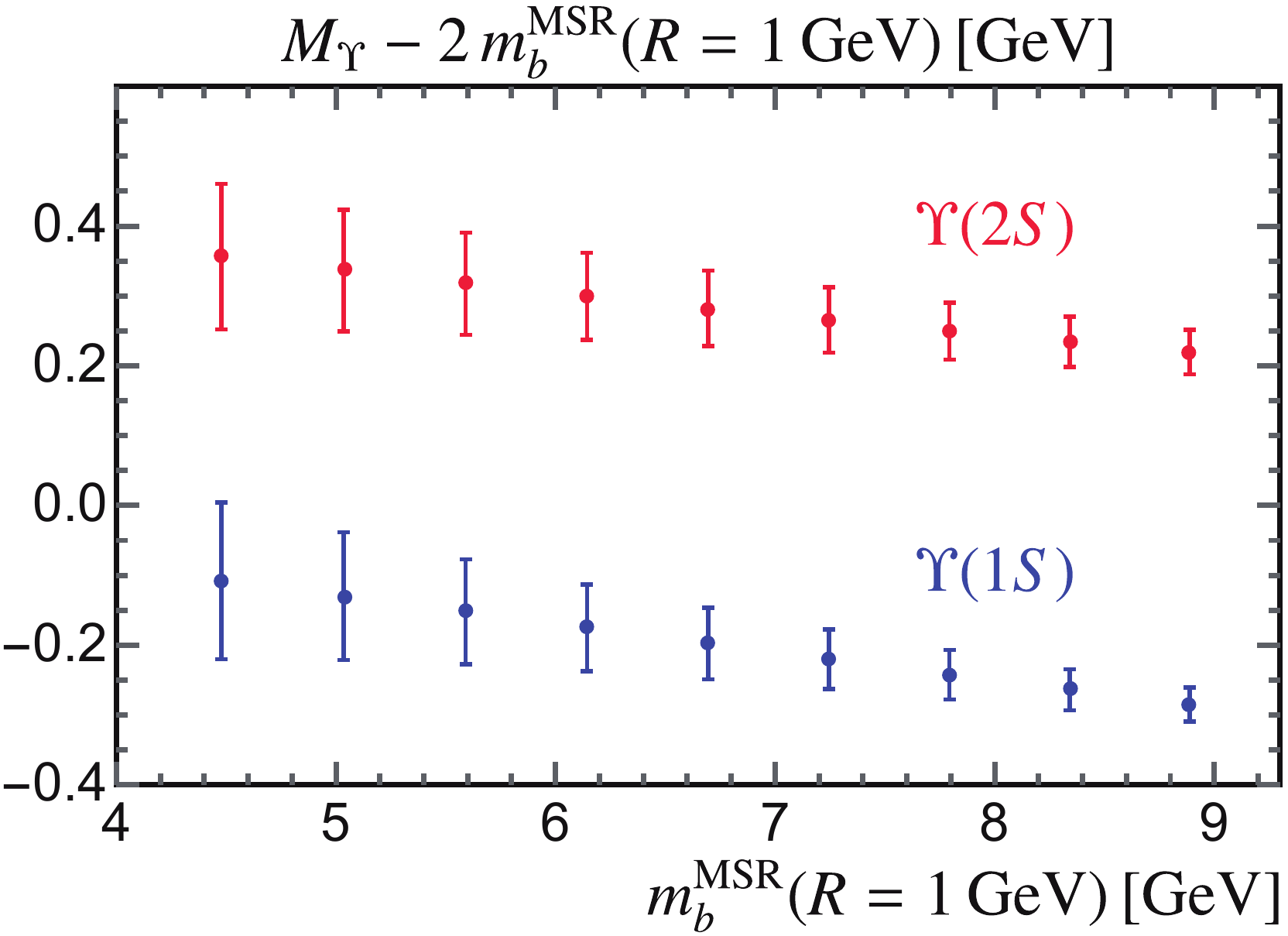}
\caption{Dependence of the $\Upsilon(1S)$ (blue) and $\Upsilon(2S)$ masses with the bottom mass. We subtract from the
hadron masses twice the MSR bottom mass at the scale $R = 1\,$GeV.}
\label{fig:mb-scan-Upsilon}
\end{figure}

We generate QCD predictions for the bottomonium spectrum varying the two renormalization scales $\mu_{n_p}$ and
$R_{n_p}$ which depend only on the principal quantum number. For a given value of $n_p$, the two scales
are varied independently, but $\mu$'s and $R$'s for various $n_p$ values are correlated, as explained in
Ref.~\cite{Mateu:2017hlz}. This makes sure theoretical correlations are properly
propagated but avoids the so called d'Agostini bias~\cite{DAgostini:1993arp}. The range in which $\mu_{n_p}$
is varied comes from analyzing the argument of the logarithm in Eq.~\eqref{eq:NRlog} in the MSR scheme, such that
its value varies between $1/2$ and $2$. Such range will depend on the value of the bottom mass, and therefore we will
adapt the results of Ref.~\cite{Mateu:2017hlz} for the masses covered in our scan.  Since the mass subtraction
involves powers of logs of $\mu_{n_p}/R_{n_p}$ that should be $\mathcal{O}(1)$, we take the same variation for $\mu_{n_p}$ and
$R_{n_p}$. We find that the upper limit of $\mu_{n_p}$ is independent of both $m_b$ and $n_p$, and hence we fix
$\mu^{\rm max}_{n_p} = 4\,$GeV. We find that $\mu^{\rm min}_{n_p}$ depends on $n_p$ and increases approximately
linearly with $m_b$. We parameterize them with the following approximate expressions\,:
\begin{align}\label{eq:mui}
\mu_1^{\rm min} & = 0.638\,{\rm GeV} + 0.209\,\overline{m}_b\,,\\
\mu_2^{\rm min} & = 0.510\,{\rm GeV} + 0.120\,\overline{m}_b\,.\nonumber
\end{align}
Since both values have a positive slope, the range in which scales are varied decreases as $m_b$ increases, and larger
scales are probed. This renders smaller perturbative uncertainties for larger values of the bottom mass, as expected.
In Fig.~\ref{fig:mb-scan-Upsilon} the dependence of the mass of the two lowest-lying vector resonances with $m_b$ is
shown graphically. We remove its main contribution, namely twice the bottom mass in the MSR scheme 
at the scale $R = 1\,$GeV to better see how the uncertainty shrinks as we increase $m_b$. We also observe that these
residual masses slightly decrease as the QCD bottom mass increases.

Once we have generated our ensemble of bottomonium masses for a given bottom mass, we will determine the parameters
of the Cornell model that best reproduce the QCD prediction. We generate QCD pseudo-data at LO, NLO, N$^2$LO and
N$^3$LO, which can be viewed as a set of highly correlated experimental measurements. This makes a regular $\chi^2$
fit with a non-diagonal covariance matrix impossible due to the d'Agostini bias. But if we do not include the theoretical
covariance matrix there are no other uncertainties left and it is not even possible to write down a $\chi^2$ function. Therefore
we will simply make a statistical regression analysis, which will provide us with a dispersion uncertainty. The strategy is
then very similar to that of Ref.~\cite{Mateu:2017hlz}\,: the QCD renormalization scales are varied in a correlated
way in terms of two dimensionless variables $\mu_{n_p} = \mu_{n_p}^{\rm min} + x\,(4\,{\rm GeV} - \mu_{n_p}^{\rm min})$,
\mbox{$R_{n_p} = \mu_{n_p}^{\rm min} + y\,(4\,{\rm GeV} - \mu_{n_p}^{\rm min})$}, with $\mu_i$ defined in Eq.~\eqref{eq:mui} and
$n_p = \{1, 2\}$, $0\le \{x,y\}\le1$. We define our regression $\chi^2$ function as follows\,:
\begin{align}\label{eq:chi2Reg}
\chi^2(x,y,\mbar_b)=\frac{\sum_i\big(M_i^{\rm QCD}(x,y,\mbar_b)-M_i^{\rm Cornell}\big)^{2}}
{\frac{\chi^2_{\rm min}(x,y)}{\rm d.o.f.}}\,,
\end{align}
where in practice $\chi^2_{\rm min}(x,y)$ is only known after the minimization is carried out, but renders the $\chi^2$
function dimensionless, and yields the right dimensions to the covariance matrix and parameter uncertainties. With this
definition, the reduced $\chi^2$ is exactly one at the minimum. For a given value of $\mbar_b$ we scan over all possibles
values of $\{x,y\}$, and for each pair we determine the parameters of the Cornell model. The average of the best-fit values
and regression uncertainties become the central value and fit uncertainty, respectively, while the semi-sum of the largest
and smaller values attained for each parameter in the scan is taken as the theoretical uncertainty. The latter dominates over
the fit uncertainty. After this procedure is repeated for all values of $\mbar_b$ and $\alpha_s$, we obtain the Cornell model
parameters as functions of these QCD fundamental quantities.

\section{Results of the Calibration}\label{sec:results}
\begin{figure*}[tbh!]
\center
\subfigure[]
{\label{fig:N3LO}\includegraphics[width=0.47\textwidth]{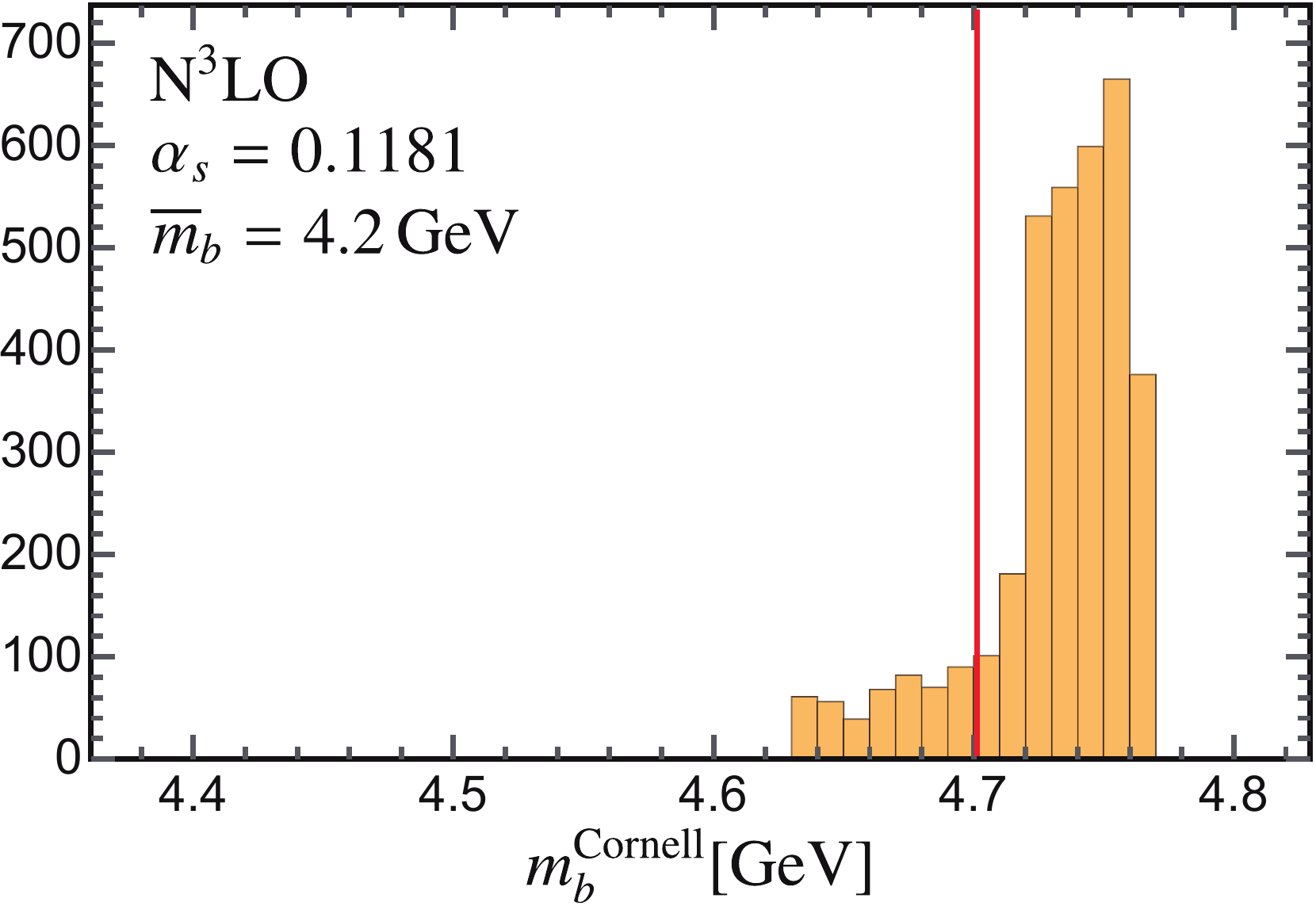}~~}
\subfigure[]
{\label{fig:N2LO}\includegraphics[width=0.47\textwidth]{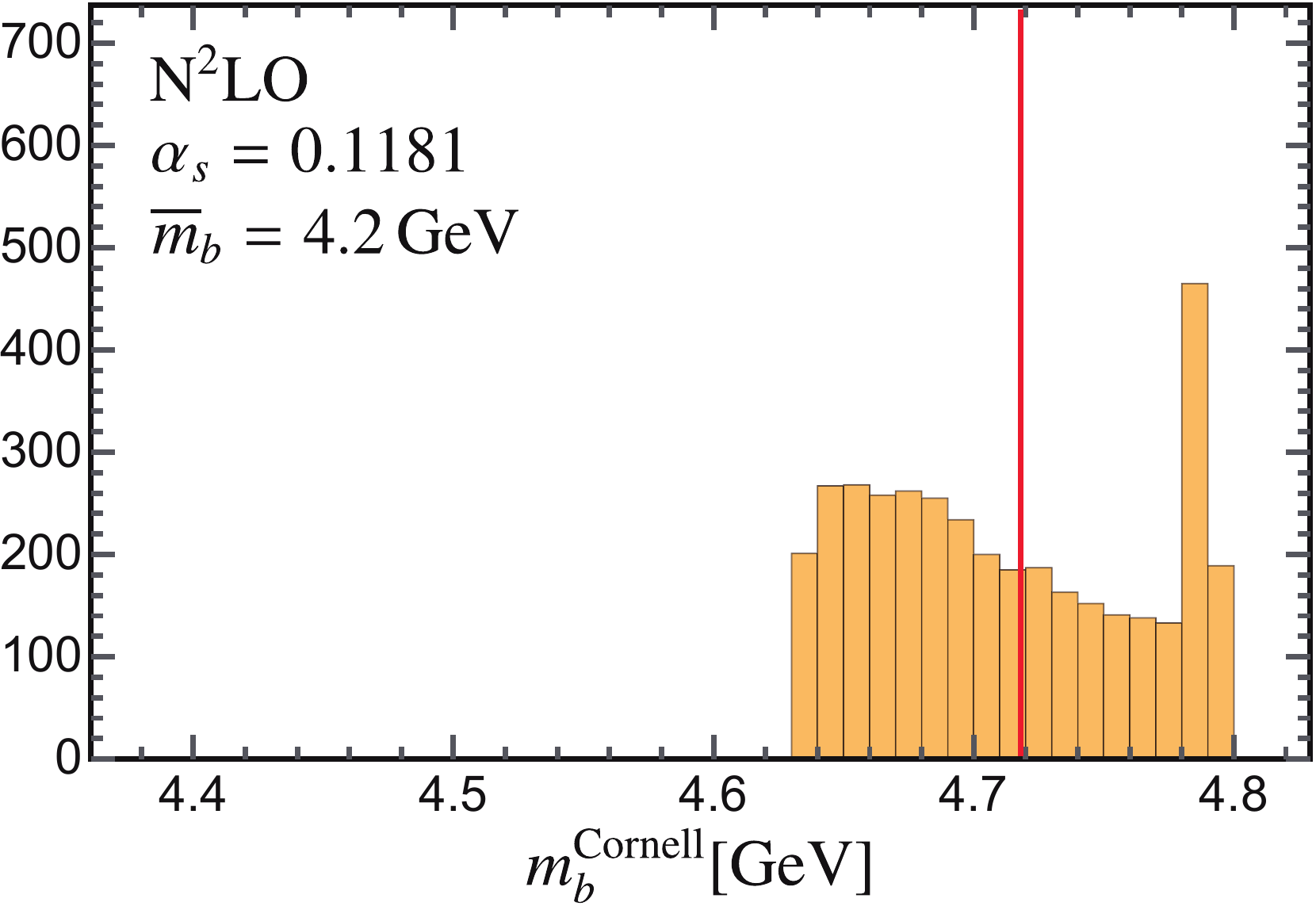} }
\subfigure[]
{\label{fig:NLO}\includegraphics[width=0.47\textwidth]{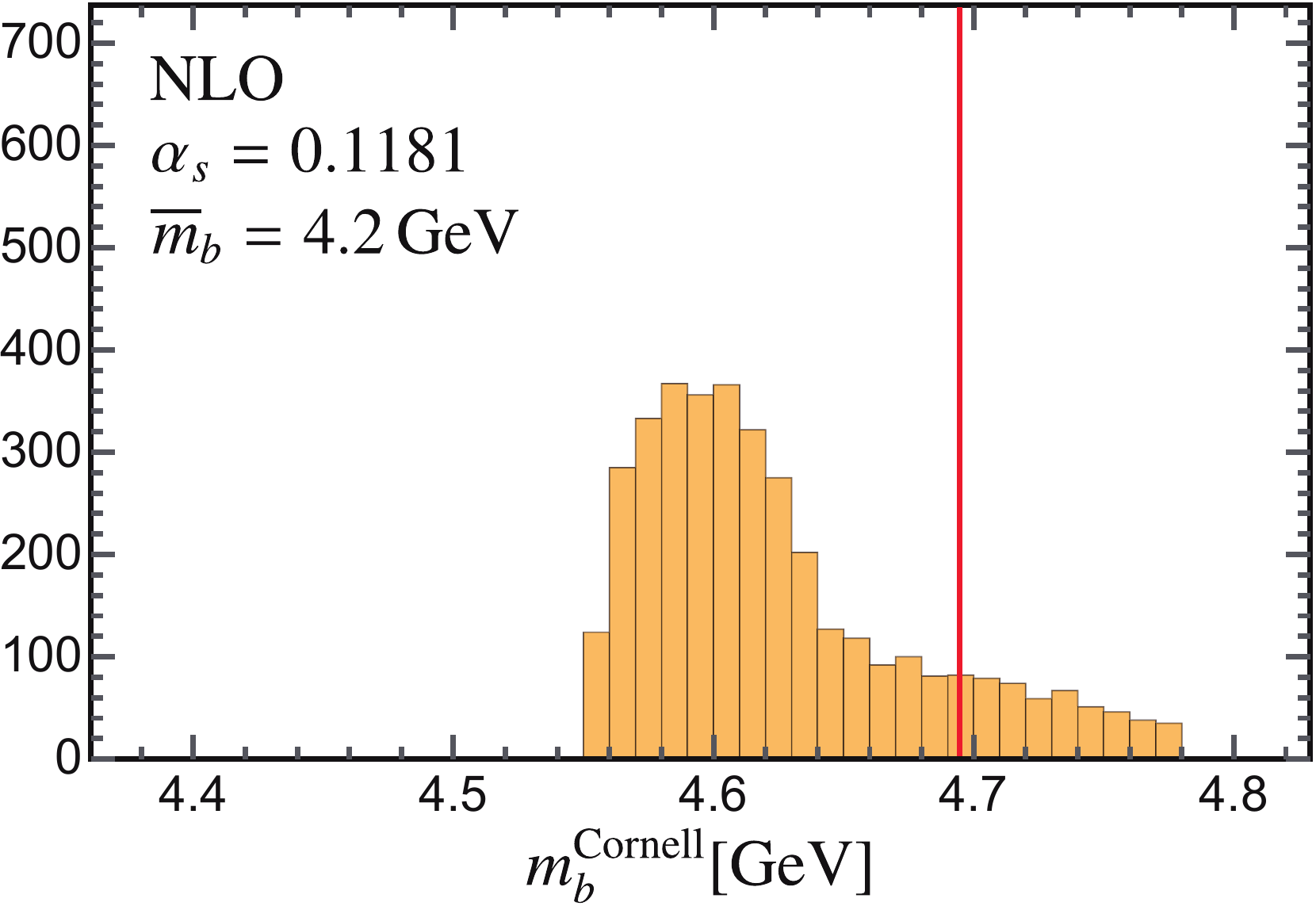}~~}
\subfigure[]
{\label{fig:LO}\includegraphics[width=0.47\textwidth]{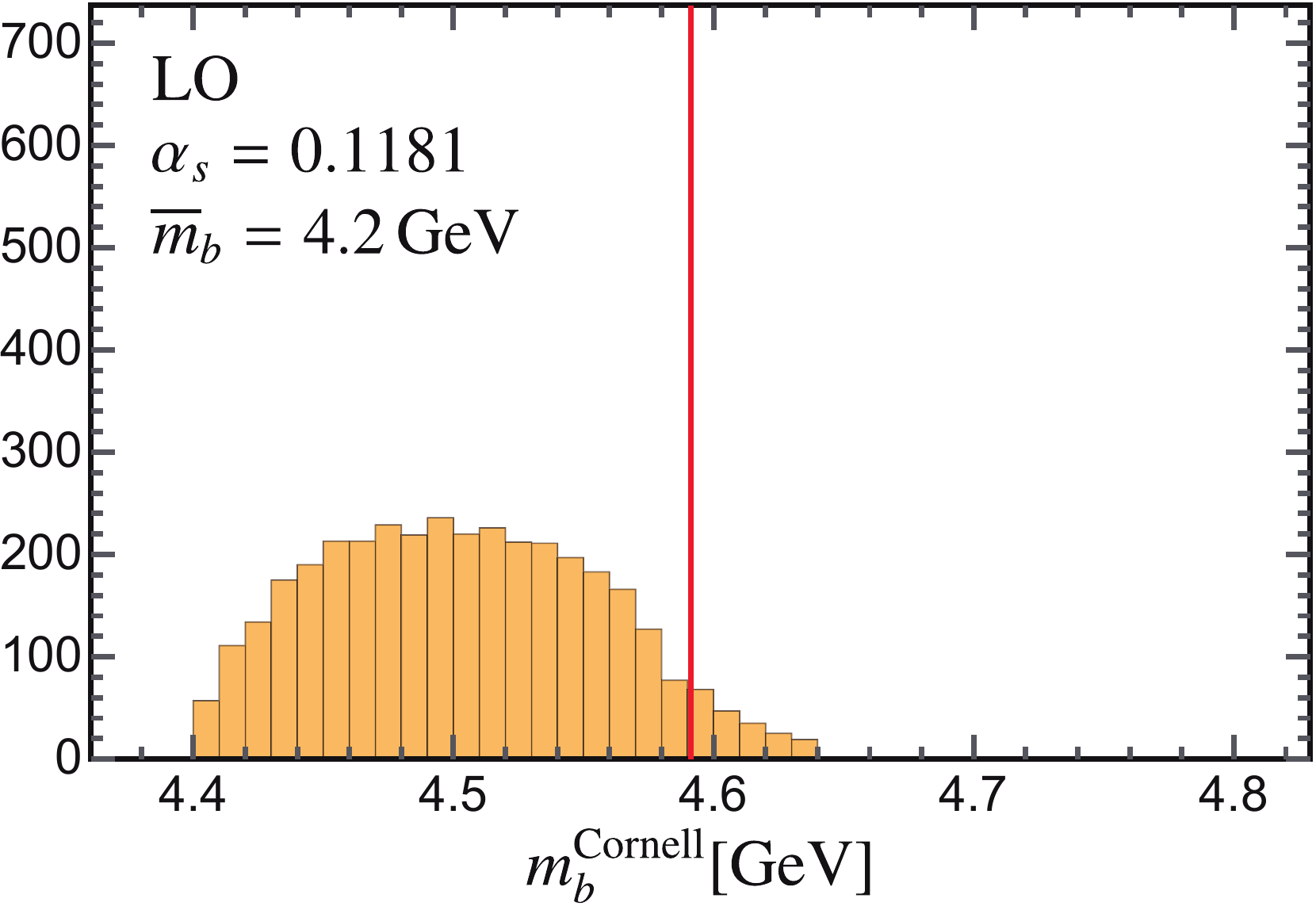} }
\caption{\label{fig:histograms} Distribution of Cornell masses in the theory renormalization scale scan for
\mbox{$\mbar_b = 4.2\,$GeV} using the world average value for the strong coupling constant when compared to QCD results
at N$^3$LO (a), N$^2$LO (b), NLO (c) and LO (d). The red vertical line shows the MSR mass at $R=1\,$GeV, matched and
run from $\mbar_b$ at the same order in the $\Upsilon$-expansion than the meson masses.}
\end{figure*}
We start by showing details of the calibration for a given value of the $\MSb$ bottom mass, using the world average value
for $\alpha_s$. We take the value \mbox{$\mbar_b = 4.2\,$GeV}, as obtained from NRQCD fits to bottomonium states,
which corresponds to \mbox{$m_b^{\rm MSR}(R = {1\,\rm GeV}) = 4.701\,$GeV} at four loops. Given these values of the
bottom mass
and the strong coupling constant, at each order we get $3719$ triads of best-fit values. A useful way of showing these
results is through histograms, in which, after appropriate binning, one can see how often a given value of a parameter
is generated. For the Cornell mass parameter, we choose our bins equally spaced, with bin-size $10\,$MeV. A closer
look into the histograms reveals that there are large, nearly unpopulated tails extending towards low mass values. This
would be of no concern if the renormalization scales were parameters that could be varied in a Gaussian way, since the
average and standard deviation automatically dump such effects. This is not our case, and therefore these tails make the
uncertainties unnaturally large, and can bias the central values. Therefore we cut off bins whose height is less than
$8\,\%$ of the highest bin. This translates into discarding between $4\,\%$ and $15\,\%$ of the best-fit values for each order,
leaving always ensembles of more than $3000$ elements. The resulting histograms can be viewed in Fig.~\ref{fig:histograms},
together with the corresponding value of the MSR mass at the reference value of $1\,$GeV, which happens to be always within
the covered values of the Cornell mass. One can see a very clear peak at the highest order, which gets somehow broader
towards lower orders. At N$^2$LO we observe two maximums, one of them much narrower and higher than the other. Except at
lowest order, the distributions are not symmetric. However, the semi-sum of the maximum and minimum values attained in the
scan and the average of all points are very close, being the difference much smaller than perturbative uncertainty. For
simplicity we will use the average. Finally, we use the average of the individual (rescaled) fit uncertainties as the
global fit uncertainty. The results of this procedure at various orders is shown in Table.~\ref{tab:Calibration}.
\begin{table}[tbh!]
\centering
\caption{Results of the calibration for $\mbar_b = 4.2\,$GeV using the world average value of $\alpha_s$. Rows second
to fifth correspond to N$^3$LO, N$^2$LO, NLO and LO, respectively. The second and third column show the central value
and uncertainty due to the theory scan, while in the fourth and fifth column the fit and total uncertainties are
depicted.}
\label{tab:Calibration}        
\begin{tabular}{lcccc}
\hline \noalign{\smallskip}
order & central & perturbative & fit & total \\
\noalign{\smallskip}
\hline\noalign{\smallskip}
N$^3$LO & $4.731$ & $0.068$ & $0.071$ & $0.099$\\
N$^2$LO & $4.712$ & $0.085$ & $0.054$ & $0.100$\\
NLO     & $4.624$ & $0.114$ & $0.097$ & $0.149$\\
LO      & $4.503$ & $0.120$ & $0.198$ & $0.231$\\
\hline
\end{tabular}
\end{table}
\begin{figure}[tbh!]\centering
\includegraphics[width=0.85\linewidth]{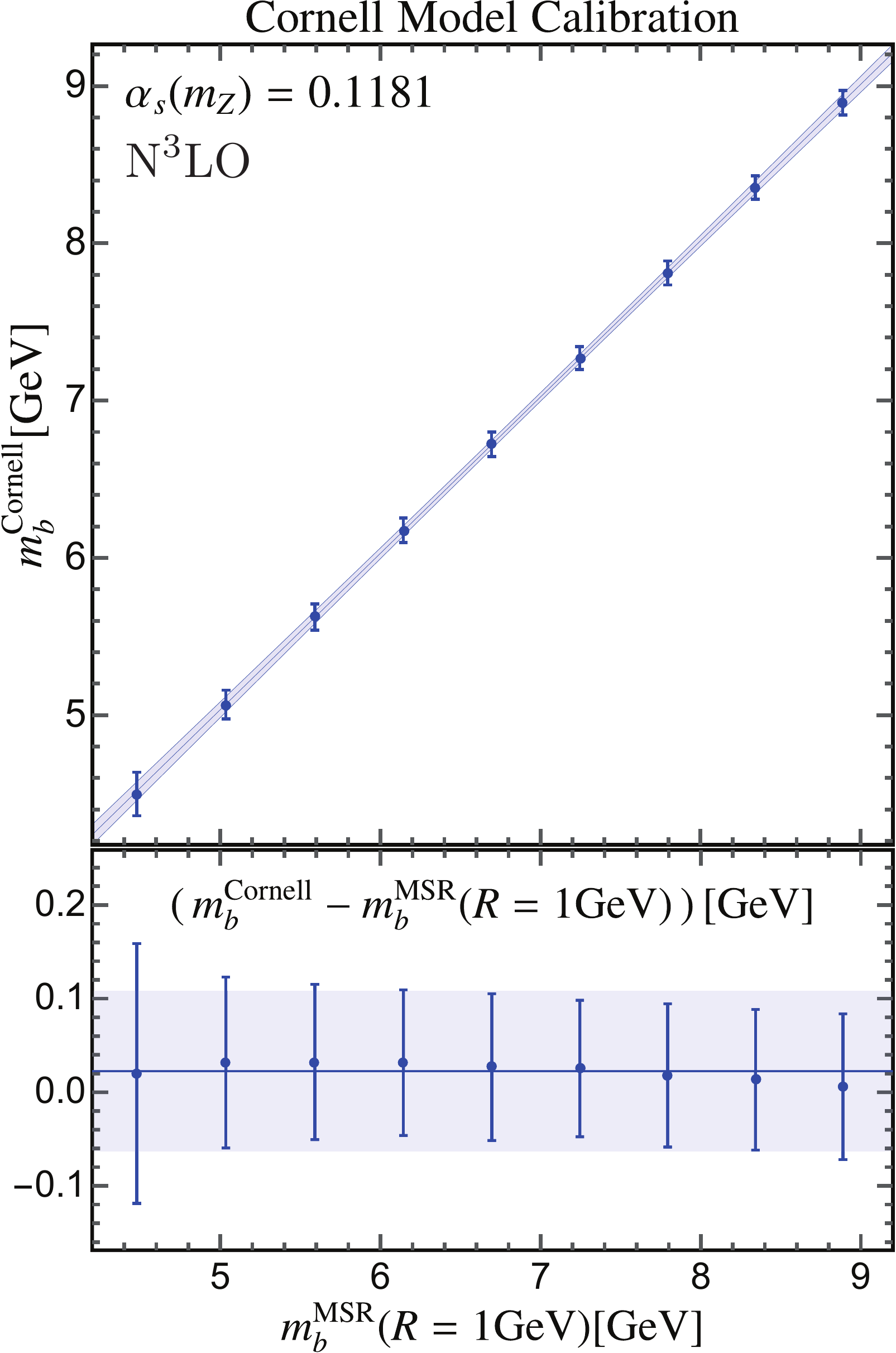}
\caption{Upper part of the plot\,: Dependence of the Cornell mass parameter with the MSR bottom mass at
$R=1\,$GeV at N$^3$LO, employing the world average value for $\alpha_s$. Lower part of the plot\,: Difference of the
Cornell and MSR masses as a function of the latter. The error bars show the quadratic sum of fit and perturbative
uncertainties. The blue line and faint blue band in the upper plot correspond to a linear fit to the points taken the individual
uncertainties as uncorrelated, while in the lower plot they show the weighted average of the differences and the regular
average of the uncertainties, respectively.}
\label{fig:Money}
\end{figure}

\begin{figure*}[tbh!]
\center
\subfigure[]
{\label{fig:orders}\includegraphics[width=0.4\textwidth]{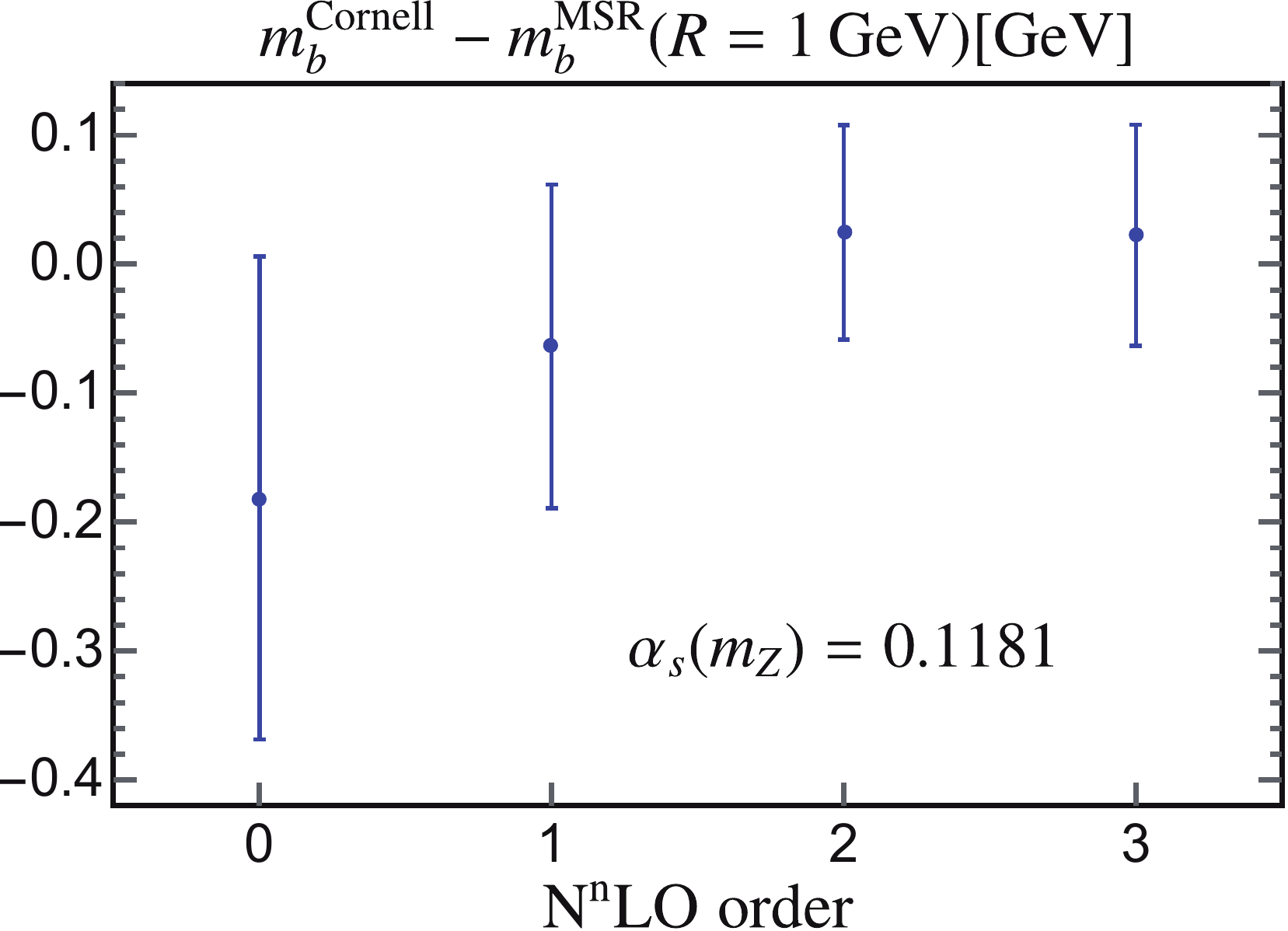}~~}
\subfigure[]
{\label{fig:alphaS}\includegraphics[width=0.41\textwidth]{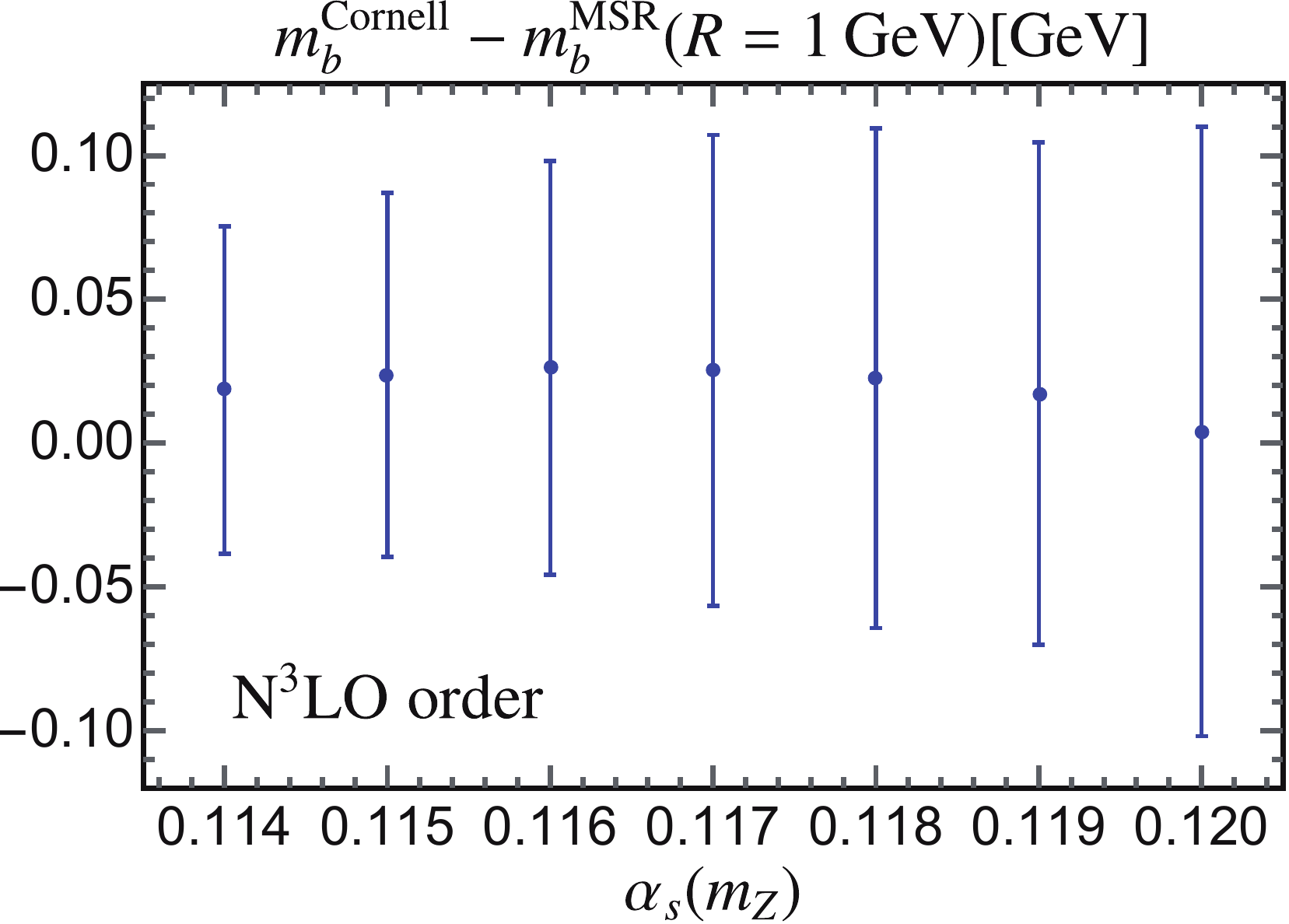} }
\caption{Dependence of the difference between the Cornell mass parameter and the MSR mass at the reference scale
$R=1\,$GeV on (a)~the perturbative order and (b)~the value of the strong coupling constant at the Z-boson mass.
Error bars include perturbative and fit uncertainties added in quadrature. Results are valid for any value of
the MSR mass between $4\,$GeV and $8\,$GeV.}
\label{fig:intersect}
\end{figure*}

We move to the (arguably) most interesting result of this article, the dependence of the Cornell mass parameter with
a short-distance QCD mass. Even though (for convenience) we have generated our QCD predictions in terms of the
$\MSb$ mass, this scheme is far from a kinematic mass, and should rather be thought as a coupling constant.
On the other hand the MSR mass for small values of $R$ is a kinetic mass free from renormalon ambiguities, and given the
results of Secs.~\ref{sec:StaticPot} and \ref{sec:lattice}, we will calibrate the Cornell mass against
\mbox{$m_b^{\rm MSR}(R={1\rm GeV})$}. As expected, we observe
a linear relation between these two parameters, and we find that the slope is very close to (and compatible with)
unity\,: $0.995\pm 0.026$ with an intersect compatible with zero within uncertainties\,: $0.05 \pm 0.19$.\,\footnote{For
simplicity we computed these uncertainties assuming uncorrelated errors for each mass value.} This pattern is
found for every value of $\alpha_s$ and also at various orders, but for simplicity we show the linear relation only at
N$^3$LO and for the world average in Fig.~\ref{fig:Money}. Rather than the intersect with zero, our final result for
the difference of the Cornell model mass parameter and the MSR mass is computed as the weighted average of the individual
$m_b^{\rm Cornell} - m_b^{\rm MSR}(1\,{\rm GeV})$ values, and for its incertitude we simply take the regular average
of the individual uncertainties, finding one of the most important results of this article\,:
\begin{align}\label{eq:main-result}
\!\!\!\!\!m_b^{\rm Cornell} \!= m_b^{\rm MSR}(R = \!1{\rm GeV}) + [\,0.023 \pm 0.086\,{\rm GeV}\,].
\end{align}
We believe this procedure is justified since individual uncertainties are almost
$100\,\%$ correlated. We have checked that the difference of the regular and weighted average of the individual central
values is of the order of half an MeV. In Fig.~\ref{fig:orders} we show the value of the difference between the
Cornell and MSR masses at various orders. We find nice convergence and decreasing error bars as the perturbative
information is increased, while at each order the result is compatible with zero. Similarly, Fig.~\ref{fig:alphaS}
makes it clear that for a wide range of values of the strong coupling constant the difference between the two masses
is compatible with zero, with larger uncertainties for higher $\alpha_s$ values.\footnote{We show in the horizontal axis
$\alpha_s^{(n_f=5)}(m_Z)$ rather than $\alpha_s^{(n_f=4)}(m_c)$, where the former is understood as the value that one
obtains running from the latter using threshold matching relations for $\mbar_b=4.2\,$GeV. }
At this point we have indirect evidence that appears to address the question raised at the end of
Sec.~\ref{sec:StaticPot}, namely whether the difference of the Cornell and QCD potentials in the UV can be absorbed in the
short-distance definition of the quark mass. The analysis carried out in this section, streamlined in the result shown in Eq.~\ref{eq:main-result}, seems to indicate this indeed happens, well within our uncertainties, if the MSR mass is
employed.\footnote{Similar conclusions could be drawn with other low-scale masses.}
It has the following physical interpretation\,:~the linear rising term of the Cornell potential incorporates in an effective way
medium-distance (perturbative) quantum fluctuations. For distances smaller than $1\,{\rm GeV}^{-1}$  Cornell sees only the
classical Coulomb-like potential, therefore we can assume that energy fluctuations above $1\,$GeV are absorbed into
the quark mass. This interpretation matches up with the definition of the MSR mass with $R=1\,$GeV, hence legitimating
our initial motivation. In other words, even though the Cornell and pQCD potentials disagree at short distances, the mass
spectrum is not overly determined by this region, and a suitable scheme choice makes the two approaches compatible at
the observable level. For short-distance-dominated observables, it could have been impossible to reconcile the two approaches.
A similar conclusion was reached in Ref.~\cite{Butenschoen:2016lpz}, where the MSR top quark mass with
$R=1\,$GeV was found to agree within uncertainties with the top quark parameter used  in \textsc{Pythia}.

We close this section by showing the dependence of the other two Cornell parameters, $\alpha_s^{\rm Cornell}$
and  $\sigma$ with the QCD quantities $\alpha_s$ and $m_b^{\rm MSR}$. Histograms for these two parameters look
less peaky than in Fig.~\ref{fig:histograms}, and therefore we trim only those values which appear less frequently than
$2\,\%$ of the highest peak. Figs.~\ref{fig:alpha-vs-mass}
and \ref{fig:alpha-vs-alpha} show this dependence for the $\alpha_s$ Cornell parameter as blue dots with error bands,
together with the QCD coupling, depicted as a red solid line, evaluated at the non-relativistic scale $\mu_{\rm NR}$. This
scale is chosen such that the argument of the logarithm in Eq.~\eqref{eq:NRlog}, once the pole mass is expressed in
terms of the MSR scheme, becomes one. Since our fit includes $n_p =1,\,2$ states, we take the average value $\bar n_p=3/2$.
Also we choose $R=\mu$ and determine $\mu_{\rm NR}$ by solving numerically the following equation\,:
\begin{figure*}[tbh!]
\center
\subfigure[]
{\label{fig:alpha-vs-mass}\includegraphics[width=0.4\textwidth]{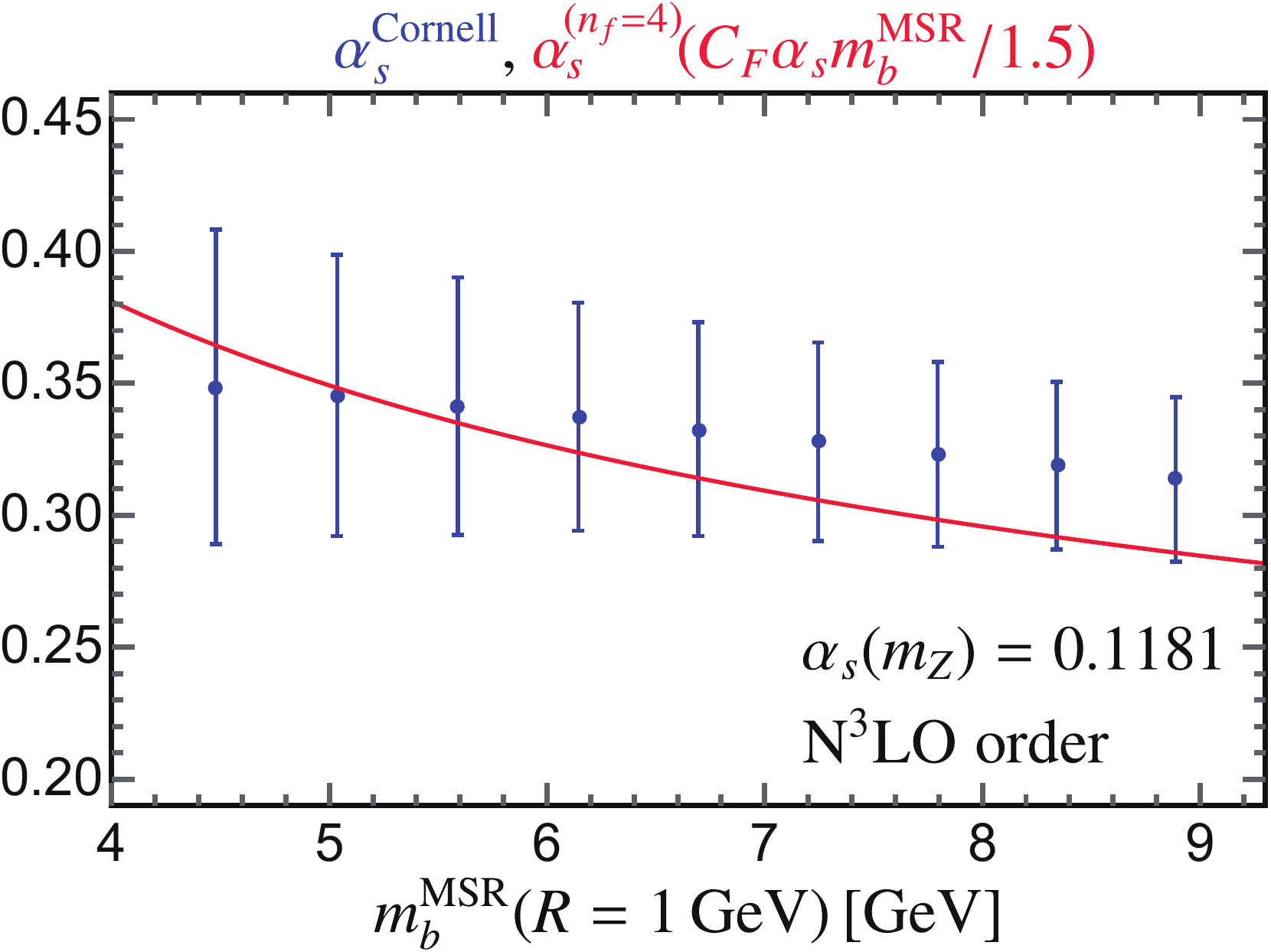}~~~~}
\subfigure[]
{\label{fig:alpha-vs-alpha}\includegraphics[width=0.41\textwidth]{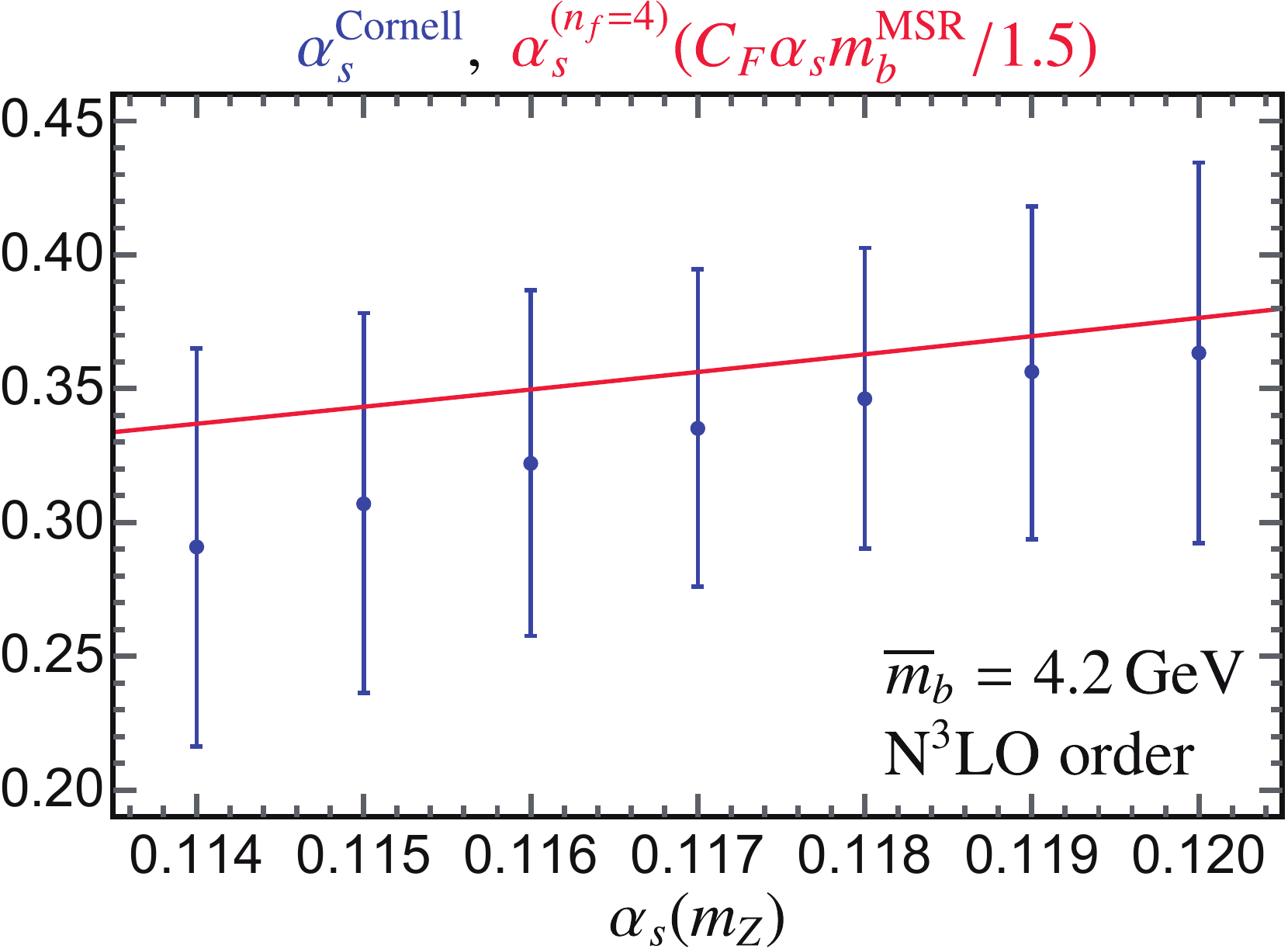} }
\vspace*{-0.3cm}
\subfigure[]
{\label{fig:sigma-vs-mass}\includegraphics[width=0.4\textwidth]{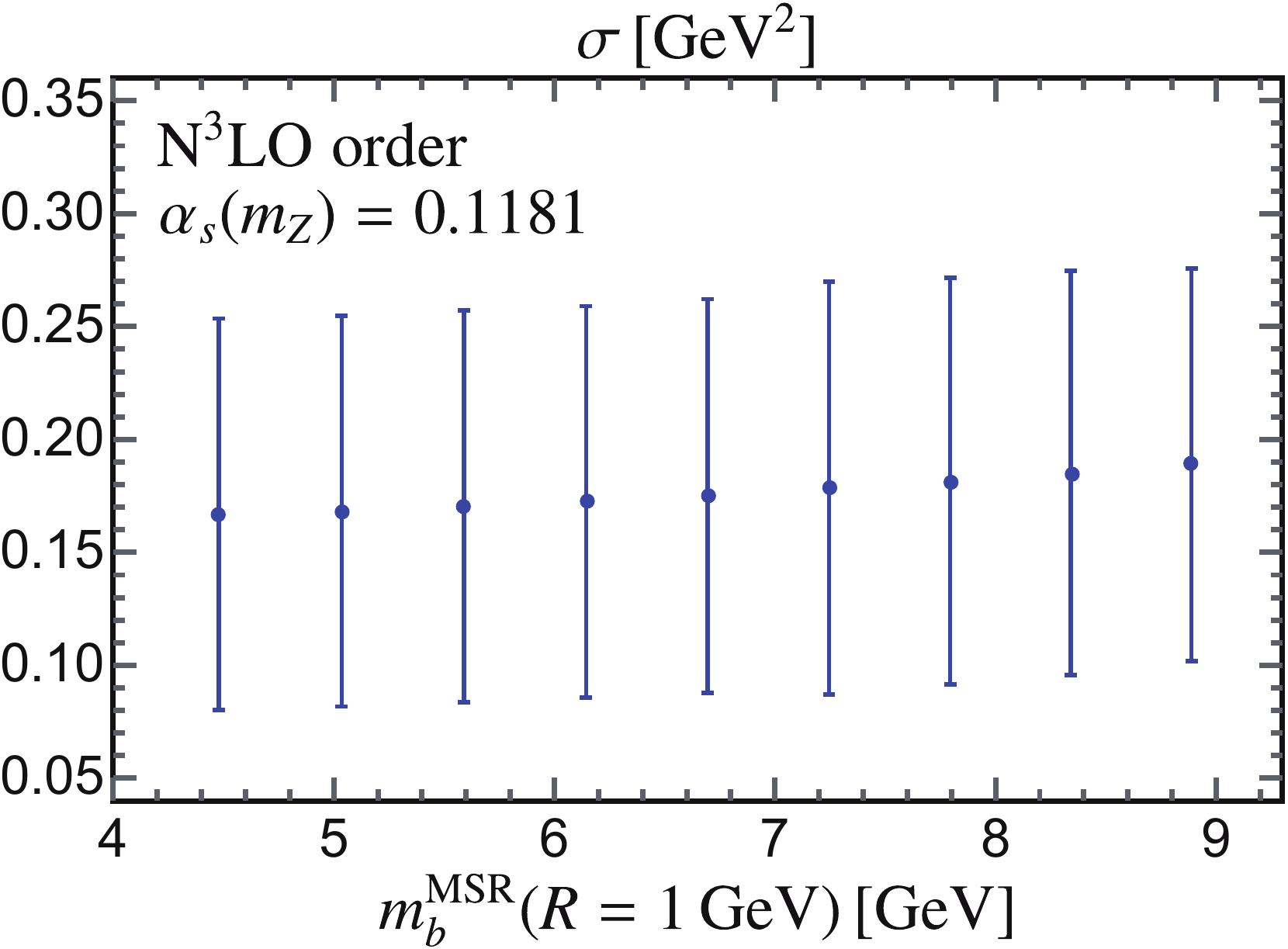}~~~~}
\subfigure[]
{\label{fig:sigma-vs-alpha}\includegraphics[width=0.41\textwidth]{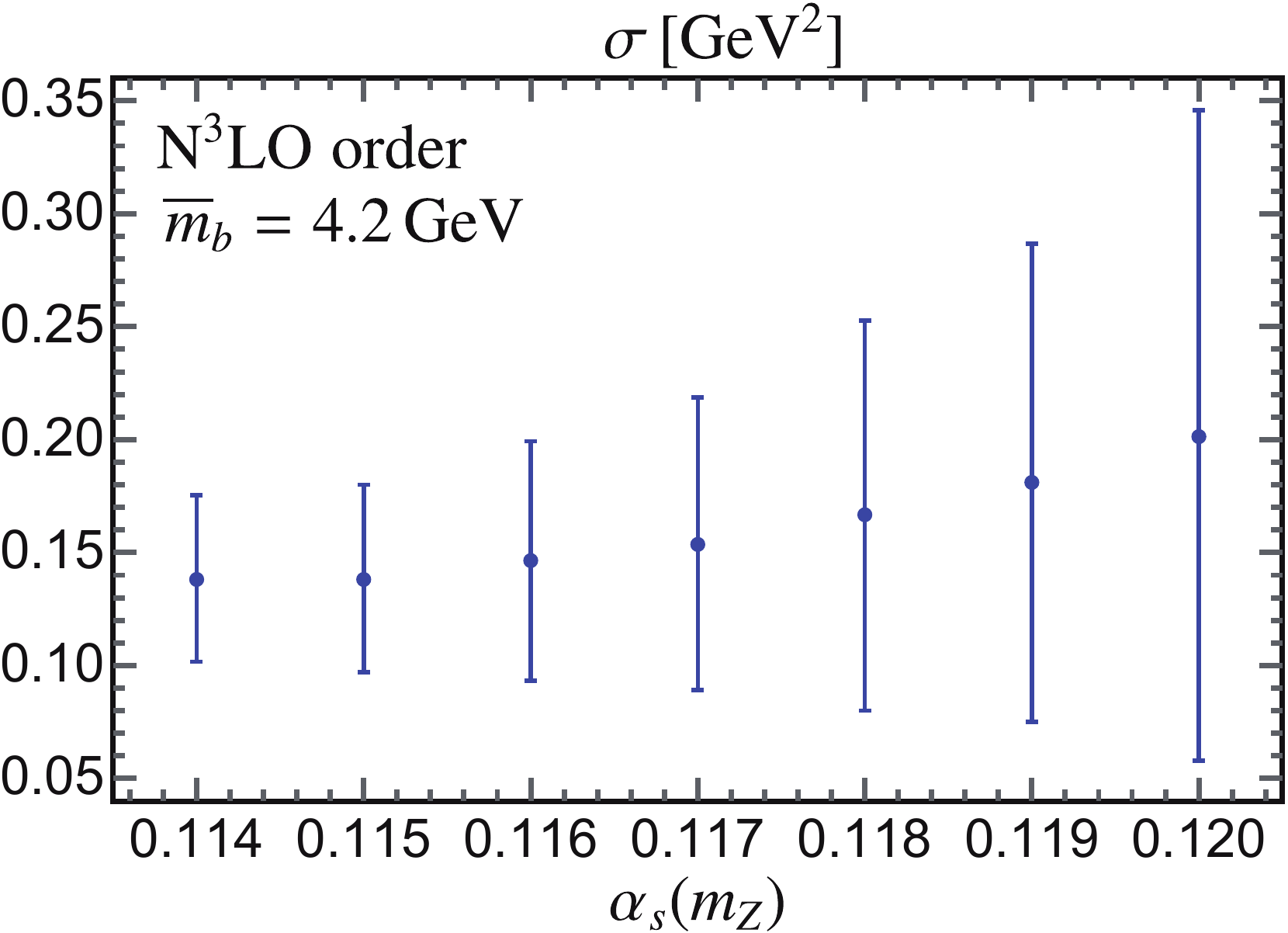} }
\caption{Dependence of the Cornell parameters $\alpha_s^{\rm Cornell}$ (upper two panels)
and $\sigma$ (lower two panels) with the MSR bottom mass (leftmost two panels) and the QCD coupling constant at the
Z-pole (rightmost two panels). The two upper plots also show, with a solid red line, the strong coupling constant evaluated
at a characteristic non-relativistic scale.}
\label{fig:alpha-sigma}
\end{figure*}
\begin{align}\label{eq:muNR}
C_F\alpha_s^{(n_f=4)}(\mu_{\rm NR})\,m_b^{\rm MSR}(\mu_{\rm NR}) = \frac{3}{2}\,\mu_{\rm NR}\,.
\end{align}
Eq.~\eqref{eq:muNR} is solved for different values of $\alpha_s^{(n_f=5)}(m_Z)$ and the bottom mass. A remarkable
agreement between $\alpha_s^{\rm Cornell}$ and $\alpha_s^{(n_f=4)}(\mu_{\rm NR})$ within perturbative uncertainties is
observed when scanning either over the bottom mass or the strong coupling constant reference value. Not only the order
of magnitude is the same, but also its dependence on $m_b$ and $\alpha_s(m_Z)$ follows the same pattern (although within
uncertainties the dependence on $m_b$ could be considered flat). If we take $n_p = 1(2)$ the value of $\alpha_s$
decreases (increases) roughly by $10\,\%$, leaving our conclusions unchanged. Needles to say, expecting a one-to-one
correspondence between these parameters is too na\"\i ve, but this simple analysis disfavors quark model analyses in
which $\alpha_s^{\rm Cornell}$ is assigned a QCD-like running, evaluated at the reduced mass of the $Q\overline Q$ pair.
However, given that the QCD running is logarithmic and depends only on the ratio of initial and final scales, as long as
the bottom pair reduced mass is taken as the boundary condition to obtain $\alpha_s^{\rm Cornell}$ for a charmonium
analysis (or vice-versa), no serious mistake is committed.

We show the dependence of $\sigma$ with the bottom quark mass and the strong coupling constant reference value in
Figs.~\ref{fig:sigma-vs-mass} and \ref{fig:sigma-vs-alpha}. The dependence of $\sigma$ with the former
shows exactly what one would expect from a static potential parameter\,: there is no mass dependence at all and
one could simply take the average of all points, obtaining $\sigma^{\rm Cal.} = 0.176 \pm 0.088\,$GeV$^2$.
This value compares well with the one obtained from the fit to the bottomonium experimental data in
Eq.~\eqref{eq:Cornell-Fit} \mbox{$\sigma^{\rm fit} = 0.207 \pm 0.011\,{\rm GeV}^2$}. It appears there is
some non-flat dependence on the value of $\alpha_s^{(n_f=5)}(m_Z)$, although drawing strong conclusions is
not possible given the
size of the uncertainties, which grow for larger values of $\alpha_s$. In any case one expects some dependence
of $\sigma$ with $\alpha_s$, since as argued in Refs.~\cite{Sumino:2003yp,Sumino:2004ht}, the linear rising term in the
static potential is of perturbative nature.

\section{Conclusions}\label{sec:conclusions}
In this article we have confronted a simple version of the Cornell model with both experimental data and QCD. This
model contains three terms, with one parameter associated to each one of them\,: a constituent quark mass, a linear
raising potential, and a Coulomb-like potential. We have solved numerically the Schr\"odinger equation for the Cornell
model using the Numerov algorithm, performing several checks to make sure the uncertainty of the approximation is much
smaller than any other uncertainty involved. The Cornell model includes the static approximation (solved exactly) plus
the leading non-relativistic correction, in the form of (angular-momentum-dependent) $1/m_Q^2$ suppressed potentials,
whereas in QCD we include as many non-relativistic corrections as necessary to achieve up to N$^3$LO accuracy.

As a warm-up exercise we have determined the Cornell potential parameters for the bottomonium and charmonium systems
from fits to the $8$ states with the lowest masses. We have confirmed that this simple version of the Cornell model fails to
predict states with larger mass, possibly because it does not include string-breaking effects, but gives reasonable
post-dictions of the masses of the states that enter the fit. NRQCD can predict the first $8$ bottomonium states and
the first $2$ charmonium states within perturbation theory, and therefore it is possible to calibrate the Cornell model
using bottomonium QCD predictions. Of course this only makes sense if the Cornell model is solely of perturbative nature.

The QCD static potential suffers form a $u=1/2$ renormalon, identical to that of the quark pole mass up to a factor
of $2$ and a sign. Therefore one can cancel the renormalon in the static energy (the sum of the
static potential and twice the pole quark mass) by expressing the quark mass in a short-distance scheme. Since we are
dealing with a threshold-like problem (bound states of a quark-antiquark pair), a low-scale short-distance mass should be
used. To keep logarithms small in the static potential and in the subtraction series relating
the pole and short-distance masses, it is compulsory to use a scheme with a tunable subtraction scale. The MSR mass
satisfies these two criteria, and has already been used in the context of quarkonium. Therefore we also
employ this scheme to make predictions for the static energy. Furthermore, we use \mbox{R-evolution}
to sum-up large renormalon-type logarithms in the static potential, whose argument is the ratio of the subtraction and
renormalization scales. In this way, we define an R-improved MSR static potential, which shows nice order-by-order
convergence, and has the same linear rising behavior as the Cornell model. In fact, if the renormalon subtraction
scale is chosen close to $1\,$GeV, the Cornell and R-improved static potential at N$^3$LO nicely agree for moderate
and large values of $r$. As expected, both potentials disagree in the UV regime probed by small values of $r$.
Since the linear raising potential is of perturbative nature and agrees with that of the
Cornell model, we conclude that all ingredients in this model for bottomonium are perturbative and a calibration
against QCD makes sense.

We have compared our R-improved MSR static potential with lattice simulations of the same quantity, performing a fit for
the strong coupling constant and the renormalon subtraction reference scale. After the fit, we find very nice agreement
with lattice QCD results for the entire set of $r$ values covered in the simulation, which includes distances as
large as $0.84\,$fm. To make this agreement possible, it is essential to use a ``profile function'' for the
renormalization scale $\mu$, which freezes to $1\,$GeV for values of the radius larger than $1\,$fm.

To calibrate the Cornell model we generate templates for the $8$ lightest bottomonium bound states with NRQCD. These
predictions depend on two scales\,: the renormalization scale $\mu$ and the renormalon subtraction scale $R$. These
take different values for the various bound states, but are varied in a correlated way. For a given value of the
bottom quark mass and strong coupling constant we generate templates at LO, NLO, N$^2$LO and N$^3$LO, which contain
the QCD prediction for the masses of the bound states in a two-dimensional grid of $\mu$ and $R$.
To avoid the d'Agostini bias, we adjust the Cornell model parameters to every entry on each template,
effectively scanning over the two renormalization scales. Since there are no uncertainties in the template that can be
used to construct a $\chi^2$ function, we use a regression algorithm to assign ``fit-incertitudes'' to the adjusted
parameters, normalizing the $\chi^2$ such that equals the number of degrees of freedom at its minimum.
On top of these, there are theoretical uncertainties from the scale scan. To figure these out, we first trim away values
of the Cornell model parameters that are found in the regression less often, discarding those that are less than
$8\,\%$ ($2\,\%$) less frequent than the most likely occurrence for $m_b^{\rm Cornell}$ ($\sigma$,
$\alpha_s^{\rm Cornell}$). After applying this procedure, we take the average of the remaining values as the central value for
the parameter, and half the sum of the maximum and minimum values as the perturbative uncertainty. We also take the
average of the regression errors as our final fit uncertainty.

We find an almost perfect linear relation between the Cornell and MSR masses with slope compatible with $1$ within
$0.19$ standard deviations, and if one chooses the MSR reference scale $R = 1\,$GeV the intersect of this relation is
compatible with zero within $0.26$ standard deviations. This pattern is replicated in a wide range of $\alpha_s$ values
and for all perturbative orders considered, and nicely complies with the agreement found between the Cornell and static
potentials for this particular choice of the reference scale. This seems to indicate that the difference of the
Cornell and Static potentials in the UV can be absorbed in the short-distance definition of the quark mass.
Our calibration exercise also reveals that the confining
parameter $\sigma$ does not depend on the value of the QCD quark bottom mass, as expected, but the precision of the
analysis cannot discard some dependence of this parameter with $\alpha_s(m_Z)$. On the other hand, the Coulomb-like
Cornell parameter $\alpha_s^{\rm Cornell}$ is found to agree within uncertainties with $\alpha_s^{(n_f=4)}(\mu_{\rm NR})$,
that is with the QCD strong coupling constant evaluated at a typical non-relativistic scale.

Our analysis could be extended in several directions\,: On the QCD side one could consider more refined predictions,
for instance using pNRQCD resummation as done in Ref.~\cite{Peset:2018jkf}, or including non-perturbative effects as
in Ref.~\cite{Rauh:2018vsv}. On the Cornell side, one could consider more sophisticated models, for example incorporating
string breaking effects or coupled channels. The calibration itself could be carried out for mixed $b\bar{c}$ states
in which both masses are varied such that their ratio remains constant. As for the comparison with lattice QCD results on the
static potential, the next step is studying perturbative uncertainties, order and dataset dependence, incorporating results
for other lattice spacings, and taking into account the lattice correlation matrices once they are known.

We close this article raising a concern. We have shown in Sec.~\ref{sec:Numerov} that the confining part of the Cornell
potential cannot be treated by any means as a perturbation of the Coulomb potential. We have also seen in
Sec.~\ref{sec:StaticPot} that the $\sigma$ term can be entirely described in perturbative QCD once the renormalon has
been canceled. Finally we have presented an analytic formula for the $Q\overline{Q}$ mass in Sec.~\ref{sec:massless},
which is based on solving the Schr\"odinger equation perturbatively around the lowest order result, that is, the Coulomb
potential, including both radiative and relativistic corrections. Therefore one could call into question
this perturbative treatment of the static QCD potential since it is known to fail for the not-so-different Cornell
potential. On the other hand, pNRQCD power counts perturbative and non-relativistic corrections on an equal footing, as
can be seen in Eq.~\eqref{eq:EXpole}, where there is no distinction between the former and the latter. One way of
shedding light on this apparent puzzle is though a numerical, exact, solution of the Schr\"odinger equation for the
QCD static potential. A step in this direction has been taken in Refs.~\cite{Pineda:2013lta,Peset:2018ria,Peset:2018jkf}.

\section*{Acknowledgments}
We thank X.~G.~Tormo for providing a computer readable file with the QCD static potential. We thank P.~Petreczky
for providing us with lattice results for the static potential.
This work has been partially funded by the Spanish MINECO {\it Ram\'on y  Cajal program} (RYC-2014-16022), the MECD
grants FPA2016-78645-P, FPA2016-77177-C2-2-P and the IFT {\it Centro de Excelencia Severo Ochoa} Program under Grant
SEV-2012-0249.
P.G.O. acknowledges the financial support from {\it Junta de Castilla y Le\'on} and European Regional Development
Funds (ERDF) under Contract no. SA041U16, and by Spanish MINECO's {\it Juan de la Cierva-Incorporaci\'on} program
with grant agreement no.~IJCI-2016-28525.

\appendix
\section{Numerical solution of the Cornell Potential}
\label{sec:NumerovAppendix}
In this work, the Numerov algorithm~\cite{Numerov1,Numerov2} (or Cowell's method) is used to solve 
numerically the Cornell potential. Such algorithm can be used to solve ordinary differential 
equations of second order in which the first derivative does not appear, and therefore is particularly 
well suited to solve the Schr\"odinger equation~\cite{koonin1998computational}. We follow the
specific implementation of Ref.~\cite{DomenechGarret:2008sr}, to which the reader is referred to for
a complete and pedagogical explanation. We outline the main points of the algorithm in what follows.
The method consist in numerically solving the ordinary differential equation in Eq.~\eqref{eq:ODE} in the range
$r\in[r_{\rm min},r_{\rm max}]$, discretized in $N + 1$ nodes with step-size $h=(r_{\rm max}-r_{\rm min})/N$.
This implies $r_n = r_{\rm min} + n\,h$ with $0\le n \le N$ and $r_{n+1} = r_n + h$.
Note that one cannot include the point $r = 0$ as starting point for a numerical solution since the potential
is singular at the origin. Nevertheless we know that for very small $r$ the solution behaves as in
Eq.~\eqref{eq:boundary1}. The reduced wave function $u(r)$ and the kernel $k(r)$ can be discretized as well,
and we use the notation $u_n=u(r_n)$, $k_n=k(r_n)$. One can Taylor expand $u^{(m)}_{n+1}$ and $u^{(m)}_{n-1}$ 
around $r = r_n$, where the superscript in parentheses means that we are taking the $m$-th derivative\,:
\begin{align}
u^{(m)}_{n\pm1} = \sum_{i = 0}\frac{(\pm1)^ih^i}{i\,!}\,u_n^{(i+m)}\,.
\end{align}
Taking the sum and the difference of the plus and minus equations we isolate even and odd terms,
respectively\,:
\begin{align}
u^{(m)}_{n+1} + u^{(m)}_{n-1} & = 2\sum_{i = 0}\frac{h^{(2i)}}{(2i)\,!}\,u_n^{(2i+m)}\,,\label{eq:even}\\
u^{(m)}_{n+1} - u^{(m)}_{n-1} & = 2\sum_{i = 0}\frac{h^{(2i+1)}}{(2\,i+1)\,!}\,u_n^{(2i+1+m)}\,.\label{eq:odd}
\end{align}
If one uses Eq.~\eqref{eq:even} with $m = 0,\,2$ and truncates at $\mathcal{O}(h^{6,4})$, respectively, obtains
\begin{align}\label{eq:par}
& u_{n+1} + u_{n-1} = u_n\,(2 - h^2\,k_n) + \frac{h^4}{12}u_n^{(4)}+\mathcal{O}(h^6)\,, \\
& k_{n-1}\,u_{n-1} + k_{n+1}\,u_{n+1} = 2\,k_n\, u_n - h^2\,u^{(4)}_n+\mathcal{O}(h^4)\,,\nonumber
\end{align}
where we have used $u_n^{(2)} = -\,k_n\,u_n$. Solving for $u_{n+1}$ from Eqs.~\eqref{eq:par} and shifting
$n\to n-1$ we obtain the following recursive formula,
\begin{align}
u_n=\frac{2\,\bigl(1-\frac{5h^2}{12}k_{n-1}\bigr)u_{n-1}-\bigl(1+\frac{h^2}{12}k_{n-2}\bigr)u_{n-2}}
{1+\frac{h^2}{12}k_n}+{\cal O}(h^6)\,.
\end{align}
The above equation gives us the wave function at any $r_n$ from the two initial values $u_0$ and $u_1$
[\,obtained from Eq.~\eqref{eq:boundary1}\,], so it builds the wave function in the forward direction
(referred to as $u_{\rm in}$). Isolating $u_{n-1}$ in Eqs.~\eqref{eq:par} we find another recursive
formula that builds the wave function in the backward direction,
\begin{align}
u_{n-1}=\frac{2\,\bigl(1-\frac{5h^2}{12}k_n\bigr)u_n-\bigl(1+\frac{h^2}{12}k_{n+1}\bigr)u_{n+1}}
{1+\frac{h^2}{12}k_{n-1}}+{\cal O}(h^6)\,,
\end{align}
which implies the knowledge of $u_N$ and $u_{N-1}$, extracted from Eq.~\eqref{eq:boundary2}. That solution
will be dubbed $u_{\rm out}$. Our method also needs the derivative of the reduced wave function, which
can be obtained truncating Eq.~\eqref{eq:odd} at order $\mathcal{O}(h^{5,3})$ with $m = 0,\,2$,
respectively\,:
\begin{align}\label{eq:impar}
& u_{n+1} - u_{n-1} = 2\,h\,u^\prime_n + \frac{h^3}{3}u_n^{(3)}+\mathcal{O}(h^5)\,, \\
& k_{n-1}\,u_{n-1} - k_{n+1}\,u_{n+1} = 2\,h\, u_n^{(3)} + \mathcal{O}(h^3)\,.\nonumber
\end{align}
One can solve for $u^\prime_n$ in the above equations to find\,:
\begin{align}
u^\prime_n\,=\, &\frac{1}{2h}\biggl[\biggl(1+\frac{h^2}{6}k_{n+1}\biggr)u_{n+1}-
\biggl(1+\frac{h^2}{6}k_{n-1}\biggr)u_{n-1}\biggr]\nonumber\\
&\,+\,{\cal O}(h^5)\,.
\end{align}
The energy eigenvalues and eigenstates are obtained when the forward and backward solutions and their
derivatives match at an intermediate position $r_c$. Since at this point the normalization of both inward and
outward solutions is arbitrary, one can simply impose the matching on the logarithmic derivative
\begin{align}\label{eq:logarderiv}
\bigg[\frac{u^\prime_{\rm in}}{u_{\rm in}}\bigg]_{r_c}=\bigg[\frac{u^\prime_{\rm out}}{u_{\rm out}}\bigg]_{r_c}\,.
\end{align}
The point $r_c$ will be chosen as the distance where $k(r_c)=0$, that is
\begin{equation}
E_{n\ell}=V(r_c)+\frac{\ell\,(\ell+1)}{m_Q\, r_c^2}\,,
\end{equation}
which marks the distance where the classically forbidden region begins and the asymptotic exponential behavior
is expected to start dominating. It depends on $\ell$ and $n$.

The condition in Eq.~\eqref{eq:logarderiv} as well as $r_c$ depend on the energy. For a given value of $\ell$ we use
the classic bisection method to find the root of Eq.~\eqref{eq:logarderiv}, requiring
an accuracy of $1\,$eV, but limiting the number of steps to $1\,000$. Once the ground state is found, the same
method can be used to find the energy of excited states. Iterating the process for various values of the
orbital angular momentum one figures out the complete (discrete) spectrum of the static Cornell potential.

Regarding the calculation of the leading relativistic corrections of the Cornell model, the radial integration 
for the $V_{\rm SS}^{\rm OGE}$ operator can be performed analytically, and we find\,:
\begin{align}\label{ec:spinspinOGE}
&\langle\, n,\ell,s\,|\,V_{\rm SS}^{\rm OGE}\, |\,n,\ell,s\,\rangle =\\
& \frac{8\alpha_s^{\rm Cornell}}{9m_Q^2}
\langle\vec S_1\cdot \vec S_2\rangle \lim_{r\to0}\frac{|u_{n\ell}(r)|^2}{r^2}\,\delta_{\ell,0}\,.\nonumber
\end{align}

Since $u_{n\ell}(r\to0)=A_{n,\ell}\, r^{\ell+1}$ this matrix element is non-zero only if $\ell = 0$, 
and the limit is simply $|A_{n,0}|^2$.  In practice we compute
$A_{n,0}$ from the value of the reduced wave function at the fist node\,: $A_{n,0} = u_{n0}(r_{\rm min})/r_{\rm min}$. For
the rest of the operators in Eq.~\eqref{eq:subleading} we perform a numerical integration between $0$ and $r_{\rm max}$.
For the integration we use a Fortran~90 implementation of the QUADPACK package~\cite{quadpack}.
We reconstruct the function $u_{n\ell}(r)$ with an interpolation using the values computed in the nodes by the
Numerov method between $r_{\rm min}$ and $r_{\rm max}$, while for $r < r_{\rm min}$ we assume it follows the same
patter as the first two nodes, $u_{n\ell}(r\to0)= u_{n\ell}(r_{\rm min})\,r^{\ell+1}/r_{\rm min}^{\ell+1}$. For the 
interpolation we use a Fortran 2008 implementation of the algorithm described in \cite{DeBoor:1428148}.

\bibliography{NRQCD}
\bibliographystyle{spphys}

\end{document}